\documentclass[11pt]{article}

\usepackage{amsmath}
\usepackage{amssymb}
\usepackage{array}
\usepackage{booktabs}
\usepackage[skip=4pt,labelfont=it,font=small]{caption}
\usepackage{subcaption}
\usepackage{float}
\usepackage{fullpage}
\usepackage{graphicx}
\graphicspath{{./figures/}}
\usepackage[version=4]{mhchem}
\usepackage[frozencache,cachedir=minted-cache]{minted}
\setminted{xleftmargin=5mm, xrightmargin=5mm}
\usepackage{multirow}
\usepackage{tocloft}   % must be before parskip, https://tex.stackexchange.com/questions/395779
\usepackage{parskip}
\usepackage{setspace}
\usepackage{siunitx}
\usepackage{titlesec}
\usepackage{upgreek}
\usepackage{xurl}
\usepackage{xcolor}
\usepackage{pdfpages}
\usepackage[hidelinks]{hyperref}
\usepackage[capitalise,noabbrev]{cleveref}
\DeclareSIUnit{\molar}{\textsc{m}}
\DeclareSIUnit{\ppm}{ppm}
\DeclareSIUnit{\gauss}{G}
\newcommand*{\Poise}{POISE}
\newcommand*{\SInf}{\textit{Supporting Information}}
\newcommand*{\carbon}{\ce{^{13}C}}
\newcommand*{\proton}{\ce{^{1}H}}

\usepackage[style=chem-acs,biblabel=dot,subentry=true]{biblatex}
\DeclareCiteCommand{\citenum}{}{\printfield{labelnumber}}{}{} % use "ref.\ (\citenum{REF_NAME})" to get "ref. (8)"
\begin{document}

% Title page {{{1
\begin{center}  % Authors, etc.
    \Large \textbf{On-the-fly, Sample-tailored Optimisation of NMR Experiments}

    \vspace{0.5cm}

    \large Jonathan R. J. Yong and Mohammadali Foroozandeh*

    \normalsize \textit{Chemistry Research Laboratory, Department of Chemistry, University of Oxford, \\ 12 Mansfield Road, Oxford, OX1 3TA, U.K.}

    * \texttt{mohammadali.foroozandeh@chem.ox.ac.uk}
\end{center} 
\vspace{0.5cm}
\begin{figure}
    \centering
    \begin{minipage}[c]{5.35cm}
        \includegraphics[width=\textwidth]{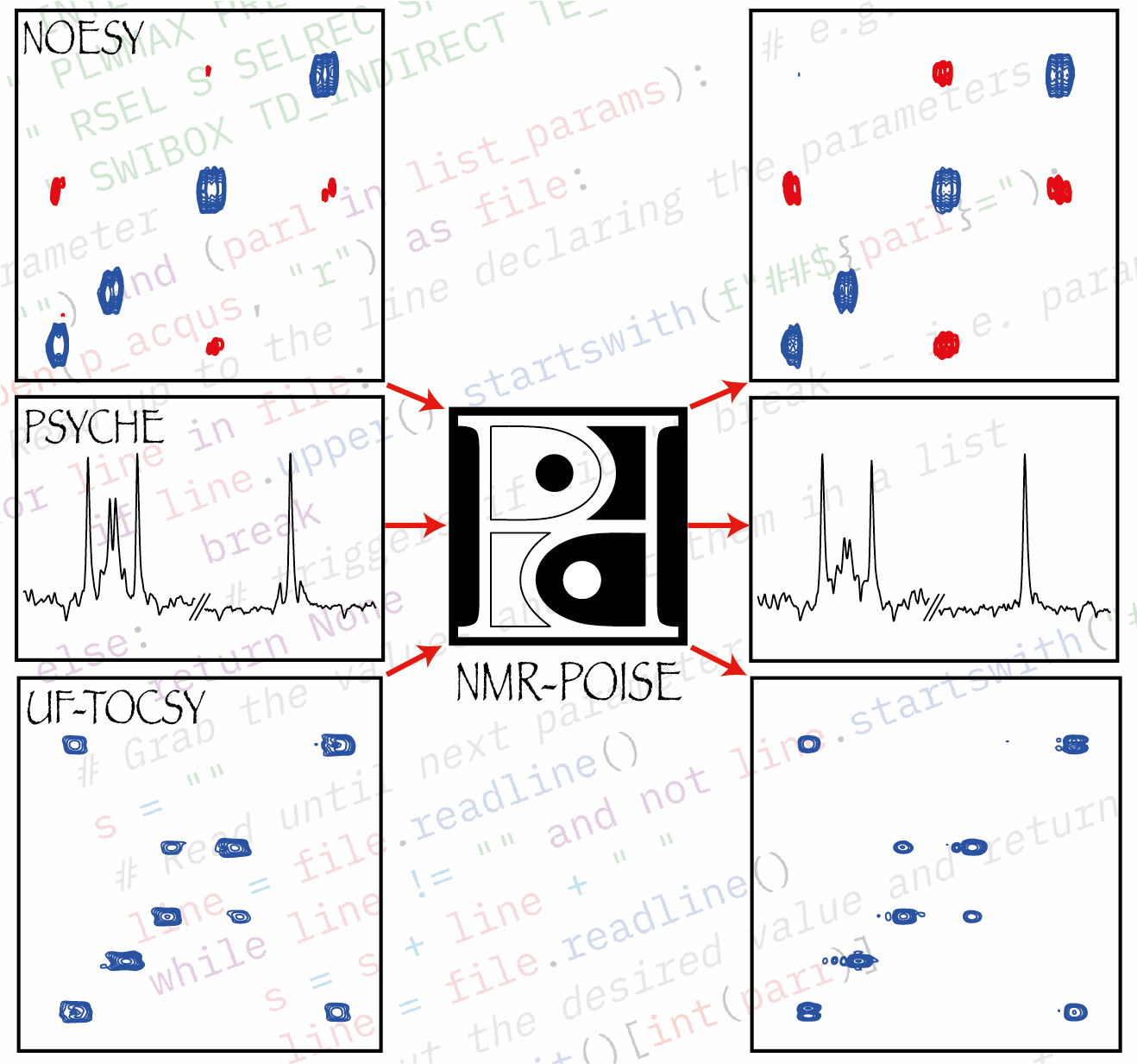}
    \end{minipage}
    \hspace{0.2cm}
    \begin{minipage}[c]{9cm}
        \caption*{
            NMR parameters must often be adjusted for maximum performance on different samples or instruments.
            We show that, using our implementation of POISE (Parameter Optimisation by Iterative Spectral Evaluation), a wide range of experiments may be optimised on the fly in a fully automated fashion, using only a few spectrum acquisitions.
        }
    \end{minipage}
     \label{toc_graphic}
\end{figure}
\begin{abstract}
    NMR experiments, indispensable to chemists in many areas of research, are often run with generic, unoptimised experimental parameters.
    This approach makes robust and automated acquisition on different samples and instruments extremely challenging.
    Here, we introduce NMR-POISE (Parameter Optimisation by Iterative Spectral Evaluation), the first demonstration of on-the-fly, sample-tailored, and fully automated optimisation of a wide range of NMR experiments.
    We illustrate how POISE maximises spectral sensitivity and quality with a diverse set of 1D and 2D examples, ranging from HSQC and NOESY experiments to ultrafast and pure shift techniques.
    Our Python implementation of POISE has an interface integrated into Bruker’s TopSpin software, one of the most widely used platforms for NMR acquisition and automation, allowing NMR optimisations to be run without direct user supervision.
    We predict that POISE will find widespread usage in academia and industry, where sample-specific and automated experiment optimisation is mandatory.
\end{abstract}
\begin{center} % Keywords
    \small \textbf{Keywords:} NMR spectroscopy $\cdot$ Analytical methods $\cdot$ NMR-POISE $\cdot$ Optimisation $\cdot$ Automation
\end{center}
% }}}1
% Introduction {{{1
Nuclear magnetic resonance (NMR) spectroscopy is firmly established as one of the most important analytical methods in chemistry\autocite{Cavanagh2007,Claridge2016}.
Although many NMR experiments are designed to be as general as possible, the sheer diversity of molecular structure means that experimental parameters must often be varied in order to achieve optimal performance for a particular sample.
Using poorly set parameters can lead to inferior spectra, as measured by (for example) the attainable signal-to-noise ratio (SNR) per unit time, or the presence of artefacts arising from unwanted coherences.
A typical way around this is to use a ``compromise'' value, chosen to provide acceptable (albeit suboptimal) performance across a range of samples.

In contrast, the direct optimisation of experimental parameters on a per-sample basis is rarely carried out.
When it is, it typically relies on trial-and-error or otherwise inefficient methods, leading to highly tedious and time-consuming processes.
For example, the current best routines (e.g. \texttt{popt} in Bruker's TopSpin software) rely on systematic evaluation of parameter values using an exhaustive grid search.
A more desirable approach would be to utilise an optimisation algorithm\autocite{Nocedal2006} which iteratively monitors the quality of a series of spectra acquired with different parameter values, eventually converging to optimal values for the sample under study.
Although there exist isolated examples of such iterative optimisations in laser\autocite{Bardeen1997CPL}, electron spin resonance\autocite{Goodwin2018JMR}, and nuclear quadrupole resonance\autocite{Monea2020JMR} spectroscopies, such a procedure has only been used in NMR by Emsley and co-workers in the design of dipolar decoupling pulses\autocite{DePaepe2003CPL,Elena2004CPL,Salager2010CPL}.
Note that the proposed method is distinct from other model-based ``optimisations'', e.g.\ for sparse sampling in multidimensional NMR\autocite{Eghbalnia2005JACS,Hansen2016ACIE} or relaxation measurements\autocite{Song2018JMR}, where data from multiple acquisitions are \textit{aggregated} until a certain threshold, such as a sufficient confidence in peak locations or extracted parameters, is reached.

Despite its apparent simplicity, iterative optimisation of NMR experiments clearly enjoys little popularity.
This can perhaps be attributed to a perception that such a process would require too much time, as well as the challenge of implementing it in a manner that can be suitably generalised for multiple disparate contexts.

\begin{figure}
    % I did this in Inkscape.
    \centering
    \includegraphics[width=8cm]{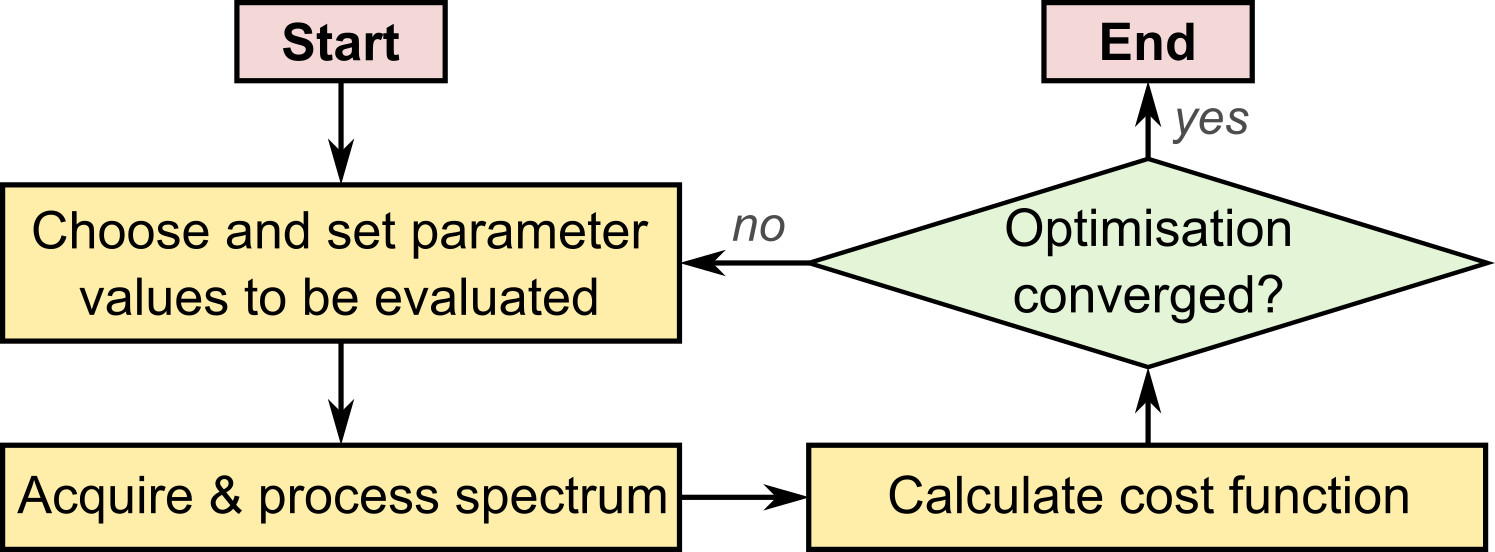}
    \caption{Flowchart depicting the main steps in a POISE optimisation.}
    \label{fig:flowchart}
\end{figure}

In this work, we dispel these notions by introducing NMR-POISE (\textit{\textbf{P}arameter \textbf{O}ptimisation by \textbf{I}terative \textbf{S}pectral \textbf{E}valuation}), a method which enables the efficient optimisation of a wide range of NMR experimental parameters and yields demonstrably better spectra in only a matter of minutes.
NMR-POISE is provided as a free Python package, with a powerful user interface allowing optimisations to be run directly from the command line of Bruker's TopSpin software.
This means that optimisations can proceed in a fully automated fashion without any user input, a benefit which is especially relevant in high-throughput settings or when non-expert users are present.
The optimisation process (\cref{fig:flowchart}) requires a \textit{cost function}, a user-defined function which measures how good a spectrum is; POISE provides a number of built-in cost functions for typical use cases, e.g.\ for maximising or minimising signal intensity.
Currently, three optimisation algorithms are provided:
Py-BOBYQA, a trust-region interpolation method\autocite{Cartis2019ACMTMS} (enabled by default);
the Nelder--Mead (NM) simplex method\autocite{Nelder1965TCJ};
and the multidimensional search (MDS) method\autocite{DennisJr1991SIAMJO}.

% }}}1
% {{{ Experimental
\textbf{Experimental section}

All NMR spectra in this work were acquired on Bruker AV III or NEO spectrometers, with field strengths ranging from 400 to \SI{700}{\MHz}.
POISE was executed on the spectrometer terminals using Python 3.8.1, 3.8.5, and 3.9.2 on Windows 10.
In general, POISE can be run on any Bruker spectrometer; the minimum version of Python needed is 3.6.
Experimental details of individual optimisations are given in the \SInf{}.

% }}}
% Case study 1: p1 {{{1
\textbf{Results and discussion}

The first example of a POISE control optimisation is the calibration of a \ang{90} \proton{} pulse, a quantity which is fundamental to practically every NMR experiment\autocite{Keifer1999CMR}.
In a simple \proton{} pulse-acquire experiment, a perfect \ang{360} pulse would generate no detectable magnetisation.
The intensity of the resulting signal can thus be used as a cost function which we seek to minimise.

\begin{figure}
    % ./figures/p1opt_graph.py
    \centering
    \includegraphics[width=8cm]{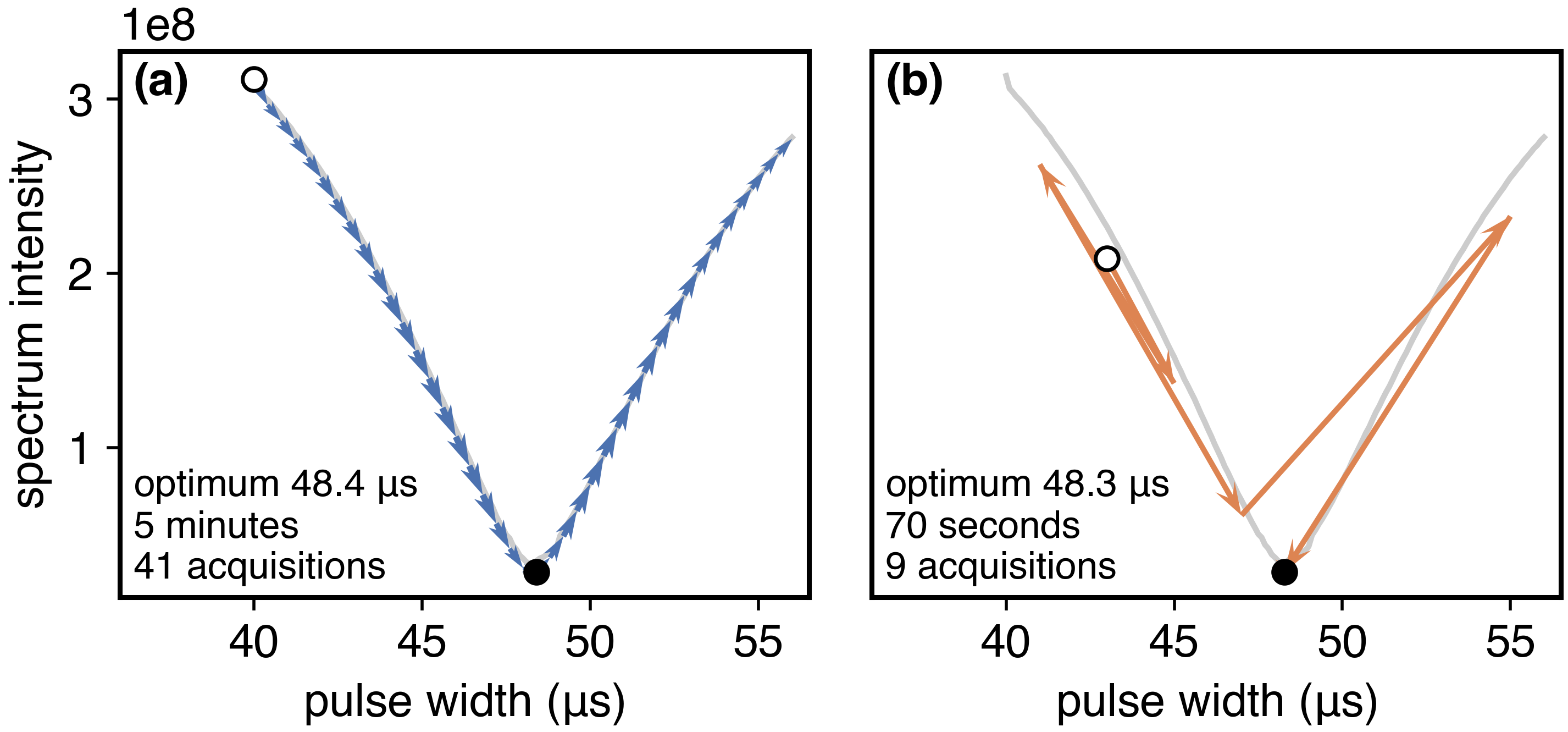}
    {\phantomsubcaption\label{fig:p1opt_graph_popt}}
    {\phantomsubcaption\label{fig:p1opt_graph_poise_bobyqa}}
    \caption{
        Optimisation trajectory for pulse width calibration on a sample of \SI{45}{\milli\molar} ferulic-acid in DMSO-$d_6$ using:
        \textbf{\subref{fig:p1opt_graph_popt}} TopSpin’s \texttt{popt}, which uses a grid search; and
        \textbf{\subref{fig:p1opt_graph_poise_bobyqa}} POISE, using the BOBYQA algorithm.
        The grey line indicates the ``true'' objective function.
        Each coloured arrow indicates a single function evaluation, unfilled dots represent the initial point, and filled dots indicate optima at which the cost function is minimised.
    }
    \label{fig:p1opt_graph}
\end{figure}

The benefits of direct optimisation, as compared to the grid search of \texttt{popt}, can be easily appreciated upon consideration of \cref{fig:p1opt_graph}.
By manually acquiring an array of spectra with different pulse widths, we first ascertained that the spectral intensity was minimised at \SI{48.4}{\us} (\cref{fig:p1scan}); knowing this allows us to evaluate the accuracy of the optimisation routines.
\texttt{popt} faithfully reproduces this optimum, but must evaluate a total of 41 spectra to search the entire parameter space, which takes 5 minutes (\cref{fig:p1opt_graph_popt}).
Although the search may in principle be terminated immediately after locating the minimum, this requires constant monitoring by the user, since the optimum point is not known \textit{a priori}.
On the other hand, POISE quickly homes in on the minimum without any user intervention: for example, using the BOBYQA algorithm, it requires only 70 seconds and 9 spectrum acquisitions to identify an accurate minimum of \SI{48.3}{\us}, even when given a suboptimal starting point of \SI{43}{\us} (\cref{fig:p1opt_graph_poise_bobyqa}).
Convergence is even faster when a good initial point is provided (\cref{tbl:p1init48,tbl:p1init43,tbl:p1init53}).
As a ``rapid'' alternative for pulse width calibration, the Bruker \texttt{pulsecal} routine (which uses a nutation experiment\autocite{Wu2005JMR}) completes in 37 seconds; however, it underestimates the \ang{360} null by almost \SI{2}{\us} (\cref{tbl:p1init48}).
For routine 1D \proton{} NMR spectra, the exact \ang{90} pulse width is often not of interest.
However, by instead searching for a \textit{maximum} in the spectral intensity, POISE can likewise be used to identify the pulse width corresponding to the Ernst angle\autocite{Ernst1966RSI} (\cref{sec:si_ernst}), which yields the maximal sensitivity for a specific repetition time.

\begin{figure}
    % Ali did this in Illustrator.
    \centering
    \includegraphics[width=8cm]{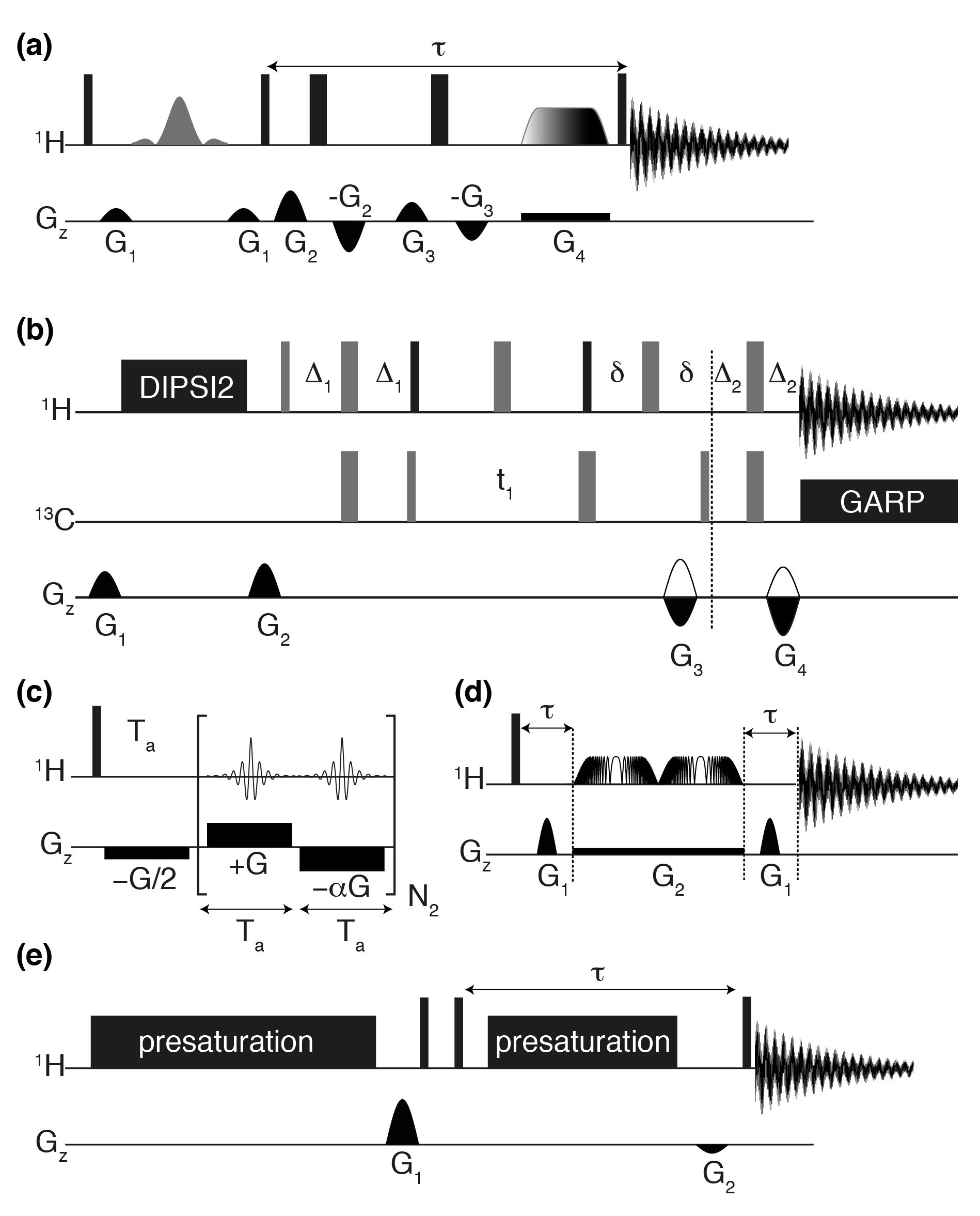}
    {\phantomsubcaption\label{fig:pprogs_noe1d}}
    {\phantomsubcaption\label{fig:pprogs_asaphsqc}}
    {\phantomsubcaption\label{fig:pprogs_epsi}}
    {\phantomsubcaption\label{fig:pprogs_psyche}}
    {\phantomsubcaption\label{fig:pprogs_solvsupp}}
    \caption{
        Pulse sequences used for POISE optimisations;
        \textbf{\subref{fig:pprogs_noe1d}} 1D selective NOESY,
        \textbf{\subref{fig:pprogs_asaphsqc}} ASAP-HSQC,\autocite{SchulzeSunninghausen2014JACS}
        \textbf{\subref{fig:pprogs_epsi}} echo-planar spectroscopic imaging (EPSI) acquisition scheme used in ultrafast NMR\autocite{Frydman2003JACS,Gouilleux2018ARNMRS}, and
        \textbf{\subref{fig:pprogs_psyche}} spin echo sequence used for evaluation of PSYCHE\autocite{Foroozandeh2014ACIE} refocusing element,
        \textbf{\subref{fig:pprogs_solvsupp}} 1D NOESY sequence with presaturation, used for solvent suppression.\autocite{Mckay2011CMR}
        Details of the optimisation processes, along with complete lists of experimental parameters used, can be found in the \SInf{}.
    }
    \label{fig:pprogs}
\end{figure}

\begin{figure}
    % ../figures/noesy_spec.py
    \centering
    \includegraphics[width=8cm]{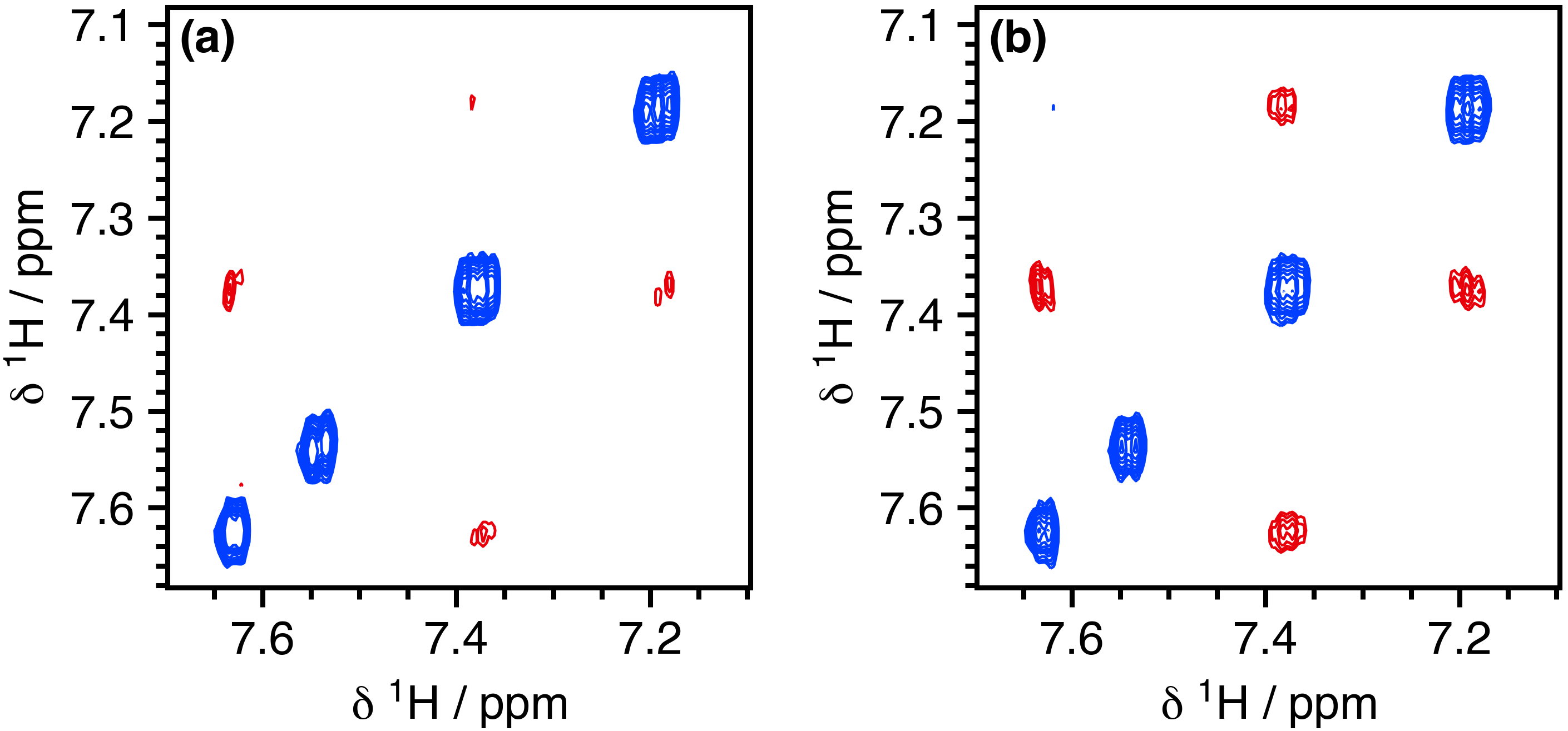}
    {\phantomsubcaption\label{fig:noesy_spec_before_opt}}
    {\phantomsubcaption\label{fig:noesy_spec_after_opt}}
    \caption{
        2D NOESY spectra of a sample of \SI{120}{\milli\molar} 3-fluorophenylboronic acid in DMSO-$d_6$, obtained 
        \textbf{\subref{fig:noesy_spec_before_opt}} before and
        \textbf{\subref{fig:noesy_spec_after_opt}} after optimising the mixing time.
        Optimisation leads to an approximate $2\times$ improvement in crosspeak sensitivities.
        In order to accomplish this sensitivity improvement, one would need to acquire the unoptimized experiment with four times as many transients.
    }
    \label{fig:noesy_spec}
\end{figure}

% }}}1
% Case study 2: NOE mixing time {{{1
As long as a suitable cost function can be defined, POISE is capable of providing rapid results with no compromise in accuracy.
Our POISE interface is thus designed to grant users great flexibility in defining their own cost functions, which allows the optimisation of a wide range of parameters which simply cannot be carried out using currently available tools.
One such example is the mixing time in nuclear Overhauser effect spectroscopy (NOESY) experiments, which typically varies from hundreds of milliseconds to several seconds.
By finding the optimal mixing time for a given sample, the sensitivity of the resulting NOESY spectrum can be maximised.
Here, we perform this optimisation using a selective 1D NOESY experiment (\cref{fig:pprogs_noe1d}), using the intensity of the spectrum excluding the selectively irradiated peak as the cost function.
Starting from an initial value of \SI{0.5}{\s}, a typical value for many organic molecules, POISE is capable of finding the optimum mixing time for the sample under study (\SI{3.5}{\s}).
This optimisation converges in around 5 minutes (\cref{tbl:noe}), leading to an approximate twofold increase in NOESY crosspeak intensity for this particular case (\cref{fig:noesy_spec}).

Where possible, POISE optimisations should be performed using 1D pulse sequences, in order to minimise the optimisation duration.
The NOE mixing time optimisation shown here, for example, takes only a fraction of the time needed for a 2D NOESY.
With more dilute samples, a larger number of transients per function evaluation may be required in order for a reliable cost function to be measured; consequently, the optimisation as a whole will take a longer time to run.
However, since the same can be said of the final 2D spectrum that is acquired, the benefits that can be reaped are correspondingly larger, making the ``return on investment'' the same.
A similar strategy can be used to search for the ``null'' in an inversion-recovery (IR) experiment: by running the optimisation on a 1D IR (i.e.\ a single increment of the full experiment), $T_1$ relaxation times can be measured in a far shorter time than the full 2D IR (\cref{sec:si_invrec}).

\begin{figure}
    % ../figures/hsqc_spec.py
    \centering
    \includegraphics[width=8cm]{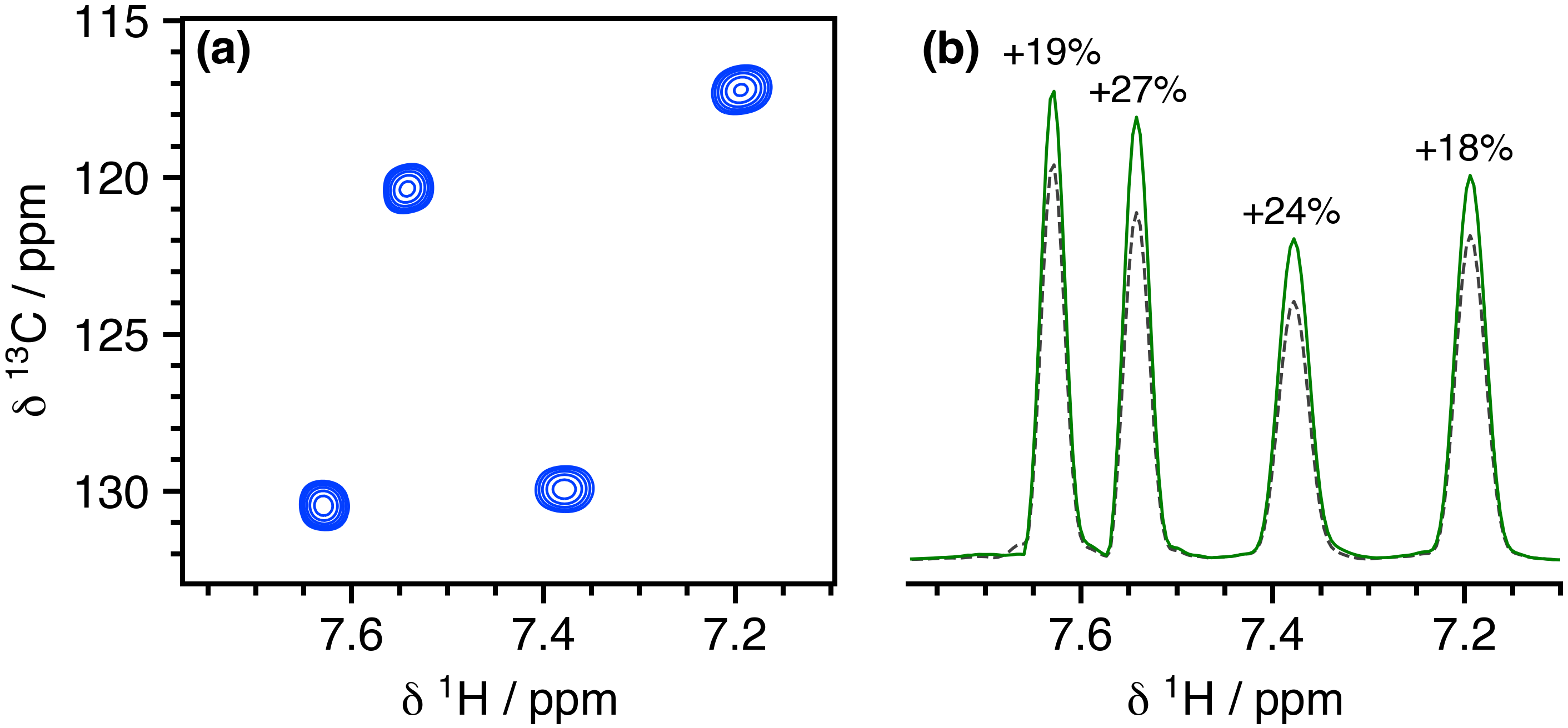}
    {\phantomsubcaption\label{fig:hsqc_spec_2d}}
    {\phantomsubcaption\label{fig:hsqc_spec_projections}}
    \caption{
        \textbf{\subref{fig:hsqc_spec_2d}} 2D ASAP-HSQC spectrum of a sample of \SI{120}{\milli\molar} 3-fluorophenylboronic acid in DMSO-$d_6$.
        \textbf{\subref{fig:hsqc_spec_projections}} $f_2$ projections of ASAP-HSQC spectra before (grey dotted line) and after POISE optimisation of the INEPT delay (green solid line).
        The sensitivity increases attained are indicated above each peak.
    }
    \label{fig:hsqc_spec}
\end{figure}

% }}}1
% Case study 3: ASAP-HSQC excitation {{{1
An example of a 2D experiment which can be optimised in a reasonable time is the ASAP-HSQC experiment (\cref{fig:pprogs_asaphsqc})\autocite{SchulzeSunninghausen2014JACS}, where the recovery delay between transients is dramatically shortened.
Decreasing the duration of the INEPT delay leads to greater sensitivity through Ernst angle-style excitation of \carbon{}-bound proton magnetisation.
Using the projection of the full 2D spectrum onto the $f_2$ dimension as a cost function, POISE is able to optimise this delay to approximately 60\% of its original length.
This takes approximately 3 minutes for the same sample as above (\cref{tbl:hsqc}) and provides an 18--27\% increase in sensitivity (\cref{fig:hsqc_spec}).
This procedure can be extended to the optimisation of other delays in 2D spectra, particularly with ``fast'' variants such as the ASAP method used here.\autocite{Schanda2009PNMRS}

% }}}1
% Case study 4: EPSI {{{1 
The logical extension of the preceding example is ultrafast 2D NMR experiments, in which the combination of spatially encoded $t_1$ evolution and the echo-planar spectroscopic imaging (EPSI) detection method enables single-scan acquisition of 2D spectra.\autocite{Frydman2003JACS,Gouilleux2018ARNMRS}
Despite the substantial efforts of Giraudeau and coworkers in the development of an accessible protocol,\autocite{Gouilleux2018ARNMRS} obtaining high-quality ultrafast spectra remains a nontrivial task.

\begin{figure}
    % ../figures/uf_spectra.py
    \centering
    \includegraphics[width=8cm]{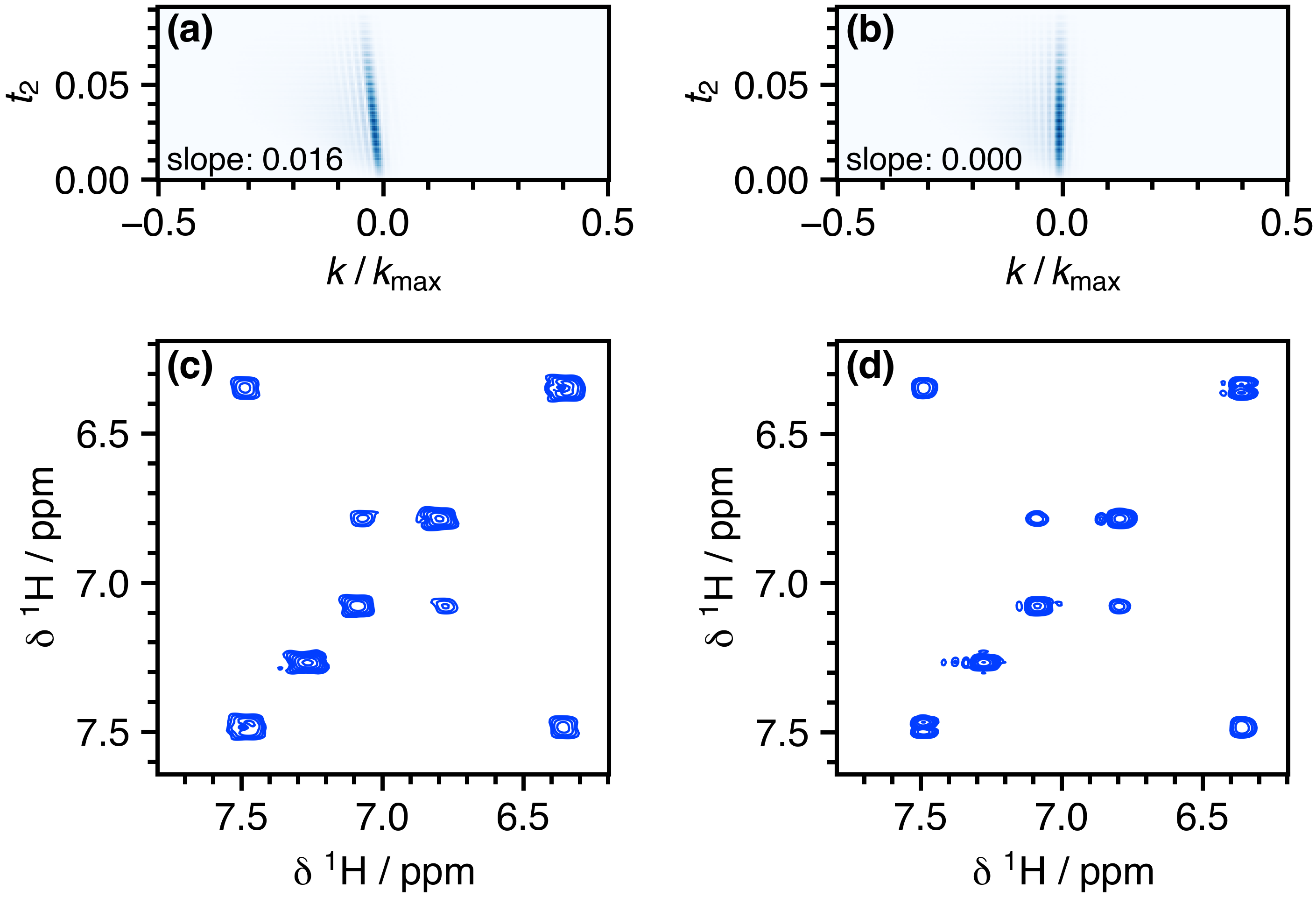}
    {\phantomsubcaption\label{fig:uf_spectra_unopt_1d}}
    {\phantomsubcaption\label{fig:uf_spectra_opt_1d}}
    {\phantomsubcaption\label{fig:uf_spectra_unopt_2d}}
    {\phantomsubcaption\label{fig:uf_spectra_opt_2d}}
    \caption{
        \textbf{\subref{fig:uf_spectra_unopt_1d}} and \textbf{\subref{fig:uf_spectra_opt_1d}}: Magnitude-mode $(k,t_2)$ data matrix obtained from the excitation--EPSI pulse sequence (\cref{fig:pprogs_epsi}), before and after optimisation of the gradient imbalance factor $\alpha$.
        Data points acquired during the negative gradients have been discarded.
        \textbf{\subref{fig:uf_spectra_unopt_2d}} and \textbf{\subref{fig:uf_spectra_opt_2d}}: The corresponding 2D ultrafast TOCSY spectra of a sample of \SI{45}{\milli\molar} ferulic acid in DMSO-$d_6$ (\SI{10}{\ms} MLEV-16 mixing), before and after optimisation.
    }
    \label{fig:uf_spectra}
\end{figure}

We illustrate here how POISE can be used to aid in the setup of ultrafast experiments, in this case by compensating for systematic imbalances in the positive and negative gradients used during EPSI acquisition.
Such an imbalance leads to a progressive shifting of echoes in the spatial frequency ($k$) domain (\cref{fig:uf_spectra_unopt_1d}), ultimately causing degradation of lineshapes in the resulting spectra.\autocite{Frydman2003JACS}
The pulse programme used for optimisation consists of \ang{90} excitation, immediately followed by EPSI acquisition (\cref{fig:pprogs_epsi}); the amplitudes of the negative gradients are multiplied by a factor $\alpha$ in order to compensate for any imbalance.
Here, the cost function is the slope generated by the echoes in a $(k, t_2)$ data matrix (\cref{fig:uf_spectra_unopt_1d,fig:uf_spectra_opt_1d}), which directly reflects the extent of gradient imbalance.
On our system, POISE finds the optimal value of $\alpha$ to be $1.00038$ in under a minute (\cref{tbl:epsi_opt}).
Such a small imbalance may appear inconsequential, but in fact leads to marked improvements in the resulting lineshapes of ultrafast spectra, such as the TOCSY spectra shown in \cref{fig:uf_spectra_unopt_2d,fig:uf_spectra_opt_2d}.

% }}}1
% Case study 5: PSYCHE {{{1
Finally, another major advantage of POISE is that its algorithms are capable of simultaneously optimising multiple parameters in an efficient manner.
Although this is possible with a grid search, the time required scales exponentially with the number of parameters, quickly rendering such a procedure unfeasible.
We show how this capability of POISE can be used to search for optimal performance in PSYCHE broadband homonuclear decoupling\autocite{Foroozandeh2014ACIE}.
Here, homodecoupling is accomplished by the application of a J-refocusing element (JRE), comprised of two saltire chirped pulses\autocite{Foroozandeh2020JMR} and a weak gradient.
These can be expressed in terms of several parameters, namely the flip angle $\beta$, gradient amplitude $g$, bandwidth $\Delta F$, and duration $\tau_\mathrm{p}$.

\begin{figure}
    % ../figures/psyche_3spectra.py
    \centering
    \includegraphics[width=8cm]{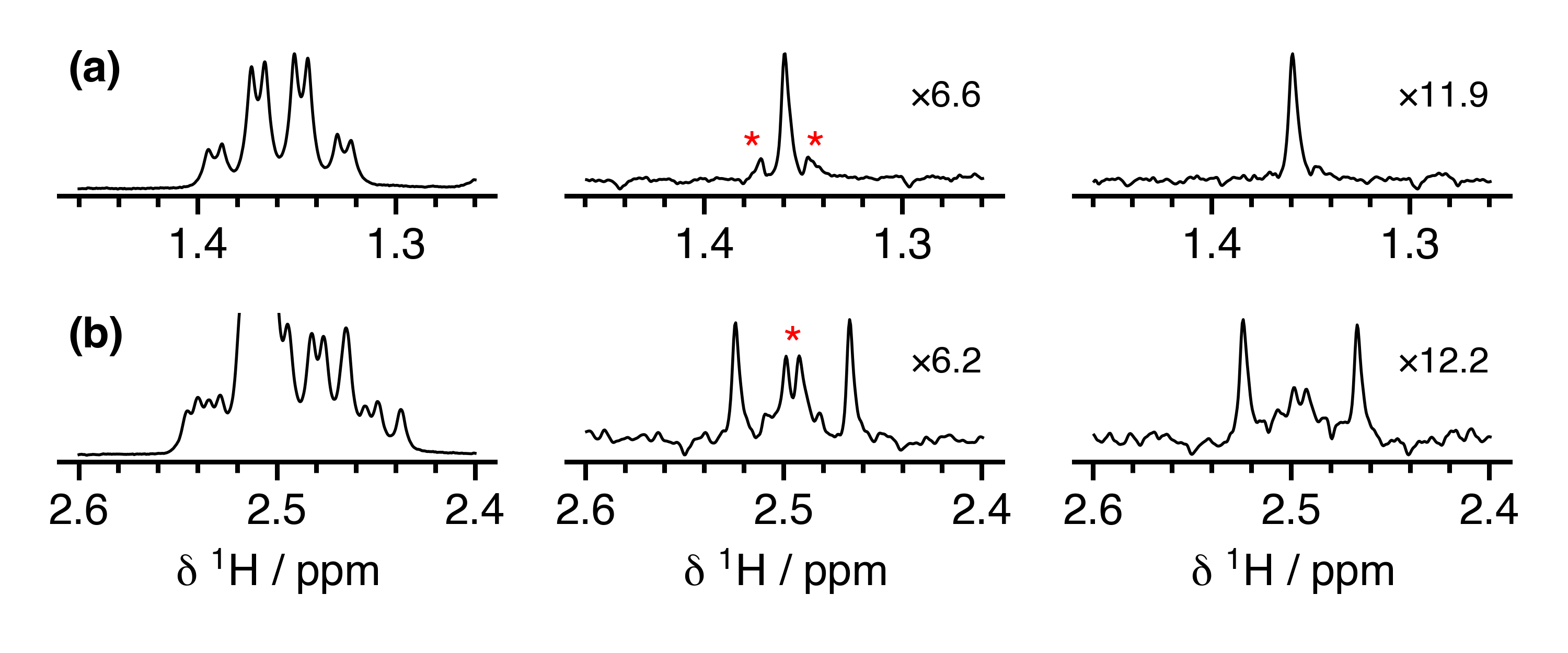}
    {\phantomsubcaption\label{fig:psyche_3spectra_1ppm}}
    {\phantomsubcaption\label{fig:psyche_3spectra_2ppm}}
    \caption{
        Insets from spectra of a sample of \SI{45}{\milli\molar} andrographolide in DMSO-$d_6$; artefacts in PSYCHE spectra are indicated with asterisks.
        \textit{Left column}: Ordinary coupled \proton{} spectrum; \textit{middle column}: unoptimised PSYCHE spectrum; \textit{right column}: PSYCHE spectrum after simultaneous optimisation of four parameters.
        \textbf{\subref{fig:psyche_3spectra_1ppm}} Peak at \SI{1.36}{\ppm}, where recoupling artefacts are eliminated through optimisation. 
        \textbf{\subref{fig:psyche_3spectra_2ppm}} Peak at \SI{2.5}{\ppm}, where optimisation leads to a substantial reduction in strong coupling artefacts.
    }
    \label{fig:psyche_3spectra}
\end{figure}

We used a 1D experiment for optimisation of these parameters, which consists of a \ang{90} excitation pulse followed by a spin echo with the JRE in the middle (\cref{fig:pprogs_psyche}).
Any non-idealities in the JRE can be captured in a ``spectral difference'' cost function (\cref{sec:si_psyche}), which compares this against a standard pulse-acquire experiment.
The simultaneous optimisation of all four parameters can be completed in around 15 minutes and leads to noticeably better decoupling performance for the sample under study, especially for the peak at \SI{1.36}{\ppm} (\cref{fig:psyche_3spectra_1ppm}).
Furthermore, the optimised spectrum exhibits improved suppression of the artefacts at \SI{2.5}{\ppm} (\cref{fig:psyche_3spectra_2ppm}), which arise from a challenging case of strong coupling.
Although the optimised spectrum has a decreased SNR, the cost function used here also penalises sensitivity losses, thereby ensuring that any SNR loss is not excessive.

\begin{figure}
    % ../figures/solvsupp_spec.py
    \centering
    \includegraphics[width=8cm]{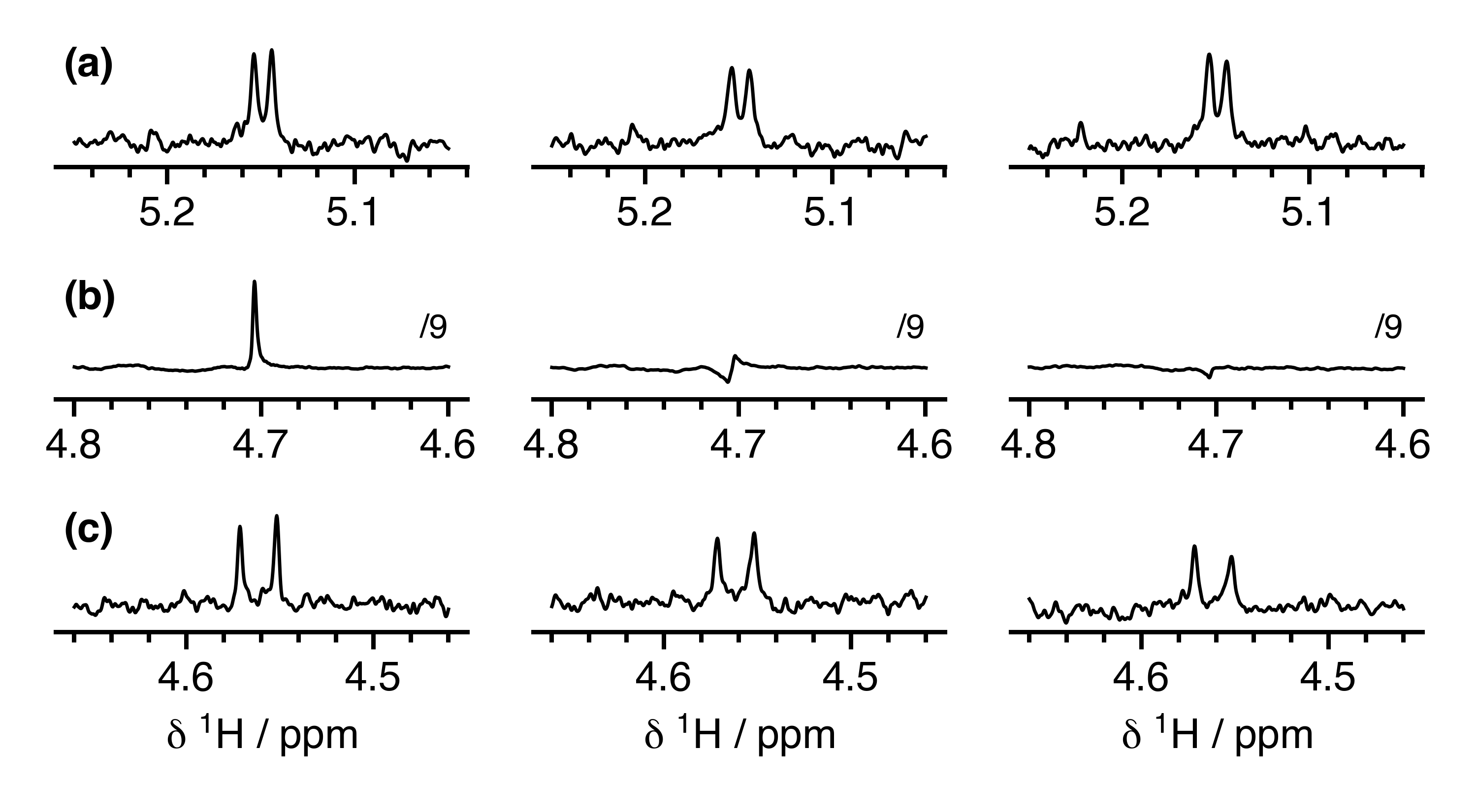}
    {\phantomsubcaption\label{fig:solvsupp_spec_5ppm}}
    {\phantomsubcaption\label{fig:solvsupp_spec_water}}
    {\phantomsubcaption\label{fig:solvsupp_spec_4ppm}}
    \caption{
        Spectra of a sample of rodent urine.
        \textit{Left column}: Unoptimised spectrum; \textit{middle column}: spectrum following optimisation of two parameters (transmitter offset and presaturation power); \textit{right column}: spectrum after optimisation of all four parameters.
        \textbf{\subref{fig:solvsupp_spec_5ppm}} Peak at \SI{5.15}{\ppm}.
        \textbf{\subref{fig:solvsupp_spec_water}} Residual water peak.
        \textbf{\subref{fig:solvsupp_spec_4ppm}} Peak at \SI{4.56}{\ppm}.
        Note that the flanking peaks in \subref{fig:solvsupp_spec_5ppm} and \subref{fig:solvsupp_spec_4ppm} are largely unaffected by the improvement in water suppression.
    }
    \label{fig:solvsupp_spec}
\end{figure}

In a similar fashion, we applied POISE to the 1D NOESY / presaturation experiment commonly used for water suppression (\cref{fig:pprogs_solvsupp})\autocite{Mckay2011CMR}, where up to four parameters (transmitter offset, presaturation power, mixing time, and presaturation duration) are optimised.
This procedure takes between 4 and 25 minutes, depending on the number of parameters under optimization (\cref{sec:si_solvsupp}).
By applying appropriate upper bounds on the presaturation power and duration, \Poise{} greatly reduces the intensity of the residual water peak, while also ensuring that nearby peaks of interest are not lost (\cref{fig:solvsupp_spec}).
Finally, a further example of multiple-parameter optimisation is presented in the context of diffusion NMR, enabling a fully automated diffusion ordered spectroscopy (DOSY) experiment (\cref{sec:si_dosy}).

% }}}1
% Conclusion {{{1
Through a series of examples, we have shown how POISE can be used to great effect in the optimisation of NMR parameters.
Each optimisation provides parameter values that are tailored to the instrument and sample under study, typically in only a few minutes and in a fully automated fashion through the TopSpin command line or acquisition scripts.
Importantly, POISE allows users to define arbitrary cost functions, making it a highly versatile tool that can be adapted to the myriad uses of NMR in modern chemistry.
POISE is fully compatible with, and will be immediately usable in, other fields such as solid-state and biomolecular NMR.
More generally, we also note that the iterative optimisation process demonstrated here can be extended to other forms of spectroscopy.

The source code of NMR-POISE is publicly hosted on GitHub: \url{https://github.com/foroozandehgroup/nmrpoise}.
Detailed instructions on installation, cost function definition, and optimisation setup may be found in the user documentation, which is included in the \SInf{} and can also be found online at \url{https://foroozandehgroup.github.io/nmrpoise}.
% }}}1

\section*{Acknowledgements}
We thank Coralia Cartis, Tim Claridge, Jean-Nicolas Dumez, David Goodwin, Iain Swan, and Chris Waudby for helpful discussions; as well as Fay Probert and Abi Yates for providing the rodent urine sample.
J.R.J.Y.\ thanks the Clarendon Fund (University of Oxford) and the EPSRC Centre for Doctoral Training in Synthesis for Biology and Medicine (EP/L015838/1) for a studentship, generously supported by AstraZeneca, Diamond Light Source, Defence Science and Technology Laboratory, Evotec, GlaxoSmithKline, Janssen, Novartis, Pfizer, Syngenta, Takeda, UCB, and Vertex.
M.F.\ thanks the Royal Society for a University Research Fellowship and a University Research Fellow Enhancement Award (Grant Nos. URF\textbackslash{}R1\textbackslash{}180233 and RGF\textbackslash{}EA\textbackslash{}181018).

\printbibliography{}
% Preamble-like commands {{{1
\singlespacing
% Move table captions to the top
\floatstyle{plaintop}
\restylefloat{table}
% Reset counters
\newcommand{\sectionbreak}{\clearpage}
\renewcommand*{\thefigure}{S\arabic{figure}}
\renewcommand*{\thetable}{S\arabic{table}}
\renewcommand*{\thepage}{S\arabic{page}}
\renewcommand*{\thesection}{S\arabic{section}}
\setcounter{page}{1}
\setcounter{figure}{0}
\setcounter{table}{0}
% Boring long strings
\newcommand{\nfev}{$n_\text{fev}$ refers to the number of function evaluations, i.e.\ the number of spectra acquired in the course of the optimisation.}
\newcommand{\fiveruns}{The results are aggregated from five separate runs for each optimisation algorithm.}
\setlength{\cftsecnumwidth}{4em}
\setlength{\cftsubsecindent}{1em}
\setlength{\cftsubsecnumwidth}{4em}
\setlength{\cftsubsubsecindent}{2em}
\setlength{\cftsubsubsecnumwidth}{4em}
% }}}1
% Title page {{{1
\hspace{0pt}
\vfill
\begin{center}  % title page
    \huge
    Supporting Information

    \vspace{1cm}

    On-the-fly, Sample-tailored Optimization of NMR Experiments

    \vspace{1cm}

    \Large
    Jonathan R.\ J.\ Yong and Mohammadali Foroozandeh*
    
    \vspace{1cm}

    \large
    \textit{Chemistry Research Laboratory, Department of Chemistry, University of Oxford,\\12 Mansfield Road, Oxford, OX1 3TA, U.K.}

    \vspace{1cm}

    * \texttt{mohammadali.foroozandeh@chem.ox.ac.uk}
\end{center}
\vfill
\hspace{0pt}
\newpage
\tableofcontents

\vspace{1cm}

\textbf{Note:} The \SInf{} sections mainly contain summaries of the results presented in this paper, as well as technical details for reproducing these results.
It does not explain how to set up \Poise{}, or to how to create your own optimisation routines.
For step-by-step instructions on these, please refer to the documentation.
A copy of it is attached at the end of the \SInf{}, but the most updated version will be accessible at \url{https://foroozandehgroup.github.io/nmrpoise} (a PDF copy is located at \url{https://foroozandehgroup.github.io/nmrpoise/poise.pdf}).

\newpage
% }}}1

\section{Raw data and reproducibility}
\label{sec:si_rawdata}

All raw data used for this paper, as well as the Python scripts to generate the plots in the main paper and SI, can be downloaded from Zenodo at the DOI \href{https://doi.org/10.5281/zenodo.4698423}{\texttt{10.5281/zenodo.4698423}}.

\subsection{Datasets}

The datasets that were used for each section are tabulated in \cref{tbl:datasets}.

\begin{table}
    \centering
    \begin{tabular}{cccc}
        \toprule
        Section               & Description                               & Folder name             & Expnos   \\
        \midrule
        \ang{90} pulse width  & Reference grid search                     & \texttt{P1\_Scan}       & 1--161   \\
                              & Optimisation (initial value \SI{48}{\us}) & \texttt{P1\_Opt}        & 4        \\
                              & Optimisation (initial value \SI{43}{\us}) & \texttt{P1\_Opt}        & 5        \\
                              & Optimisation (initial value \SI{53}{\us}) & \texttt{P1\_Opt}        & 6        \\
        Ernst angle           & 2D inversion-recovery                     & \texttt{Invrec\_Opt}    & 2        \\
                              & Optimisation (6 to \SI{8}{\ppm})          & \texttt{Ernst\_Opt}     & 1001     \\
                              & Optimisation (\SI{6.79}{\ppm} only)       & \texttt{Ernst\_Opt}     & 1002     \\
        NOE mixing time       & Reference grid search                     & \texttt{NOESY\_OptScan} & 101--159 \\
                              & Optimisation                              & \texttt{NOESY\_OptScan} & 3        \\
        Inversion-recovery    & 2D inversion-recovery                     & \texttt{Invrec\_Opt}    & 2        \\
                              & Optimisation (6 to \SI{8}{\ppm})          & \texttt{Invrec\_Opt}    & 3        \\
                              & Optimisation (\SI{7.08}{\ppm} only)       & \texttt{Invrec\_Opt}    & 4        \\
        ASAP-HSQC             & Reference grid search                     & \texttt{HSQC\_OptScan}  & 101--133 \\
                              & Optimisation                              & \texttt{HSQC\_OptScan}  & 2        \\
        EPSI                  & Reference grid search                     & \texttt{EPSI\_Scan}     & 95--116  \\
                              & Optimisation                              & \texttt{EPSI\_Opt}      & 1        \\
                              & Ultrafast 2D TOCSY                        & \texttt{EPSI\_Opt}      & 10--11   \\
        PSYCHE                & Flip angle reference grid search          & \texttt{PSYCHE\_FAScan} & 301--331 \\
                              & Optimisation (1 parameter)                & \texttt{PSYCHE\_Opt}    & 101      \\
                              & Optimisation (2 parameters)               & \texttt{PSYCHE\_Opt}    & 102      \\
                              & Optimisation (3 parameters)               & \texttt{PSYCHE\_Opt}    & 103      \\
                              & Optimisation (4 parameters)               & \texttt{PSYCHE\_Opt}    & 104      \\
                              & Pseudo-2D homodecoupled spectra           & \texttt{PSYCHE\_Opt}    & 1400--1404 \\
        Solvent suppression   & Offset reference grid search              & \texttt{Solvsupp\_O1Scan} & 101--141 \\
                              & Optimisation (1 parameter)                & \texttt{Solvsupp\_Opt}  & 11       \\
                              & Optimisation (2 parameters)               & \texttt{Solvsupp\_Opt}  & 14       \\
                              & Optimisation (3 parameters)               & \texttt{Solvsupp\_Opt}  & 15       \\
                              & Optimisation (4 parameters)               & \texttt{Solvsupp\_Opt}  & 16       \\
                              & Representative optimised spectra          & \texttt{Solvsupp\_Opt}  & 31, 35--37 \\
        DOSY                  & Optimisation (sequential)                 & \texttt{DOSY\_Opt}      & 101--102 \\
                              & Optimised Oneshot DOSY                    & \texttt{DOSY\_Opt}      & 1, 1001--1021 \\
                              & Optimisation (simultaneous)               & \texttt{DOSY\_Opt}      & 201--202 \\
        \bottomrule
    \end{tabular}
    \caption{Datasets used for each section of the Supporting Information.}
    \label{tbl:datasets}
\end{table}

All of the \Poise{} optimisation data was logged in \texttt{poise.log} files, which can be found in the corresponding raw data folders.
These log files detail the optimisation trajectory and the cost function at each point, and can be parsed using the \texttt{parse\_log()} function (please see the documentation for more information).
However, some of these date from an older development version of \Poise{} in which less information was recorded, so some details (e.g.\ of the AU programme used for acquisition) may be blank.

Full details of acquisition and processing parameters for each optimisation can be found in these folders.
For this reason, except for a few key values, they are not tabulated in the main text or here.

\subsection{Figures}

All figures are done in Python 3 and use the \texttt{penguins} package (written by J.R.J.Y.) to parse and plot NMR data.
As of the time of writing, this package is still in development; consequently, the interface and functionality may change in a backward-incompatible manner.
In order to ensure full reproducibility, please install version 0.4.1 of \texttt{penguins}, which was released specifically for this paper.
This can be done using the command

\begin{minted}{text}
pip install penguins=0.4.1
\end{minted}

After downloading the raw data, the NMR datasets themselves should be found in a directory called \texttt{datasets}, and the scripts in a directory called \texttt{figures}.
The figures themselves are provided in this folder, but can be regenerated at any point in time by re-running the scripts (the scripts also provide a ``trace'' of which data is being used to plot the data).

The scripts will be executable as long as this directory structure is preserved, as they use a relative path to locate the datasets.
If the directories are moved for any reason, the environment variable \texttt{POISE\_DATA} should be set such that it points to the \texttt{datasets} directory.
For example, in \texttt{bash} and similar shells, this can be done using (replace the path accordingly)

\begin{minted}{text}
export POISE_DATA=/path/to/nmrpoise_data/datasets
\end{minted}

\section{Routines}

In \Poise{}, routines are stored in the human-readable JSON format, and are thus provided in the sections which follow.
These JSON files should be placed in the TopSpin directory \texttt{.../exp/stan/nmr/\\py/user/poise\_backend/routines}.
Each routine consists of several entries:

\begin{itemize}
    \item \texttt{name}: The name of the routine, which can be arbitrarily chosen (although it should match the filename: thus a routine with name \texttt{my\_routine} should be stored in \texttt{my\_routine.json}).
    \item \texttt{pars}: A list of string parameters which are to be optimised. These correspond exactly to parameter names in TopSpin.
    \item \texttt{lb}: A list of numbers acting as the lower bounds for each parameter. Units are given in TopSpin's native units, i.e.\ pulses are in microseconds, delays in seconds, etc.
    \item \texttt{ub}: A list of numbers acting as the upper bounds for each parameter.
    \item \texttt{init}: A list of numbers acting as the initial values for each parameter.
    \item \texttt{tol}: A list of numbers acting as the tolerances for each parameter.
    \item \texttt{cf}: The name of the cost function. A function with this name must be defined in either of the files \texttt{poise\_backend/costfunctions.py} or \texttt{poise\_backend/costfunctions\_user.py}. For a list of pre-installed cost functions and descriptions, please refer to the \Poise{} user documentation.
    \item \texttt{au}: The name of the AU programme used for acquisition and processing. This can be left blank (see below).
\end{itemize}

If no AU programme is specified by the user upon setup, the \texttt{poise\_1d} or \texttt{poise\_2d} AU programmes will automatically be used depending on the dimensionality of the dataset.
These default AU programmes are used for many of the optimisations described in the \SInf{}, and are detailed below.

All the routines in the \SInf{} can be found in the \texttt{.../exp/stan/nmr/py/user/\\poise\_backend/example\_routines} directory, and can be directly copied to the \texttt{routines} directory if desired.

\clearpage

\subsection{poise\_1d AU programme}

\begin{minted}{c}
ZG   // acquire
EFP  // exponential apodisation, Fourier transform
APBK // automatic phase and baseline correction
QUIT
\end{minted}

\subsection{poise\_1d\_noapk AU programme}

This AU programme is not used by default, but comes installed together with POISE, used in some of the optimisations described below.

\begin{minted}{c}
ZG            // acquire
EFP           // exponential apodisation, Fourier transform
XCMD("abs n") // automatic baseline correction
QUIT
\end{minted}

\subsection{poise\_2d AU programme}

\begin{minted}{c}
ZG             // acquire
XFB            // 2D Fourier transform
XCMD("apk2d")  // automatic phase correction
ABS2           // f2 baseline correction
QUIT
\end{minted}

\section{1D \texorpdfstring{\proton{}}{1H} spectra for all samples}

As a reference, we provide here the \proton{} spectra of all samples used in this work.
These can also be found in the raw data (\cref{sec:si_rawdata}).

\begin{figure}
    % ./figures/si_proton_ferulic.py
    \centering
    \includegraphics[width=0.7\textwidth]{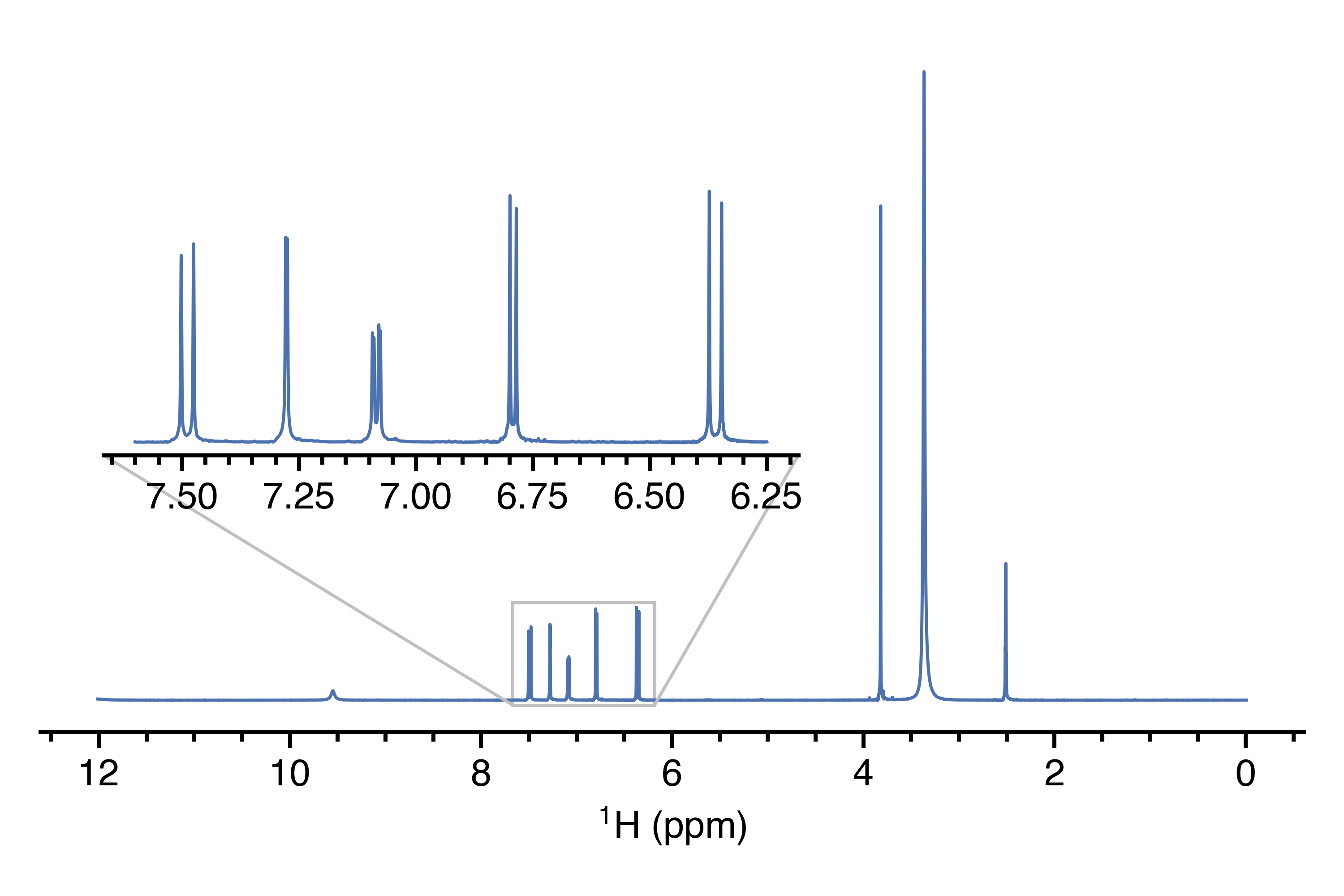}
    \caption{
        \proton{} spectrum of \SI{45}{\milli\molar} ferulic acid in DMSO-$d_6$ (\SI{500}{\MHz} resonance frequency).
    }
    \label{fig:si_proton_ferulic}
\end{figure}

\begin{figure}
    % ./figures/si_proton_3fpba.py
    \centering
    \includegraphics[width=0.7\textwidth]{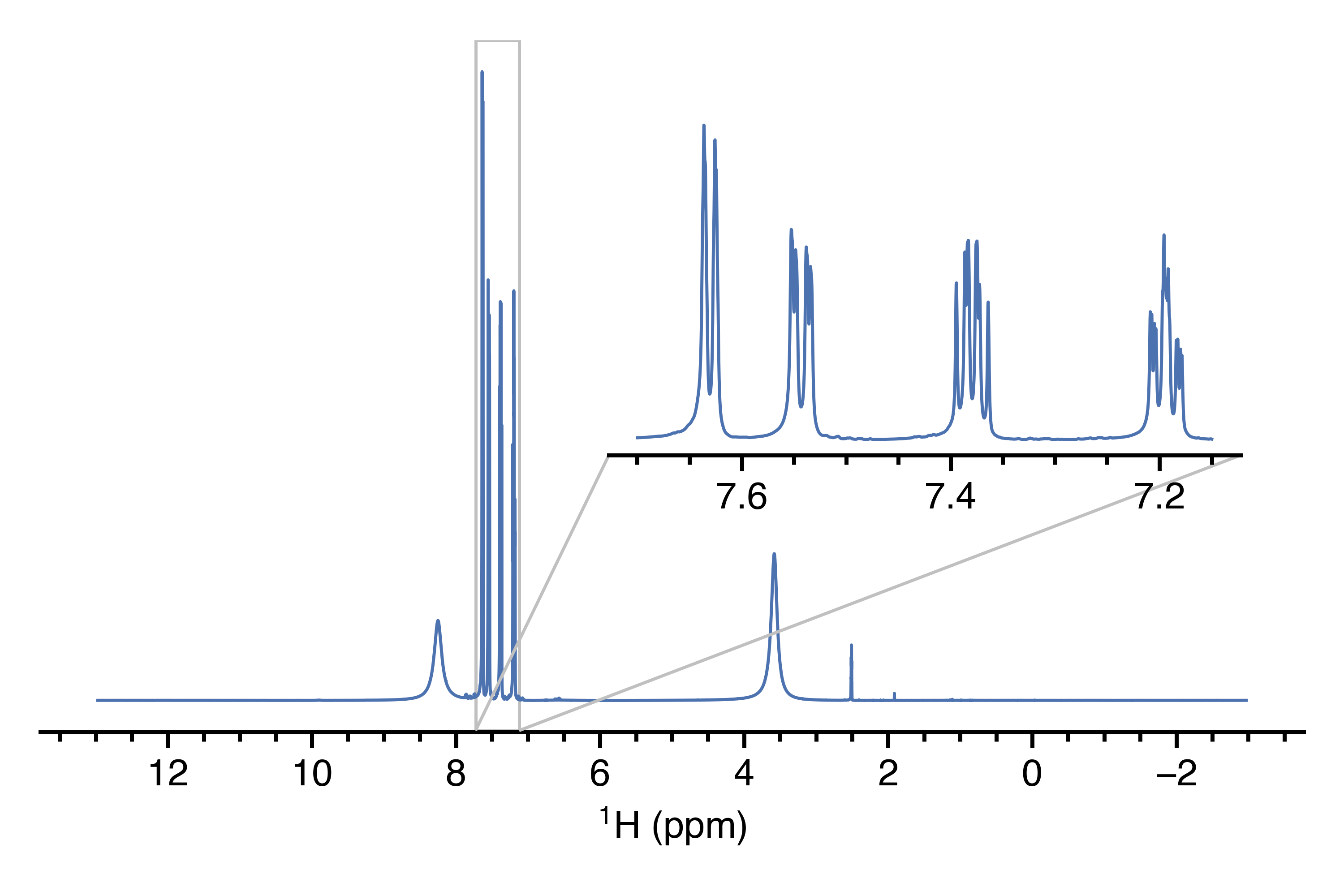}
    \caption{
        \proton{} spectrum of \SI{120}{\milli\molar} 3-fluorophenylboronic acid in DMSO-$d_6$ (\SI{700}{\MHz} resonance frequency).
    }
    \label{fig:si_proton_3fpba}
\end{figure}

\begin{figure}
    % ./figures/si_proton_andro.py
    \centering
    \includegraphics[width=0.7\textwidth]{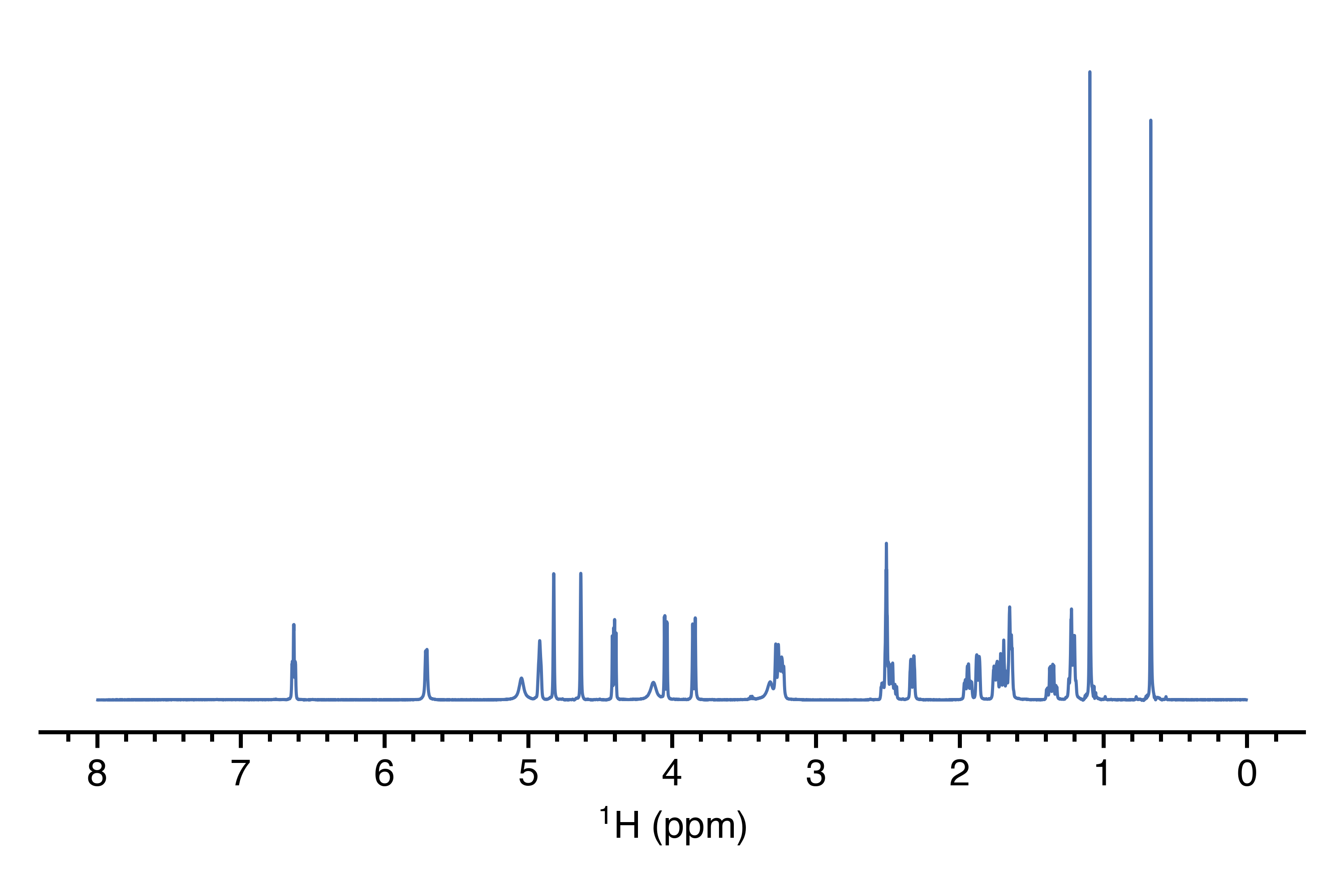}
    \caption{
        \proton{} spectrum of \SI{45}{\milli\molar} andrographolide in DMSO-$d_6$ (\SI{600}{\MHz} resonance frequency).
    }
    \label{fig:si_proton_andro}
\end{figure}

\begin{figure}
    % ./figures/si_proton_rodent.py
    \centering
    \includegraphics[width=0.7\textwidth]{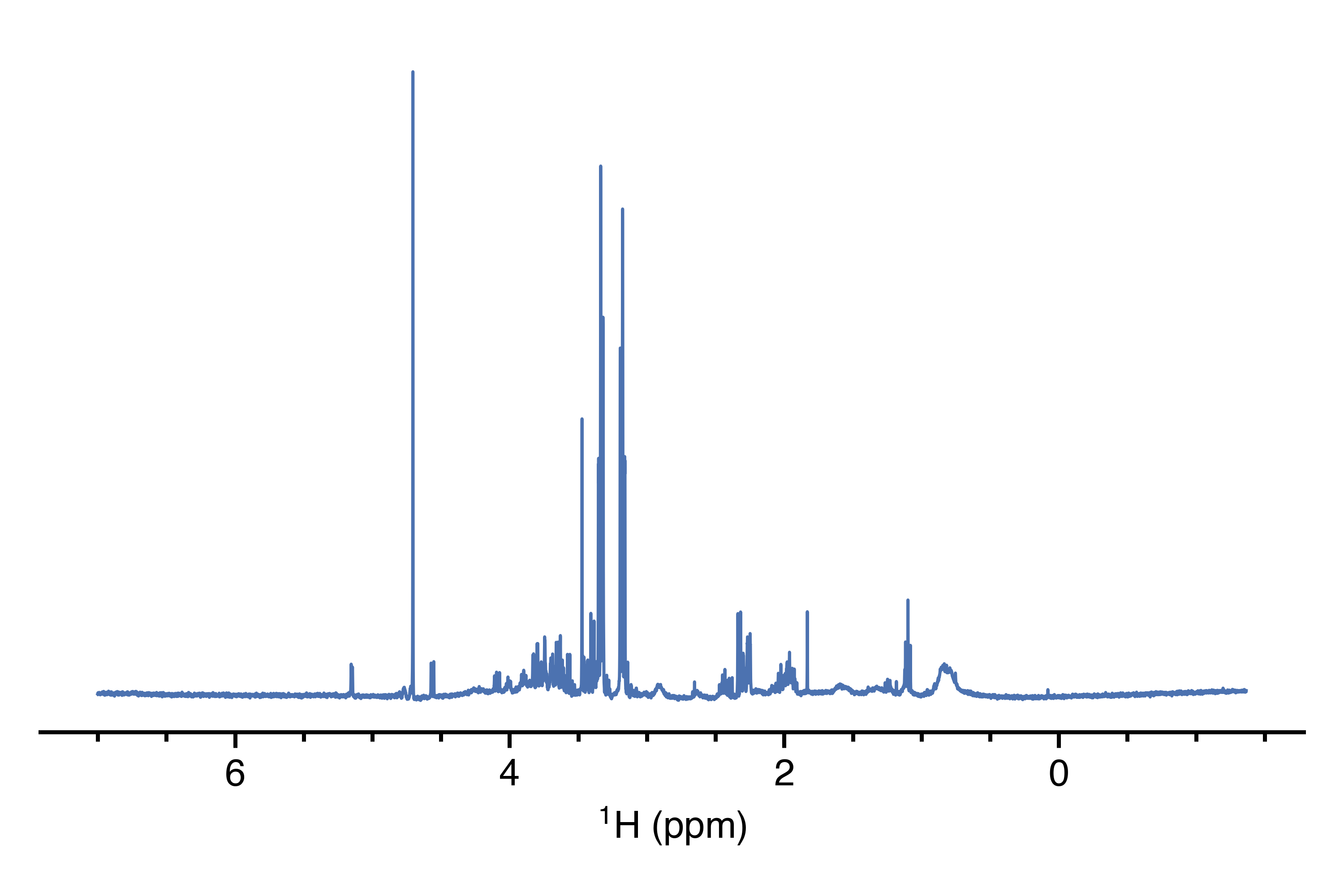}
    \caption{
        \proton{} spectrum of rodent urine in \ce{D2O} (\SI{400}{\MHz} resonance frequency).
        The water peak has been suppressed using the 1D NOESY / presaturation sequence.
    }
    \label{fig:si_proton_rodent}
\end{figure}

\section{\texorpdfstring{$\mathbf{90^\circ}$}{90°} pulse width calibration}

All spectra in this section were acquired on a \SI{600}{\MHz} Bruker AVIII spectrometer, equipped with a Prodigy \ce{N2} broadband cryoprobe with a nominal maximum $z$-gradient strength of \SI{66}{\gauss\per\cm}.
The sample used was \SI{45}{\milli\molar} ferulic acid in DMSO-$d_6$.

\subsection{Cost function}

The cost function used in this section, \texttt{minabsint}, comes pre-installed with POISE.
It can be mathematically expressed as

\begin{equation}
    f = \sum_i \sqrt{\mathbf{S}_{\text{real},i}^2 + \mathbf{S}_{\text{imag},i}^2}
\end{equation}

where $\mathbf{S}_{\text{real}}$ and $\mathbf{S}_{\text{imag}}$ are vectors of length \texttt{SI} corresponding to the real and imaginary parts of the spectrum respectively.
The elements $\mathbf{S}_i$ are the intensities of the spectrum at each point along the spectral width.

\subsection{Reference grid search}

The grid search, as used in the builtin \texttt{popt}, is a highly time-consuming method of locating the optimum.
It is used here only as a reference, i.e.\ to prove that the faster optimisation algorithms are indeed finding the true optimum.

\begin{figure}
    % ./figures/p1_scan.py
    \centering
    \includegraphics[width=0.7\textwidth]{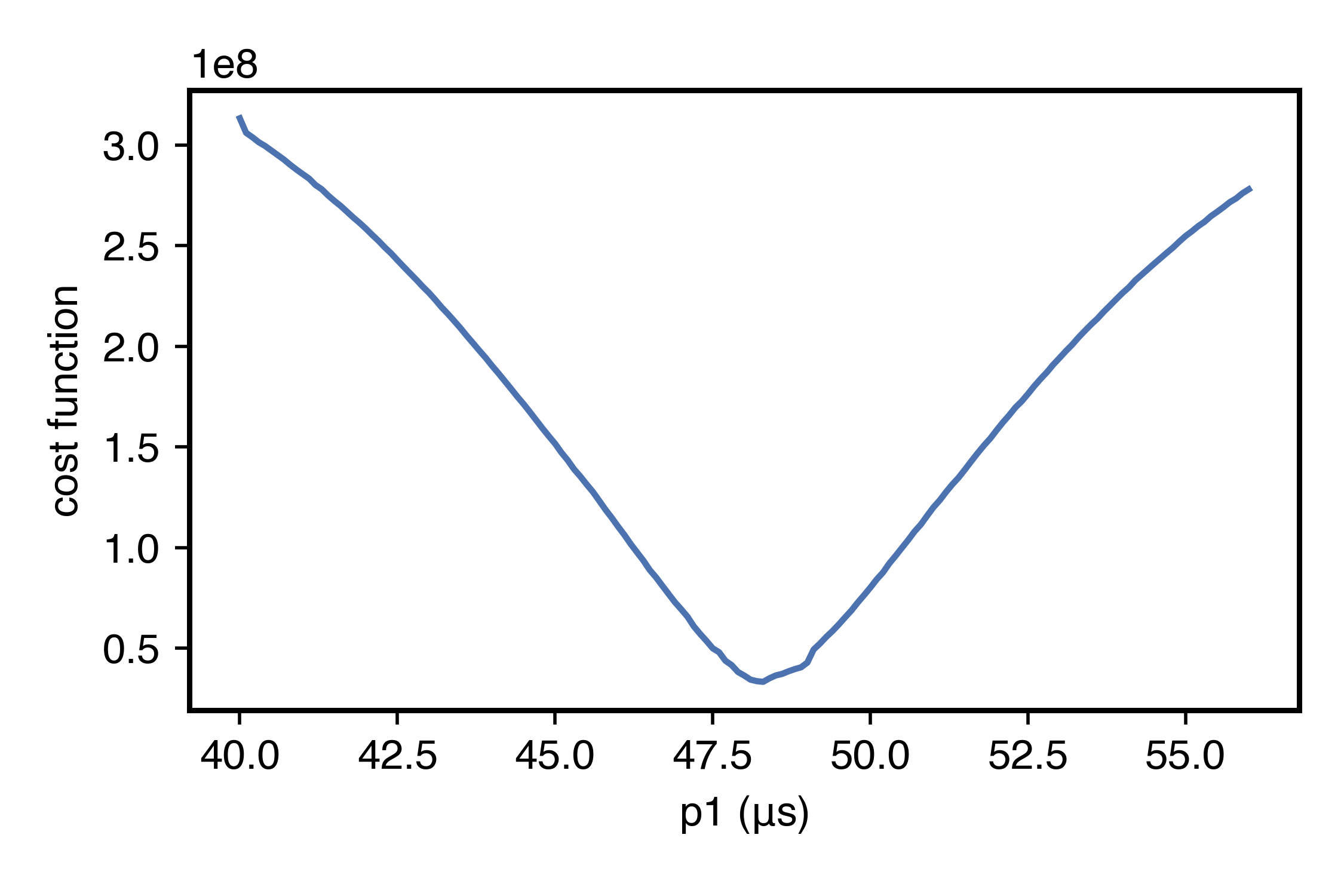}
    \caption{
        Dependence of the cost function on the pulse width (\texttt{p1}).
        The minimum value of the cost function is $3.34 \times 10^7$ and occurs at $\mathtt{p1} = \SI{48.3}{\us}$.
    }
    \label{fig:p1scan}
\end{figure}

\subsection{Optimisation}

\begin{table}
    \hbadness=10000
    \centering
    \begin{tabular}{cccccc}
        \toprule
        Entry & Method & Description & Optimum found (\si{\us}) & $n_\text{fev}$ & Time taken (\si{\s}) \\
        \midrule
        1  & \texttt{popt}     & \texttt{MAGMIN} cost function & 48.40        & 41   & 299    \\
        2  & \texttt{pulsecal} & --                            & 46.64        & --   & 37     \\
        3a & \Poise{}          & Nelder--Mead                  & 48.38        & 10   & 76--79 \\
        3b &                   & MDS                           & 48.38        & 10   & 77--80 \\
        3c &                   & BOBYQA                        & 48.29--48.41 & 6--7 & 46--54 \\
        \bottomrule
    \end{tabular}
    \caption{
        \ang{360} pulse width optimisations run with an initial point of \SI{48}{\us}.
        \texttt{popt} optimisations were run between values of \SI{40}{\us} and \SI{56}{\us}, with a linear increment of \SI{0.4}{\us}.
        The \Poise{} routine is: \mintinline[breaklines]{json}{{"name": "p1cal", "pars": ["p1"], "lb": [40.0], "ub": [56.0], "init": [48.0], "tol": [0.2], "cf": "minabsint", "au": "poise_1d"}}.
        The pulse programme used was the Bruker standard \texttt{zg}.
        Key acquisition parameters: \texttt{NS=1}, \texttt{DS=0}, \texttt{SW=12.0}, \texttt{TD=16384}, \texttt{D1=1.5}.
        \nfev{}
        \fiveruns{}
    }
    \label{tbl:p1init48}
\end{table}

The BOBYQA algorithm takes the \textit{exact value} of the cost function into consideration when deciding what points to sample.
On the other hand, the simplex-based methods only consider how the various points are ordered according to the value of the cost function at each point.
This is partly why BOBYQA tends to converge faster than the other two algorithms, which in turn explains why we have set it as the default algorithm for POISE; it also accounts for the slight variability in the optimum found (note that this range is still well within the specified tolerance of \SI{0.2}{\us}, so it does not reflect a failing).
However, in edge cases when the cost function is extremely unbalanced or noisy, we find that BOBYQA can be unduly biased by the resulting cost function values (this is not the case here).
We find that the simplex-based NM and MDS methods, although slower on average, can be slightly more reliable in such situations.

The next two tables describe optimisations starting from an inaccurate initial point.

\begin{table}
    \centering
    \begin{tabular}{cccccc}
        \toprule
        Entry & Method & Description & Optimum found (\si{\us}) & $n_\text{fev}$ & Time taken (\si{\s}) \\
        \midrule
        1a & \Poise{}          & Nelder--Mead                  & 48.38        & 14 & 109--114 \\
        1b &                   & MDS                           & 48.38        & 14 & 108--112  \\
        1c &                   & BOBYQA                        & 48.27--48.33 & 9  & 70       \\
        \bottomrule
    \end{tabular}
    \caption{
        Pulse width optimisations with an initial point of \SI{43}{\us}.
        The \Poise{} routine is the same as in \cref{tbl:p1init48}, except with \texttt{"init":[43.0]}.
        All other acquisition parameters are identical.
        \fiveruns{}
    }
    \label{tbl:p1init43}
\end{table}

\begin{table}
    \centering
    \begin{tabular}{cccccc}
        \toprule
        Entry & Method & Description & Optimum found (\si{\us}) & $n_\text{fev}$ & Time taken (\si{\s}) \\
        \midrule
        1a & \Poise{}          & Nelder--Mead                  & 48.38        & 14 & 110--114 \\
        1b &                   & MDS                           & 48.25--48.38 & 16 & 123--126 \\
        1c &                   & BOBYQA                        & 48.26--48.33 & 9  & 69--70   \\
        \bottomrule
    \end{tabular}
    \caption{
        Pulse width optimisations with an initial point of \SI{53}{\us}.
        The \Poise{} routine is the same as in \cref{tbl:p1init48}, except with \texttt{"init":[53.0]}.
        All other acquisittion parameters are identical.
        \fiveruns{}
    }
    \label{tbl:p1init53}
\end{table}

As can be seen, \Poise{} is capable of ``recovering'' from a bad start. Although this requires more time, even in the worst-case scenario, the time taken is still less than half of that needed for the full grid search (299 seconds, cf.\ Entry 1, \cref{tbl:p1init48}).

\subsection{Using POISE under automation}

\Poise{} provides a command-line interface which allows it to be called from within AU and Python scripts once the appropriate routines and cost functions have been set up.
Thus, for example, a typical optimisation `procedure' would involve the following steps:
\begin{enumerate}
    \item Create a new dataset specially for the optimisation.
    \item Read in a parameter set, where the appropriate pulse programme for optimisation, as well as all other unoptimised parameters, have been chosen.
    \item Run the \Poise{} optimisation with \texttt{poise <routine\_name> <options>}. (For the available options please see the user documentation.)
    \item Transfer the optimised parameter values back to the original dataset, and run the optimised experiment.
\end{enumerate}

All of this can be easily automated using either AU or Python scripts.
Here we provide examples of both AU and Python scripts, called \texttt{poisecal}, which perform a similar role to the existing \texttt{pulsecal} AU programme in TopSpin.
Running either of these will carry out a \Poise{} optimisation in expno 99999, then set the optimised value of \texttt{p1} to the currently active dataset.
(These examples are also included in the documentation with extra explanation.)
\textbf{Please note that both of these can be downloaded from GitHub; the links are near the top of the source code.}

In general, the AU programme should be preferred as it contains some extra functionality (namely the option to dynamically specify the region of the spectrum to optimise on: please see the header of the AU programme for more information).

\subsubsection{AU programme}

\begin{minted}{c}
/******************************************************************************
 * poisecal
 * --------
 * 
 * Download from: https://git.io/JUHOk
 *
 * AU script which automates p1 optimisation using POISE. Behaves similarly to
 * pulsecal: optimises the value of p1 in expno 99999, then sets the optimised
 * value to the current dataset and shows a message.
 *
 * To optimise on only one particular part of the spectrum, either:
 * 
 *  1) Use the command `dpl` to specify a particular region, then run
 *     `poisecal`; or
 *  2) use the command 
 *
 *         poisecal LEFT RIGHT
 *
 *     where LEFT and RIGHT are the edges of the desired spectral window in
 *     ppm. The order in which these are specified does not matter, so
 *     `poisecal 1 4` and `poisecal 4 1` behave exactly the same way.
 *
 * To optimise only one end of a spectrum (e.g. above 4 ppm only), use option 2
 * and pass a large value as the other argument, e.g. `poisecal 4 1000`.
 *
 * SPDX-License-Identifier: GPL-3.0-or-later
 *****************************************************************************/

GETCURDATA
int old_expno = expno;
double f1p, f2p;
// Argument parsing
if (i_argc == 4) {  // argv[0] is always the programme name, argv[1] is 'exec'
    double d1, d2;
    char *cp;
    d1 = strtod(i_argv[2], &cp);
    if (*cp) STOPMSG("Invalid arguments. Usage: poisecal [left right]")
    d2 = strtod(i_argv[3], &cp);
    if (*cp) STOPMSG("Invalid arguments. Usage: poisecal [left right]")
    f1p = d1 > d2 ? d1 : d2;
    f2p = d1 > d2 ? d2 : d1;
}
else if (i_argc == 2) {
    FETCHPAR("F1P", &f1p)
    FETCHPAR("F2P", &f2p)
}
else STOPMSG("Invalid arguments. Usage: poisecal [left right]")
// Use EXPNO 99999 in the current folder for optimisation.
DATASET(name, 99999, procno, disk, user)
// Set some key parameters. Notice that these lines can be substantially cut
// if an appropriate parameter set is set up beforehand.
RPAR("PROTON", "all")
GETPROSOL
STOREPAR("PULPROG", "zg")
STOREPAR("NS", 1)
STOREPAR("DS", 0)
STOREPAR("D 1", 1.0)
STOREPAR("RG", 1)
STOREPAR("F1P", f1p)
STOREPAR("F2P", f2p)
STOREPARS("F1P", f1p)
STOREPARS("F2P", f2p)
// Run optimisation (uses BOBYQA by default).
XCMD("sendgui xpy poise p1cal -q")
// POISE stores the optimised value in p1 after it's done. We can retrieve it
// here. Don't try to get the *status* parameter, since that is not the
// optimised value (it is the value used for the last function evaluation!)
float p1opt;
FETCHPAR("P 1", &p1opt)
p1opt = p1opt/4;
// Move back to old dataset and set p1 to optimised value.
DATASET(name, old_expno, procno, disk, user)
VIEWDATA_SAMEWIN  // not strictly necessary, just re-focuses the original spectrum
STOREPAR("P 1", p1opt)
Proc_err(INFO_OPT, "Optimised value of p1: %.3f", p1opt);
// (Optional) Run acquisition.
// ZG
QUIT
// vim: ft=c
\end{minted}

\subsubsection{Python script}

\begin{minted}{python}
"""
poisecalpy.py
-------------

Download from: https://git.io/JUHOU

Python script which automates p1 optimisation using POISE. Behaves similarly to
pulsecal: optimises the value of p1 in expno 99999, then sets the optimised
value to the current dataset and shows a message.

This does not have the command-line argument functionality of the poisecal AU
programme, and tends to suffer from some edge-case bugs which are not present
in poisecal. This is generally a result of TopSpin's Python programmes being
more buggy than AU programmes. Nonetheless, as a result of this, we recommend
using the poisecal AU programme instead of this (which is retained mainly to
serve as a illustrative example in the documentation).

SPDX-License-Identifier: GPL-3.0-or-later
"""

# Read in F1P and F2P from current dataset.
f1p = GETPAR("F1P")
f2p = GETPAR("F2P")
# Use EXPNO 99999 in the current folder for optimisation.
old_dataset = CURDATA()
opt_dataset = CURDATA()
opt_dataset[1] = "99999"
# Create new dataset and move to it.
NEWDATASET(opt_dataset, None, "PROTON")
RE(opt_dataset)
# Set some key parameters. Notice that these lines can be cut if an appropriate
# parameter set is set up beforehand (and loaded using NEWDATASET()).
XCMD("getprosol")
PUTPAR("PULPROG", "zg")
PUTPAR("NS", "1")
PUTPAR("DS", "0")
PUTPAR("D 1", "1")
PUTPAR("RG", "1")
XCMD("s f1p {}".format(f1p))  # PUTPAR("status F1P") doesn't work.
XCMD("s f2p {}".format(f2p))  # Even though the documentation says it should.
# Run optimisation (uses BOBYQA by default).
XCMD("poise p1cal -q")
# POISE stores the optimised value in p1 after it's done. We can retrieve it
# here. Don't try to get the *status* parameter, since that is not the
# optimised value (it is the value used for the last function evaluation!)
p1opt = float(GETPAR("P 1"))/4
# Move back to old dataset and set p1 to optimised value.
RE(old_dataset)
PUTPAR("P 1", str(p1opt))
ERRMSG("Optimised value of p1: {:.3f}".format(p1opt))
# A TopSpin quirk: using MSG() will block subsequent commands until user hits
# "OK". You can use ERRMSG(), as is done here. There are other ways around
# this, see MSG_nonmodal() in the POISE frontend script.
# (Optional) Run acquisition.
# ZG()
\end{minted}

\section{Ernst angle}
\label{sec:si_ernst}

All spectra in this section were acquired on a \SI{500}{\MHz} Bruker AVIII spectrometer, equipped with a triple-resonance TBO probe with a nominal maximum $z$-gradient strength of \SI{50}{\gauss\per\cm}.
The sample used was \SI{45}{\milli\molar} ferulic acid in DMSO-$d_6$.

\subsection{Cost function}

The cost function used in this section is \texttt{maxrealint}, which corresponds simply to a summation of all peaks in the real part of the spectrum:

\begin{equation}
    f = -\sum_i \mathbf{S}_{\mathrm{real},i}
\end{equation}

The minus sign is necessary because optimisation algorithms always seek to \textit{minimise} the cost function.
Thus, in this case, a larger peak intensity corresponds to a more negative cost function, i.e.\ a ``better'' spectrum.

\subsection{Theoretical optima via \texorpdfstring{$T_1$}{T1}}
\label{sec:si_ernst_invrec}

In lieu of a reference grid search, we sought to measure the longitudinal relaxation times ($T_1$) of each peak in the sample.
The Ernst angle $\theta$ which maximises the sensitivity for a given repetition time $\tau_\mathrm{r}$ can then be calculated from this via
\begin{equation}
    \cos \theta = \exp\left(-\frac{\tau_\mathrm{r}}{T_1}\right)
\end{equation}
thus giving us the theoretical optimal value, without having to go through a grid search.
The $T_1$ values extracted from a gradient-enhanced inversion-recovery experiment, and the corresponding Ernst angles (for a repetition time of $\tau_\mathrm{r} = \SI{1.20}{\s}$, as used in the optimisation), are summarised in \cref{tbl:ernst_2dinvrec}.

\begin{table}
    \hbadness=10000
    \centering
    \begin{tabular}{cccc}
        \toprule
        Peak & \proton{} chemical shift (ppm) & $T_1$ (\si{\s}) & $\theta$ ($^\circ$) \\
        \midrule
        1 & 7.49 & 1.750 & 59.8 \\
        2 & 7.27 & 0.977 & 73.0 \\
        3 & 7.08 & 1.279 & 67.0 \\
        4 & 6.79 & 1.615 & 61.6 \\
        5 & 6.36 & 1.415 & 64.6 \\
        6 & 3.49 & 0.949 & 73.6 \\
        \bottomrule
    \end{tabular}
    \caption{
        $T_1$ (at \SI{500}{\MHz}) and Ernst angles for each peak in ferulic acid, calculated for a repetition time of \SI{1.20}{\s}.
    }
    \label{tbl:ernst_2dinvrec}
\end{table}

\subsection{Optimisation}

It is possible to show, both theoretically and experimentally, that in order to maximise the sensitivity per unit time, the relaxation delay should be set to zero and the Ernst angle applied with $\tau_\mathrm{r}$ simply equal to the acquisition time \texttt{AQ}.
See, e.g., \textit{Concepts Magn.\ Reson.}\ \textbf{1992,} \textit{4} (2), 153--160 (DOI: \href{https://doi.org/10.1002/cmr.1820040204}{\texttt{10.1002/cmr.1820040204}}) and \textit{J.\ Mol.\ Spectrosc.}\ \textbf{1970,} \textit{35} (2), 298--305 (DOI: \href{https://doi.org/10.1016/0022-2852(70)90205-5}{\texttt{10.1016/0022-2852(70)90205-5}}).

Furthermore, in order to make an Ernst angle optimisation worthwhile, the repetition time must be on the order of $T_1$.
With longer repetition times on the order of $3T_1$ or so, the Ernst angle is very close to \ang{90}, which transforms the issue into the \ang{90} calibration already described in the previous section.
In the event, we elected to use a repetition time of \SI{1.20}{\s}, which is on the order of the $T_1$ times shown in \cref{tbl:ernst_2dinvrec}.

In the first optimisation, we set the region under optimisation to be just the aromatic and olefinic peaks between 6 and \SI{8}{\ppm} (i.e.\ peaks 1--5).
Instead of changing the actual spectral window, this is most easily done using the \texttt{dpl} command in TopSpin; for more information, please refer to the user documentation.

\begin{table}
    \hbadness=10000
    \centering
    \begin{tabular}{ccccc}
        \toprule
        Entry & Algorithm    & Optimum found ($^\circ$) & $n_\text{fev}$ & Time taken (\si{\s}) \\
        \midrule
        1     & Nelder--Mead & 67.5--73.1               & 9--13          & 91--132              \\
        2     & MDS          & 67.5--73.1               & 9              & 90--92               \\
        3     & BOBYQA       & 70.1--70.7               & 7              & 70--71               \\
        \bottomrule
    \end{tabular}
    \caption{
        Ernst angle optimisations on all aromatic and olefinic peaks in ferulic acid (between 6 and \SI{8}{\ppm}).
        The \Poise{} routine used here is: \mintinline[breaklines]{json}{{"name": "ernst", "pars": ["cnst20"], "lb": [10.0], "ub": [90.0], "init": [30.0], "tol": [3.0], "cf": "maxrealint", "au": "poise_1d"}}.
        The pulse programme used was \texttt{zgvfa\_jy} (\cref{sec:si_ernst_pp}).
        Key acquisition parameters: \texttt{NS=1}, \texttt{DS=4}, \texttt{D1=0}, \texttt{SW=14}, \texttt{TD=16384}, \texttt{AQ=1.17}.
        \nfev{}
        \fiveruns{}
    }
    \label{tbl:ernst_fivepeaks}
\end{table}

Of course, the user may select the spectral bounds themselves in order to maximise the overall sensitivity for the entire spectrum, a particular region, or even a specific peak of interest.
In the second case, we ran an optimisation on only the peak at \SI{6.79}{\ppm}.

\begin{table}
    \hbadness=10000
    \centering
    \begin{tabular}{ccccc}
        \toprule
        Entry & Algorithm    & Optimum found (\si{\s}) & $n_\text{fev}$ & Time taken (\si{\s}) \\
        \midrule
        1     & Nelder--Mead & 60.0--67.5              & 9--11          & 91--111              \\
        2     & MDS          & 65.6--67.5              & 11             & 110--111             \\
        3     & BOBYQA       & 60.0--65.2              & 6--7           & 59--71               \\
        \bottomrule
    \end{tabular}
    \caption{
        Ernst angle optimisations on the peak at \SI{6.79}{\ppm} in ferulic acid.
        The POISE routine and all acquisition parameters are the same as in \cref{tbl:ernst_fivepeaks}, but the spectral region under optimisation was set to be 6.71--\SI{6.87}{\ppm} via the \texttt{dpl} TopSpin command.
        \nfev{}
        \fiveruns{}
    }
    \label{tbl:ernst_onepeak}
\end{table}

\subsection{Variable flip angle pulse-acquire pulse programme}
\label{sec:si_ernst_pp}

Note that the following pulse programme, which requires \texttt{p1} to be the calibrated \ang{90} pulse width and sets \texttt{cnst20} to the flip angle, should only be used if the \textit{actual flip angle} is of interest.

If the sole aim is to maximise sensitivity, then the standard Bruker \texttt{zg} pulse programme suffices.
The routine should be modified to optimise \texttt{p1} directly instead of the flip angle itself.
In essence, this means that we do not care what the flip angle is, we merely want the pulse width which maximises the sensitivity.
This approach has an advantage in that \texttt{p1} does not need to be calibrated beforehand.

\begin{minted}{text}
;zgvfa_jy
;
;download from: https://git.io/JnokS
;
;pulse-acquire with variable flip angle (set by cnst20)
;set p1 to be calibrated 90 degree pulse width
;Jonathan Yong - University of Oxford - 20 June 2021
;
;$CLASS=HighRes
;$DIM=1D
;$TYPE=
;$SUBTYPE=
;$COMMENT=


#include <Avance.incl>


"p20=p1*cnst20/90"

"acqt0=-p20*2/3.1416"


1 ze
2 30m
  d1
  p20 ph1
  go=2 ph31
  30m mc #0 to 2 F0(zd)
exit


ph1=0 2 2 0 1 3 3 1
ph31=0 2 2 0 1 3 3 1


;pl1 : f1 channel - power level for pulse (default)
;p1 : f1 channel -  high power pulse
;d1 : relaxation delay; 1-5 * T1
;ns: 1 * n, total number of scans: NS * TD0
;cnst20: excitation flip angle in degrees
\end{minted}

\section{NOE mixing time}

All spectra in this section were acquired on a \SI{700}{\MHz} Bruker AVIII spectrometer, equipped with a TCI H/C/N cryoprobe with a nominal maximum $z$-gradient strength of \SI{53}{\gauss\per\cm}.
The sample used was \SI{120}{\milli\molar} 3-fluorophenylboronic acid in DMSO-$d_6$.

\subsection{Cost function}

The cost function used for the optimisation here, named \texttt{noe\_1d}, is pre-installed with POISE.
It is only applicable to 1D selective NOE spectra.
In short, it takes the processed spectrum, removes a region of \SI{50}{\Hz} around the selectively excited peak, and calculates the intensity of the remaining parts of the spectrum.
Since we aim to \textit{maximise} this intensity, the cost function must return the \textit{negative} of this intensity.
Further details are best explained through the comments in the source code, which is reproduced here:

\begin{minted}{python}
def noe_1d():
    """
    Measures the intensity of peaks in the spectrum, *except* for anything
    within 25 Hz of the selectively excited peak.
    """
    # SPOFFS2 plus O1 is the frequency of the selective pulse, in Hz.
    # We can get the values of both using getpar(), which automatically
    # converts numeric parameters to floats.
    f = getpar("SPOFFS2") + getpar("O1")
    # Define the (rough) bandwidth of the selective pulse. We will ignore the
    # region of the spectrum that is bw Hz wide and centred on the selective
    # pulse.
    bw = 50.0
    # To get the spectrum *without* the ignored region, we split up the
    # spectrum into two "portions". We need to construct the 'bounds' parameter
    # for both portions.
    sfo1 = getpar("SFO1")
    lowerhalf = f"{(f + bw/2)/sfo1:.3f}.."   # frequencies (f + bw)/2 and above
    upperhalf = f"..{(f - bw/2)/sfo1:.3f}"   # frequencies up to (f - bw)/2
    # We use get1d_real() to get both portions of the spectrum, passing the
    # bounds as appropriate. The intensities of the spectrum are obtained by
    # summing all points (np.sum).
    # Note that manually passing the bounds parameter will override the F1P/F2P
    # parameters if set by the user (this is probably a good thing).
    upper_integral = np.sum(get1d_real(bounds=upperhalf))
    lower_integral = np.sum(get1d_real(bounds=lowerhalf))
    # Lastly, we take the negative absolute value of the resulting sum, in
    # order to make sure that the resulting cost function is negative
    # regardless of how the NOE peaks are phased (positive or negative).
    return -abs(upper_integral + lower_integral)
\end{minted}

\subsection{Reference grid search}

The grid search was carried out using actual 2D NOESY spectra; this has the advantage of proving that the mixing time optimised using 1D spectra does in fact accurately represent the best mixing time for 2D spectra.
The intensity of each crosspeak here refers to the intensity of the most negative point within a \SI{0.12}{\ppm} by \SI{0.12}{\ppm} box centred on the crosspeak.
A broad minimum around a mixing time of 3--4 seconds can be seen.

\begin{figure}
    % ./figures/noesy_scan.py
    \centering
    \includegraphics[width=0.7\textwidth]{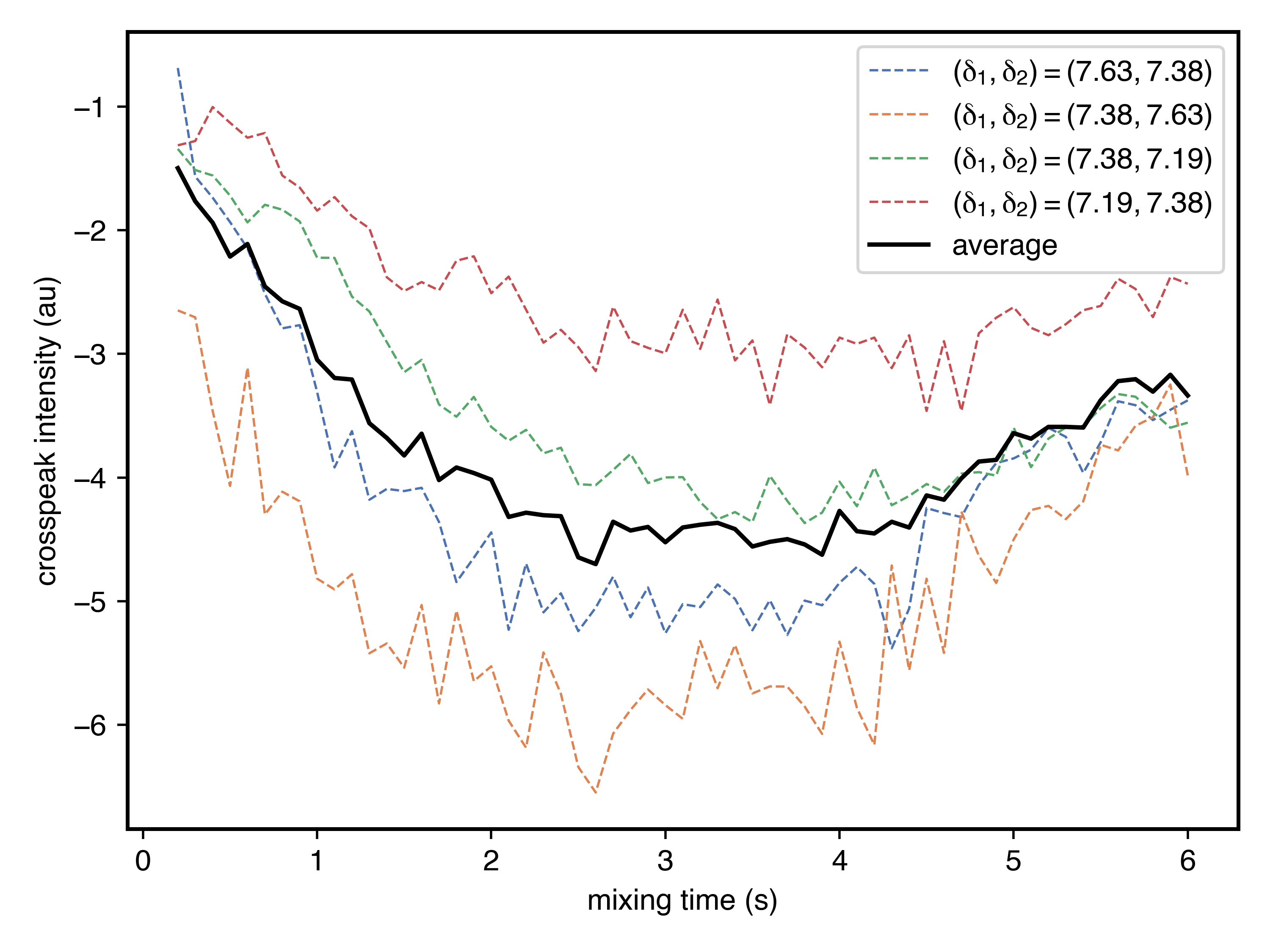}
    \caption{Dependence of crosspeak intensity on the NOE mixing time (\texttt{d8}).}
    \label{fig:noesy_scan}
\end{figure}

\subsection{Optimisation}

\begin{table}
    \hbadness=10000
    \centering
    \begin{tabular}{ccccc}
        \toprule
        Entry & Algorithm    & Optimum found (\si{\s}) & $n_\text{fev}$ & Time taken (\si{\s}) \\
        \midrule
        1     & Nelder--Mead & 3.25--3.88              & 16--18         & 268--312             \\
        2     & MDS          & 3.63--3.75              & 16--18         & 269--305             \\
        3     & BOBYQA       & 3.38--3.80              & 6--10          & 88--162              \\
        \bottomrule
    \end{tabular}
    \caption{
        NOE mixing time optimisations.
        The \Poise{} routine used here is: \mintinline[breaklines]{json}{{"name": "1dnoe", "pars": ["d8"], "lb": [0.2], "ub": [6.0], "init": [0.5], "tol": [0.1], "cf": "noe_1d", "au": "poise_1d"}}.
        The pulse programme used here is \texttt{spfgsenoezs2.1\_jy} (\cref{sec:si_noesy_pp}).
        Key acquisition parameters: \texttt{NS=2}, \texttt{DS=1}, \texttt{D1=1}.
        \nfev{}
        \fiveruns{}
    }
    \label{tbl:noe}
\end{table}

\subsection{1D NOE pulse programme}
\label{sec:si_noesy_pp}

The 1D NOE spectra were acquired with a modified pulse programme with two inversion pulses during the NOE mixing time to minimise artefacts arising from relaxation.
(In principle, these artefacts do not contribute to the cost function as they ideally sum to zero due to their antiphase nature.
However, it is almost always a good idea to try to minimise noise in the cost function where possible.)
The pulse programme is attached below, and can also be found in the main \Poise{} repository.
For most compounds, the existing family of Bruker pulse sequences \texttt{selnogpzs} should suffice.

\begin{minted}{text}
;spfgsenoezs2.1_jy
;
;download from: https://git.io/JUE0P
;
;Tim Claridge & Jonathan Yong - University of Oxford
;1D NOESY using selective refocussing with two shaped pulses
;dipolar coupling may be due to noe or chemical exchange.
;H. Kessler, H. Oschkinat, C. Griesinger & W. Bermel,
;   J. Magn. Reson. 70, 106 (1986)
;J. Stonehouse, P. Adell, J. Keeler & A.J. Shaka, J. Am. Chem. Soc 116,
;   6037 (1994)
;K. Stott, J. Stonehouse, J. Keeler, T.L. Hwang & A.J. Shaka, 
;   J. Am. Chem. Soc 117, 4199-4200 (1995)

;Tim C 2/2008 Using two nulling 180s and ZQC filter at end of mixing time
;Includes adiabatic pulse/gradient combination for ZQC suppression
;Thrippleton & Keeler, Angew. Chem. Int. Ed., 2003, 42, 3938-3941.
;Tim C 10/2007 updated to match Bruker format
;Tim C 11/2008 matches Bruker ZQS nomenclature 
;Jon Y 21/07/2020 added CALC_SPOFFS acquisition flag

;$CLASS=HighRes
;$DIM=1D
;$TYPE=
;$SUBTYPE=
;$COMMENT=

#include <Avance.incl>
#include <Grad.incl>

"p2=p1*2"
"d21=d8*0.3-p16-d16-3u"
"d22=d8*0.5-p16*2-d16*2-6u"
"d23=d8*0.2-p16-d16*2-p32-53u"

#ifdef CALC_SPOFFS
"spoff2=bf1*(cnst21/1000000)-o1"
#else
#endif /*CALC_SPOFFS*/

1 ze
2 30m
  20u BLKGRAD
  d1
  50u UNBLKGRAD
  (p1 ph1):f1
  3u
  p16:gp1
  d16 pl0:f1
  3u
  p12:sp2:f1 ph2:r
  3u
  p16:gp1
  d16 pl1:f1
  3u
  (p1 ph4):f1
  d21
  p16:gp3
  d16
  3u
  (p2 ph5):f1
  3u
  p16:gp3*-1
  d16
  d22
  p16:gp4
  d16
  3u
  (p2 ph5):f1
  3u
  p16:gp4*-1
  d16
  d23
  20u pl0:f1
  10u gron0		;ZQC suppression
  (p32:sp29 ph7):f1
  20u groff
  d16 pl1:f1
  (p1 ph6):f1
  go=2 ph31
  30m mc #0 to 2 F0(zd)
  20u BLKGRAD
exit


ph1=0 2 
ph2=0 0 1 1 2 2 3 3
ph4=0
ph5=0
ph6=0
ph7=0
ph31=0 2 2 0 0 2 2 0 


;pl0 : 120dB
;pl1 : f1 channel - power level for pulse (default)
;sp2: f1 channel - shaped pulse
;sp29: f1 channel - shaped pulse (adiabatic sweep)
;p1 : f1 channel -  90 degree high power pulse
;p2 : f1 channel - 180 degree high power pulse
;p12: f1 channel - 180 degree shaped pulse
;p16: homospoil/gradient pulse                            [1 msec]
;p32: f1 channel - 180 shaped adiabatic pulse
;		smoothed CHIRP
;d1 : relaxation delay; 1-5 * T1
;d8 : mixing time
;d16: delay for homospoil/gradient recovery
;d21: d8*0.3-p16-d16-3u
;d22: d8*0.5-p16*2-d16*2-6u
;d23: d8*0.2-p16-d16*2-p32-53u
;NS: 2 * n
;DS: 4 

;cnst21: chemical shift for selective pulse (offset, in ppm)


;choose p12 according to desired selectivity
;the flip-angle is determined by the amplitude
;set O1 on resonance on the multiplet to be excited or use spoffs

;use gradient ratio:   gp 0 : gp 1 : gp 3 : gp 4
;                      ~10  :  17  :  40  :  25

;for z-only gradients:
;gpz0: 10% or so
;gpz1: 17%
;gpz3: 40%
;gpz4: 25%

;use gradient files:   
;gpnam1: SINE.100
;gpnam2: SINE.100
;gpnam3: SINE.100
;gpnam4: SINE.100
\end{minted}

\section{Inversion-recovery}
\label{sec:si_invrec}

All spectra in this section were acquired on a \SI{500}{\MHz} Bruker AVIII spectrometer, equipped with a triple-resonance TBO probe with a nominal maximum $z$-gradient strength of \SI{50}{\gauss\per\cm}.
The sample used was \SI{45}{\milli\molar} ferulic acid in DMSO-$d_6$.

\subsection{Cost function}

In this optimisation, we sought to make the real part of the spectrum as close to zero as possible: this indicates the ``null'' in the inversion-recovery experiment where the delay $\tau$ is exactly $T_1 \ln 2$.
This is accomplished by the \texttt{zerorealint} cost function, which comes installed with POISE:

\begin{equation}
    f = \sum_i \left\lvert \mathbf{S}_{\mathrm{real},i} \right\rvert
\end{equation}

\subsection{2D inversion-recovery}

$T_1$ for each peak was measured via a gradient-selected inversion-recovery experiment.
The data presented here is the same as in \cref{tbl:ernst_2dinvrec}, but additionally includes the theoretical optimum of $T_1 \ln 2$ for each peak.

\begin{table}
    \hbadness=10000
    \centering
    \begin{tabular}{cccc}
        \toprule
        Peak & \proton{} chemical shift (ppm) & $T_1$ (\si{\s}) & $T_1\ln 2$ (\si{\s}) \\
        \midrule
        1 & 7.49 & 1.750 & 1.213 \\
        2 & 7.27 & 0.977 & 0.677 \\
        3 & 7.08 & 1.279 & 0.887 \\
        4 & 6.79 & 1.615 & 1.119 \\
        5 & 6.36 & 1.415 & 0.981 \\
        6 & 3.49 & 0.949 & 0.658 \\
        \bottomrule
    \end{tabular}
    \caption{
        $T_1$ (at \SI{500}{\MHz}) and the corresponding theoretical null in the inversion-recovery profile, given by $T_1\ln 2$, for each peak in ferulic acid.
    }
    \label{tbl:invrec_2dinvrec}
\end{table}

\subsection{Optimisation}

As previously mentioned, it is possible to use the \texttt{dpl} TopSpin command to restrict the spectral region that the cost function acts on.
In the first optimisation, we set this region to be between 6 and \SI{8}{\ppm}.
This provides an ``averaged'' value of $T_1 \ln 2$, which can be useful in e.g.\ choosing relaxation delays for 2D experiments, which invariably need to be a compromise between multiple peaks.

\begin{table}
    \hbadness=10000
    \centering
    \begin{tabular}{ccccc}
        \toprule
        Entry & Algorithm    & Optimum found (\si{\s}) & $n_\text{fev}$ & Time taken (\si{\s}) \\
        \midrule
        1     & Nelder--Mead & 0.938--0.969            & 14--16         & 204--235             \\
        2     & MDS          & 0.956--0.975            & 16             & 233--235             \\
        3     & BOBYQA       & 0.953--0.971            & 9--11          & 130--160             \\
        \bottomrule
    \end{tabular}
    \caption{
        Inversion-recovery optimisations on all aromatic and olefinic peaks in ferulic acid (between 6 and \SI{8}{\ppm}.
        The \Poise{} routine used here is: \mintinline[breaklines]{json}{{"name": "invrec", "pars": ["d27"], "lb": [0.35], "ub": [1.75], "init": [0.6], "tol": [0.01], "cf": "zerorealint", "au": "poise_1d"}}.
        The pulse programme used was \texttt{t1irgp1d\_jy} (\cref{sec:si_invrec_pp}).
        Key acquisition parameters: \texttt{NS=1}, \texttt{DS=0}, \texttt{D1=5}, \texttt{AQ=3.6} (note that there is also an additional ca.\ 5 seconds between experiments due to pulse programme compilation, etc.).
        \nfev{}
        \fiveruns{}
    }
    \label{tbl:invrec_fivepeaks}
\end{table}

In the second optimisation we focused on only the peak at \SI{7.08}{\ppm}.
This provides a relatively ``fast'' method of determining $T_1$ for one specific peak in a spectrum.

\begin{table}
    \hbadness=10000
    \centering
    \begin{tabular}{ccccc}
        \toprule
        Entry & Algorithm    & Optimum found (\si{\s}) & $n_\text{fev}$ & Time taken (\si{\s}) \\
        \midrule
        1     & Nelder--Mead & 0.863--0.875            & 14             & 202--205             \\
        2     & MDS          & 0.863--0.869            & 14             & 203--204            \\
        3     & BOBYQA       & 0.862--0.873            & 9--10          & 128--145             \\
        \bottomrule
    \end{tabular}
    \caption{
        Inversion-recovery optimisations on only the peak at \SI{7.08}{\ppm}.
        The POISE routine and all acquisition parameters are the same as in \cref{tbl:invrec_fivepeaks}, but the spectral region under optimisation was set to be 7.02--\SI{7.15}{\ppm} via the \texttt{dpl} TopSpin command.
        \nfev{}
        \fiveruns{}
    }
    \label{tbl:invrec_onepeak}
\end{table}

\subsection{1D inversion-recovery pulse programme}
\label{sec:si_invrec_pp}

\begin{minted}{text}
;t1irgp1d_jy
;
;download from: https://git.io/Jnok6
;
;T1 measurement using inversion recovery 
;using broadband 1H inversion pulse and field gradients
;1D version with only one increment
;set inversion-recovery delay to d27
;
;Tim Claridge & Jonathan Yong - University of Oxford - 20 June 2021
;
;$CLASS=HighRes
;$DIM=1D
;$TYPE=
;$SUBTYPE=
;$COMMENT=

#include <Avance.incl>
#include <Delay.incl>
#include <Grad.incl>

"acqt0=-p1*2/3.1416"

1 ze
2 50m BLKGRAD
  d1 
  "TAU=(d27-p19-d16-4u)"
  if "TAU > 0.96"
  {
  "DELTA=TAU-0.45"
  }
  50u UNBLKGRAD
  4u pl0:f1
  (p44:sp30 ph1):f1
  p19:gp1 
  d16 
  if "TAU > 0.96"
  { 
  0.4s 
  50m BLKGRAD
  DELTA pl1:f1
  }
  else
  {
  TAU pl1:f1
  } 
  p1 ph2
  go=2 ph31
  50m mc #0 to 2 F0(zd)
50m BLKGRAD
exit

ph1=0 2 
ph2=0 0 2 2 1 1 3 3
ph31=0 0 2 2 1 1 3 3

;pl1 : f1 channel - power level for pulse (default)
;p1 : f1 channel -  90 degree high power pulse
;d1 : relaxation delay; 1-5 * T1
;ns: 1
;ds: 0
;spnam30: 1H inversion [Bip720,50,20.1]
;p44: broadband inversion pulse [240 us]
;d27: inversion-recovery delay
\end{minted}

\section{ASAP-HSQC}

All spectra in this section were acquired on a \SI{700}{\MHz} Bruker AVIII spectrometer, equipped with a TCI H/C/N cryoprobe with a nominal maximum $z$-gradient strength of \SI{53}{\gauss\per\cm}.
The sample used was \SI{120}{\milli\molar} 3-fluorophenylboronic acid in DMSO-$d_6$.

\subsection{Cost function}

The cost function used here, \texttt{asaphsqc}, simply takes the $f_2$ projection of the acquired 2D spectrum and calculates the total intensity by summing up all points.
Unlike the previous cost functions, it is not builtin: to enable this cost function, please uncomment the \texttt{asaphsqc()} function in \texttt{poise\_backend/costfunctions.py}.

\subsection{Reference grid search}

Crosspeak intensities here refer to the maximum peak height, i.e.\ the intensity of the highest point found within a \SI{3}{\ppm} ($f_1$) by \SI{0.12}{\ppm} ($f_2$) box centred on the crosspeak.

\begin{figure}
    % ./figures/hsqc_scan.py
    \centering
    \includegraphics[width=0.7\textwidth]{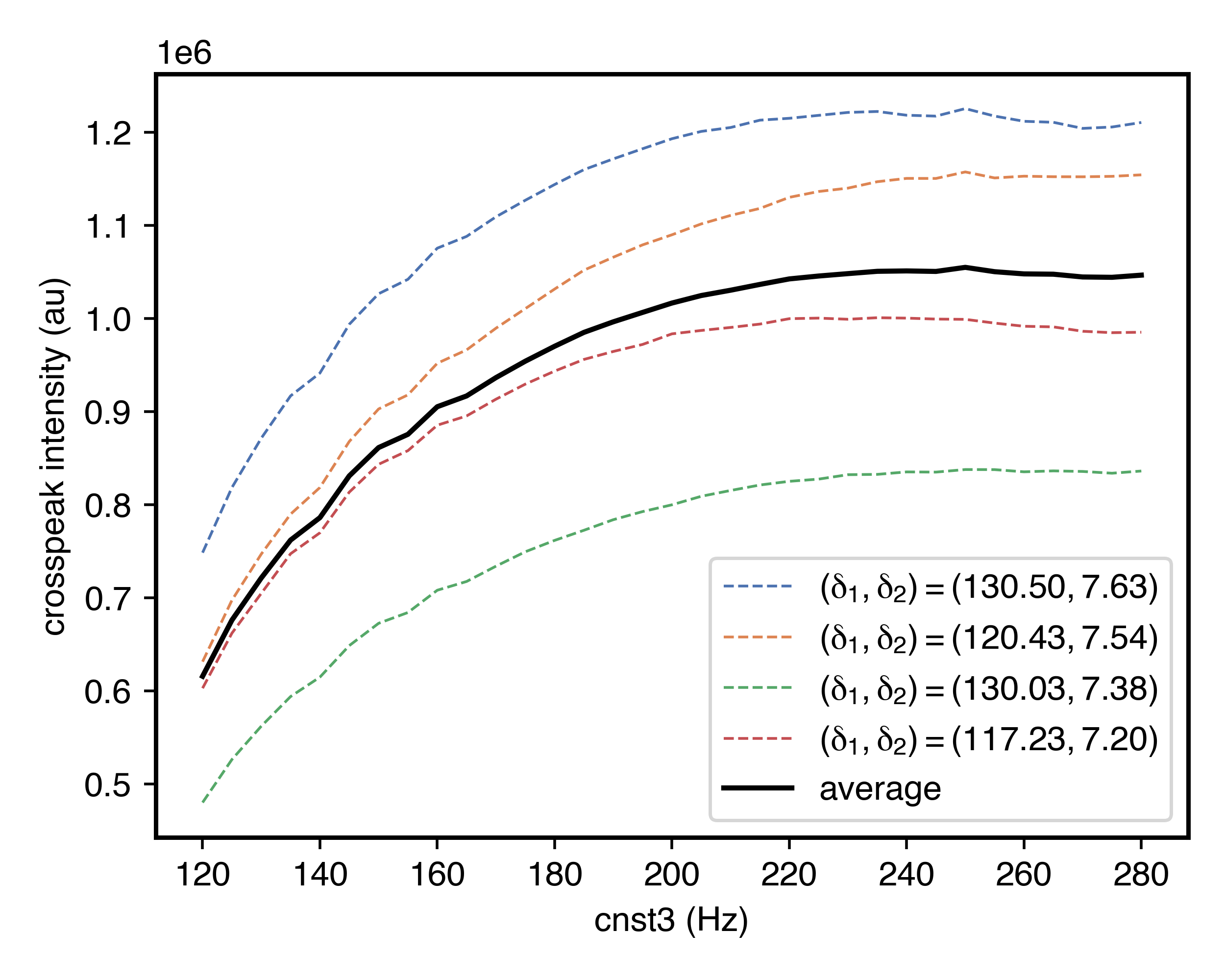}
    \caption{Dependence of crosspeak intensity on the INEPT delay $\Delta = (1 / 4 \cdot \mathtt{cnst3})$.}
    \label{fig:hsqcscan}
\end{figure}

The exact point of maximum intensity is not obvious (and not really useful in this case, as essentially \textit{any} point near the maximum will give essentially the same result).
However, broadly speaking, it can be seen that the intensity reaches a maximum at around \SI{230}{Hz}, after which it plateaus off.

\subsection{Optimisation}

\begin{table}
    \hbadness=10000
    \centering
    \begin{tabular}{ccccc}
        \hline
        Entry & Algorithm    & Optimum found & $n_\text{fev}$ & Time taken (\si{\s}) \\
        \hline
        1     & Nelder--Mead & 250.0--256.3  & 8--9           & 169--195             \\
        2     & MDS          & 237.5--243.8  & 8              & 171--179             \\
        3     & BOBYQA       & 229.8--245.6  & 4--7           & 114--157             \\
        \hline
    \end{tabular}
    \caption{
        ASAP-HSQC INEPT delay optimisations.
        The \Poise{} routine used was: \mintinline[breaklines]{json}{{"name": "asaphsqc", "pars": ["cnst3"], "lb": [120.0], "ub": [280.0], "init": [150.0], "tol": [10.0], "cf": "asaphsqc", "au": "poise_2d"}}.
        The pulse programme used here is \texttt{asaphsqc\_jy} (\cref{sec:si_hsqc_pp}).
        Key acquisition parameters: \texttt{NS=1}, \texttt{DS=2}, \texttt{TD1=36}, \texttt{TD2=512}, \texttt{D1=0.1}.
        \nfev{}
        \fiveruns{}
    }
    \label{tbl:hsqc}
\end{table}

\subsection{ASAP-HSQC pulse programme}
\label{sec:si_hsqc_pp}

The pulse programme used here is almost identical to the published version but includes several minor changes for convenience.

\begin{minted}{text}
;asaphsqc_jy
;
;download from: https://git.io/JUHOo
;
;modified from version by Luy et al. to include internal power calculations for
;DIPSI-2 mixing, as well as URPs
;order of gradients also switched round to correct frequencies
;added extra delay d30 before experiment starts
;Jonathan Yong - University of Oxford - 26 August 2020
;
;ASAP-HSQC
;2D H-1/X correlation via double inept transfer
;phase sensitive using Echo/Antiecho-TPPI gradient selection
;with decoupling during acquisition
;with DIPSI-2 mixing after acquisition for ASAP
;using shaped pulses on f1 and f2.
;David Schulze-Sunninghausen, Johanna Becker, Burkhard Luy
;"Rapid Heteronuclear Single Quantum Correlation NMR Spectra at Natural Abundance"
;J. Am. Chem. Soc. 2014, 136 (4), 1242-1245

#include <Avance.incl>
#include <Grad.incl>
#include <Delay.incl>

"cnst12=1/(1+sfo2/sfo1)"
"cnst13=(1-sfo2/sfo1)"
define list<gradient> EA1 = { 1 cnst12 }
define list<gradient> EA2 = { cnst13 1 }
"p2=p1*2"
"d0=3u"
"d11=30m"
"d16=200u"
"d20=1m"
"FACTOR1=(d9/(p6*115.112))/2" 
"l1=FACTOR1*2" ; TOCSY loop counter                 
"d0=3u"
"in0=inf1/2"
"d4=1s/(4*cnst4)" ; 1/(4J)XH 
"d3=1s/(4*cnst3)" ; reduced d4 
"DELTA=0.5*(2*d3+p21-p1)"
"DELTA1=0.5*(2*d3-p21+p1)"
"DELTA4=d4-p16-d16-3u"
"DELTA3=p16+3u+d16+p21-p25+p1"
"DELTA2=0.5*(DELTA3+p16+d16+3u-2*d0)"

"cnst21=20000"
"spw21=plw1*(cnst21/(1000000/(4*p1)))*(cnst21/(1000000/(4*p1)))"
"spw22=spw21"
"cnst23=10000"
"spw23=plw2*(cnst23/(1000000/(4*p3)))*(cnst23/(1000000/(4*p3)))" 
"spw24=spw23"
"spw25=spw23"
"spw26=spw23"

; calculate mixing length
"cnst9=p6*115.112*l1/1000"
"cnst9=cnst9" ; so that it's displayed in ased
; calculate 90deg pulse duration & power level for mixing
"p6=(1000000/(4*cnst10))"
"plw10=plw1*(p1/p6)*(p1/p6)"

1 ze
  d30
  d11 pl12:f2 
2 d1 do:f2
3 d20 UNBLKGRAD 
  p16:gp2
  d16 pl10:f1
						;begin DIPSI2
4 p6*3.556 ph23
  p6*4.556 ph25
  p6*3.222 ph23
  p6*3.167 ph25
  p6*0.333 ph23
  p6*2.722 ph25
  p6*4.167 ph23
  p6*2.944 ph25
  p6*4.111 ph23
  
  p6*3.556 ph25
  p6*4.556 ph23
  p6*3.222 ph25
  p6*3.167 ph23
  p6*0.333 ph25
  p6*2.722 ph23
  p6*4.167 ph25
  p6*2.944 ph23
  p6*4.111 ph25

  p6*3.556 ph25
  p6*4.556 ph23
  p6*3.222 ph25
  p6*3.167 ph23
  p6*0.333 ph25
  p6*2.722 ph23
  p6*4.167 ph25
  p6*2.944 ph23
  p6*4.111 ph25

  p6*3.556 ph23
  p6*4.556 ph25
  p6*3.222 ph23
  p6*3.167 ph25
  p6*0.333 ph23
  p6*2.722 ph25
  p6*4.167 ph23
  p6*2.944 ph25
  p6*4.111 ph23
 lo to 4 times l1
						;end DIPSI2
  p16:gp3
  d16 
  (p21:sp21 ph1):f1
  DELTA   
  (center (p22:sp22 ph1):f1 (p22:sp23 ph6):f2)
  DELTA1 pl1:f1
  (ralign (p1 ph1):f1 (p21:sp24 ph9):f2)
  DELTA2                                          
  d0
  (p22:sp22 ph5):f1
  d0
  DELTA2 pl1:f1 
  (lalign (p1 ph7):f1 (p25:sp25 ph10):f2)
  DELTA3 
  (p22:sp22 ph1):f1
  3u
  p16:gp1*EA1
  d16 
  (p21:sp26 ph10):f2
  d4
  (center (p22:sp22 ph1):f1 (p22:sp23 ph6):f2)
  3u
  p16:gp1*EA2
  d16 pl12:f2 
  DELTA4 BLKGRAD
  go=2 ph31 cpd2:f2
  d1 do:f2 mc #0 to 2 
  F1EA(igrad EA1 & igrad EA2, id0 & ip9*2 & ip6*2 & ip31*2)
exit

ph1=0 
ph2=1
ph5=0 0 2 2
ph6=0
ph7=3
ph8=3
ph9=2 0
ph10=0 0 0 0 2 2 2 2
ph23=3
ph25=1
ph31=0 2 0 2 2 0 2 0
  
;pl0 : 0W
;pl1 : f1 channel - power level for pulse (default)
;pl2 : f2 channel - power level for pulse (default)
;pl10 : f1 channel - power level for mixing sequence (DIPSI-2)
;pl12: f2 channel - power level for CPD/BB decoupling
;p1 : f1 channel -  90 degree high power pulse
;p2 : f1 channel -  180 degree high power pulse
;p6 : f1 channel - 90 degree pulse for mixing sequence (DIPSI-2)
;p21 : 550u excitation pulses
;p22 : 600u UR and Inversion pulses
;p25 : 1100u 13C refocusing pulse
;p16: homospoil/gradient pulse
;spnam21 : jc01_BEBOP_zx_550u_BW10_RF20_pm20_Hc0.99997119.pul
;spnam22 : jc02_BURBOP_x_600u_BW10_RF20_pm20_matched.pul
;spnam23 : jc03_BIBOP_600u_BW37.5_RF10_pm5_matched.pul
;spnam24 : jc05_BEBOP_zy_550u_BW37.5_RF10_pm5_matched.pul
;spnam25 : jc07_BURBOP_y_1100u_BW37.5_RF10_pm5_Hc0.999876221.pul
;spnam26 : jc09_BEBOP_-yz_550u_BW37.5_RF10_pm5_matched.pul
;sp21 : 20 kHz Rf Amplitude
;sp22 : 20 kHz Rf Amplitude
;sp23 : 10 kHz Rf Amplitude
;sp24 : 10 kHz Rf Amplitude
;sp25 : 10 kHz Rf Amplitude
;sp26 : 10 kHz Rf Amplitude
;d0 : incremented delay (2D)                [3 usec]
;d1 : relaxation delay; 1-5 * T1
;d4 : 1/(4J)XH   
;d3 : reduced d4
;d9 : TOCSY mixing time
;d11: delay for disk I/O                    [30 msec]
;d16: delay for homospoil/gradient recovery [200ms]
;d30: delay before experiment starts (use if running extended multizgs)
;cnst4: = J(XH)
;cnst3: reduced cnst4 ('ernst angle')
;inf1: 1/SW(X) = 2 * DW(X)
;in0: 1/(2 * SW(X)) = DW(X)
;nd0: 2
;NS: 1 * n
;DS: >= 16
;td1: number of experiments
;FnMODE: echo-antiecho
;cpd2: decoupling according to sequence defined by cpdprg2
;pcpd2: f2 channel - 90 degree pulse for decoupling sequence

;use gradient ratio:	gp 1 : gp 2 : gp 3
;                       80   : 33   : 43  

;for z-only gradients:
;gpz1: 80%
;gpz2: 33% ; gradient before mixing
;gpz3: 43% ; gradient after mixing

;use gradient files:   
;gpnam1: SMSQ10.100
;gpnam2: SMSQ10.100
;gpnam3: SMSQ10.100
\end{minted}

\section{EPSI}

All spectra in this section were acquired on a \SI{600}{\MHz} Bruker AVIII spectrometer, equipped with a Prodigy \ce{N2} broadband cryoprobe with a nominal maximum $z$-gradient strength of \SI{66}{\gauss\per\cm}.
The sample used was \SI{45}{\milli\molar} ferulic acid in DMSO-$d_6$.

\subsection{Cost function}

The cost function used for EPSI optimisations (called \texttt{epsi\_gradient\_drift}) comes pre-installed with POISE.
It is more involved than usual, as the entirety of the EPSI processing must be carried out within the cost function (\Poise{} itself only providing a function to get the FID).
Briefly, this processing consists of the following steps:

\begin{itemize}
    \item Removal of the ``group delay'' (by circularly shifting the FID).
    \item Reshaping of the 1D FID into a 2D $(k, t_2)$ matrix.
    \item Removing parts of the FID acquired under negative gradients, as well as during the delay between gradients.
    \item Absolute-value processing.
    \item Apodisation.
    \item Removing any rows (i.e.\ any values of $t_2$) for which the maximum intensity is below a certain threshold. This ensures that only parts of the $(k, t_2)$ matrix with strong signal are used for the calculation of the slope.
\end{itemize}

The slope of $k$ versus $t_2$ can then be calculated by finding, for each row, the value of $k$  where the intensity is maximum, and performing a least-squares fitting.
Technically $k/k_{\text{max}}$, not $k$, is used instead of $k$, but $k_{\text{max}}$ is merely a constant factor which has no influence on the results.

For more details, please refer to the source code which is attached below:

\begin{minted}{python}
def epsi_gradient_drift():
    """
    Calculates the amount of 'gradient drift' seen in a 1D EPSI acquisition, as
    reflected by the position of the 'echo' moving over time. This can be
    caused by imbalanced positive and negative gradients.
    """
    from numpy.polynomial.polynomial import Polynomial
    # --- Calculate key parameters -------------------------------
    fid = get1d_fid(remove_grpdly=True)
    # Number of complex points in k-space, i.e. number of points per EPSI
    # gradient. TD2 is total number of real & imag points in FID. L3 is number
    # of EPSI loops (one loop includes both pos + neg gradient). So TD2 / (2 *
    # L3) is the number of points in one EPSI gradient (i.e. only positive
    # gradient). The extra factor of 2 is because we're interested only in
    # complex points.
    td_k = int(getpar("TD") / (2 * 2 * getpar("L3")))
    # Time between consecutive EPSI positive gradients, in seconds
    td_t2 = int(getpar("L3"))
    dw_eff = getpar("AQ") / td_t2

    # --- Perform EPSI processing --------------------------------
    # Reshape into 2D matrix
    ser = fid.reshape((-1, td_k))
    # Discard the part acquired with negative gradients
    ser = ser[0::2, :]

    # Discard any part of the spectrum that was not acquired during an EPSI
    # gradient, i.e. if the delay D6 was nonzero.
    if getpar("D6") > 0:
        ineligible_epsi_points = int(1e6 * getpar("D6") / (getpar("DW") * 2))
        td_k = td_k - ineligible_epsi_points
        ser = ser[:, :td_k]

    # --- Apodisation --------------------------------------------
    # Along k-dimension (Hamming window)
    alpha_0 = 0.54
    ks = np.linspace(0, 1, td_k)
    k_winfunc = (alpha_0
                 - (1 - alpha_0) * np.cos(2 * np.pi * np.linspace(0, 1, td_k)))
    ser = ser * k_winfunc[np.newaxis, :]
    # Along direct dimension
    t2_winfunc = np.sin(np.pi * np.linspace(0, 1, td_t2))
    ser = ser * t2_winfunc[:, np.newaxis]
    abs_ser = np.abs(ser)

    # Calculate k- and t2-axes
    t2_values = np.arange(td_t2) * dw_eff
    k_values = np.linspace(-0.5, 0.5, td_k)

    # Drop all rows (i.e. all values of t2) for which the maximum is less than
    # 20% of the overall maximum.
    threshold_amp = 0.2 * np.max(abs_ser)
    maxima_along_rows = np.max(abs_ser, axis=1)
    indices_to_use = np.nonzero(maxima_along_rows > threshold_amp)
    # Locate the maxima
    maximal_indices_along_k = np.argmax(abs_ser, axis=1)
    # Get the value of k at each maximum
    k_max_values = k_values[maximal_indices_along_k]
    poly = Polynomial.fit(x=t2_values[indices_to_use],
                          y=k_max_values[indices_to_use], deg=1)
    return np.abs(poly.coef[1])
\end{minted}

\subsection{Reference grid search}

The parameter optimised here is \texttt{cnst16}, which represents the ratio of nominal negative gradient amplitude to nominal positive gradient amplitude during the EPSI acquisition.
In the main text, this is referred to as $\alpha$.
On an ideal instrument, the optimised value of \texttt{cnst16} would simply be 1.

\begin{figure}
    % ./figures/epsi_scan.py
    \centering
    \includegraphics[width=0.7\textwidth]{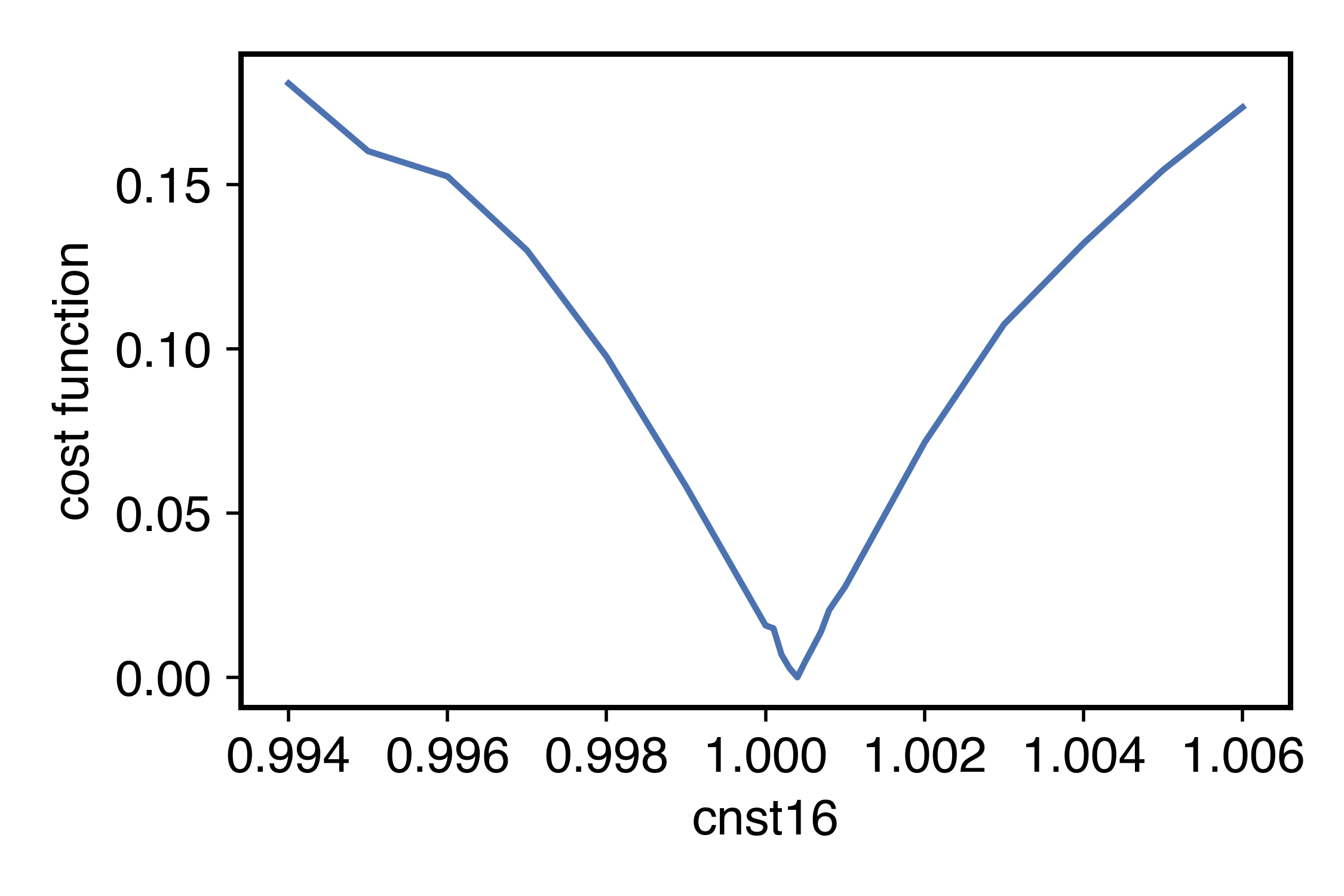}
    \caption{
        Dependence of the cost function on the value of \texttt{cnst16}.
        The minimum occurs at around \texttt{cnst16} = 1.0004.
    }
    \label{fig:epsi_scan}
\end{figure}

\subsection{Optimisation}
\begin{table}

    \hbadness=10000
    \centering
    \begin{tabular}{ccccc}
        \toprule
        Entry & Algorithm    & Optimum found      & $n_\text{fev}$ & Time taken (\si{\s}) \\
        \midrule
        1     & Nelder--Mead & 1.000375           & 10             & 49--52               \\
        2     & MDS          & 1.000375           & 10             & 49--50               \\
        3     & BOBYQA       & 1.000383--1.000388 & 7              & 34--35               \\
        \bottomrule
    \end{tabular}
    \caption{
        Results of EPSI gradient imbalance optimisations.
        The \Poise{} routine used here is: \mintinline[breaklines]{json}{{"name": "epsi", "pars": ["cnst16"], "lb": [0.99], "ub": [1.01], "init": [1.0], "tol": [0.0001], "cf": "epsi_gradient_drift", "au": "poise_1d"}}.
        The pulse programme used was \texttt{zg\_epsi.2\_jy} (\cref{sec:si_epsi_pp}).
        Key acquisition parameters: \texttt{NS=1}, \texttt{DS=0}, \texttt{L4=0}, \texttt{D1=0.5}, \texttt{DW=0.7} (note units of \texttt{DW} are \si{\us}), \texttt{AQ=0.092}, \texttt{GPZ15=35.644} (the maximum gradient amplitude on this machine is \SI{67.5}{G.cm^{-1}}).
        To avoid hardware damage \texttt{AQ} should always be kept short, ideally under \SI{100}{\ms}.
        Note that such a short value of \texttt{D1} is \textit{only} possible because there is always an extra delay between successive experiments required for pulse programme compilation, etc.
        \textbf{A short recovery delay should not be used when DS \textgreater{} 0 or NS \textgreater{} 1.}
        \nfev{}
        \fiveruns{}
    }
    \label{tbl:epsi_opt}
\end{table}

\subsection{Excitation--EPSI pulse programme}
\label{sec:si_epsi_pp}

\begin{minted}{text}
;zg_epsi.2_jy
;
;download from: https://git.io/JYGEa
;
;90 pulse followed by EPSI acquisition, with prephasing gradient

;modified from uftocsy, Giraudeau et al.
;https://madoc.univ-nantes.fr/course/view.php?id=24710&section=1
;IMPORTANT: set p15 + d6 = DW x TD(F2)/(2xL3)

;$CLASS=HighRes
;$DIM=1D
;$TYPE=
;$SUBTYPE=
;$COMMENT=

#include <Avance.incl>
#include <Grad.incl>
#include <De.incl>
#include <Delay.incl>

"d0=3u"
"d11=30m"
"p15=(td*dw/(2*l3))-d6"
"cnst17=cnst16"   ; so that cnst16 shows up in ased

1 ze
  "cnst17=cnst17" ; so that cnst16 shows up in ased
  d11
  30m zd
  ; dummy scans
4 30m
  10u
  4u
  d1 
  4u
  100u UNBLKGRAD
  20u pl1:f1
  (p1 ph1):f1
  10u
  p15:gp25
  10u
5 p15:gp15
  d6 
  p15:gp16
  d6
  lo to 5 times l3
  p27:gp27
  100u BLKGRAD
  lo to 4 times l4
  ; end of dummy scans
2 30m
  10u
  4u
  d1 
  4u
  100u UNBLKGRAD
  20u pl1:f1
  ; excitation
  (p1 ph1):f1
  ; prephasing gradient
  10u
  p15:gp25
  10u
  ; EPSI acquisition
  ACQ_START(ph30,ph31)
  1u DWELL_GEN:f1
3 p15:gp15
  d6 
  p15:gp16*cnst16
  d6
  lo to 3 times l3
  100u BLKGRAD
  rcyc=2
  100u UNBLKGRAD
  p27:gp27
  100u BLKGRAD
  30m mc #0 to 2 F0(zd)
  d17
exit

ph1=0 
ph2=0 1
ph3=0
ph29=0
ph30=0
ph31=0 2

;pl1: f1 channel - power level for pulse (default)
;p1: f1 channel - 90 degree high power pulse
;d1: relaxation delay; 1-5 * T1
;d6: delay between EPSI gradients
;d17: delay after experiment (not necessary)
;p15: EPSI gradient duration
;gpz15: EPSI positive gradient
;gpz16: EPSI negative gradient
;gpz25: prephasing gradient
;gpz27: purge gradient
;ns: 1
;l3: number of loops for EPSI acquisition
\end{minted}

\section{PSYCHE}
\label{sec:si_psyche}

All spectra in this section were acquired on a \SI{600}{\MHz} Bruker AVIII spectrometer, equipped with a Prodigy \ce{N2} broadband cryoprobe with a nominal maximum $z$-gradient strength of \SI{66}{\gauss\per\cm}.
The sample used was \SI{45}{\milli\molar} andrographolide in DMSO-$d_6$.

\subsection{Cost function}

The cost function, named \texttt{specdiff}, is given by:

\begin{equation}
    f = \left|\left|\frac{\mathbf{S}}{||\mathbf{S}||} - \frac{\mathbf{T}}{||\mathbf{T}||}\right|\right|
\end{equation}

where $\mathbf{S}$ and $\mathbf{T}$ are vectors corresponding to the spectrum under optimisation and a target spectrum respectively.
The target spectrum was acquired using a proton pulse-acquire sequence (\texttt{zg}).
The optimisation spectrum was acquired using a \ang{90}--delay--refocusing element--delay--acquire sequence, as described in the main text.
(This pulse programme is attached further below.)
The spectral region used in the optimisation was restricted to \SI{1.15}{\ppm} and above, in order to exclude methyl singlets.
This is not strictly necessary (roughly the same results will be obtained if the entire spectral window is used), but helps to reduce noise in the cost function.

\texttt{specdiff} is not pre-installed with POISE: to enable it, uncomment the \texttt{specdiff()} function in \texttt{poise\_backend/costfunctions.py}.

\subsection{Flip angle reference grid search}

Due to the large amount of time required for multiple-parameter grid searches, we chose to only run this over a single parameter, namely the flip angle $\beta$.

\begin{figure}
    % ./figures/psyche_fascan.py
    \centering
    \includegraphics[width=0.7\textwidth]{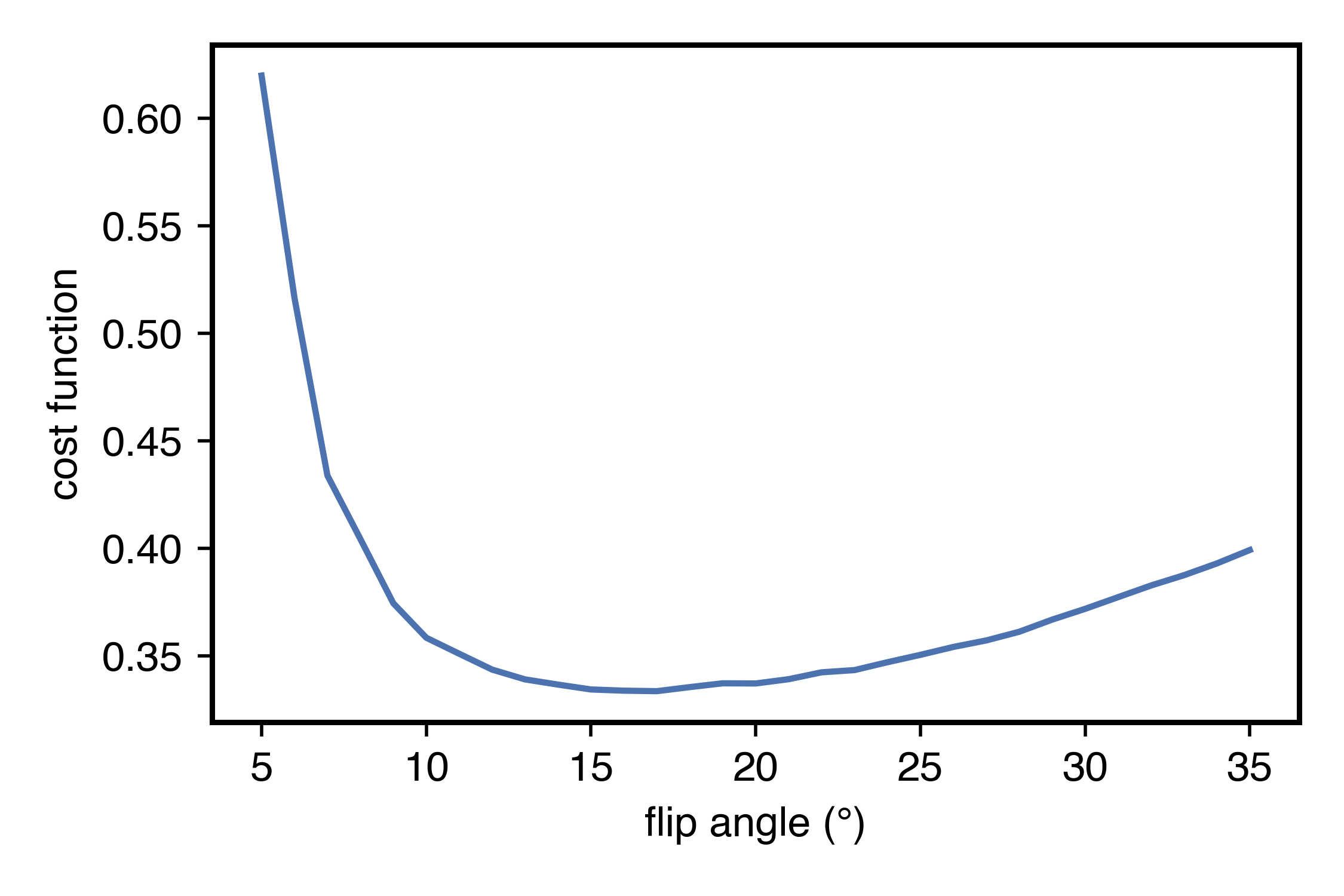}
    \caption{Dependence of cost function on PSYCHE flip angle (\texttt{cnst20}).}
    \label{fig:psychescan}
\end{figure}

\begin{figure}
    \centering
    \includegraphics[width=\textwidth]{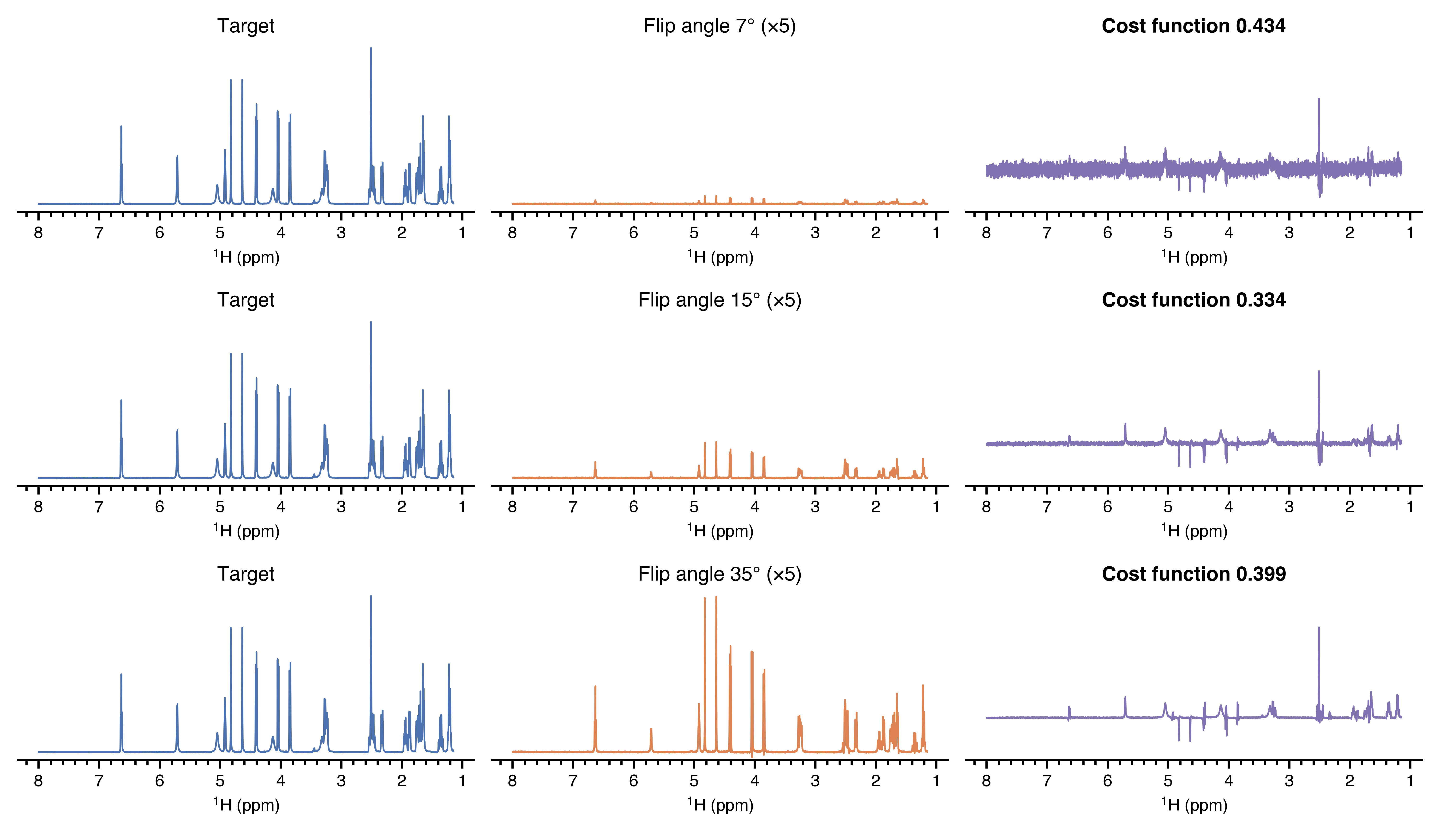}
    \caption{A closer look at the cost function at different flip angles. \textit{Left column:} The target spectrum $\mathbf{T}$ (this is the same for all three rows). \textit{Middle column:} The 1D test spectrum $\mathbf{S}$, acquired using different flip angles (the optimum is around \ang{15}). \textit{Right column:} The vector $\mathbf{S}/||\mathbf{S}|| - \mathbf{T}/||\mathbf{T}||$. One can see how at low flip angles, the main contribution to the cost function is noise. At higher flip angles, the noise contributes less and the increases in cost function are due to phase distortion.}
    \label{fig:psychedetail}
\end{figure}

The cost function is minimised at around 15--\ang{18}.
We can use this opportunity to have a closer look at exactly how this cost function works.
It penalises the low signal-to-noise seen at smaller flip angles: in this case $||\mathbf{S}||$ is small and consequently the noise in $\mathbf{S}/||\mathbf{S}||$ has a larger intensity.
This contributes to a larger value of $\mathbf{S}/||\mathbf{S}|| - \mathbf{T}/||\mathbf{T}||$ and hence a larger value of the cost function.
On the other hand, at larger flip angles this is not a problem: instead one starts to observe phase distortions in the peaks.
This causes $\mathbf{S}/||\mathbf{S}|| - \mathbf{T}/||\mathbf{T}||$ to be larger in the vicinity of the peaks, which also pushes $f$ up.
The cost function is therefore minimised in a ``compromise'' region which has acceptable signal-to-noise but also not too much phase distortion.
This is depicted in \cref{fig:psychedetail}.

\clearpage
\subsection{Optimisations}
\label{sec:si_psyche_optim}

In this section we present the overall results of all the PSYCHE optimisations ran (with 1, 2, 3, and 4 parameters being optimised at once).
From the aggregated results (\cref{tbl:psyche}), it is clear that increasing the number of parameters leads to larger decreases in the cost function, i.e.\ better performance as a refocusing element, although there is a drawback in that more time is required for convergence.
We note here that an increase in the number of parameters leads to greater complexity in the cost function.
This increases the likelihood of converging to a local minimum which may be suboptimal.
The tradeoff between various factors such as potentially improved performance, the time required for optimisation, and the chances of being caught in a shallow local minimum must be carefully weighed when deciding how complicated an optimisation to run.
In general, our opinion is that optimising two parameters (the flip angle and gradient amplitude) provides a good balance between all of these competing factors, and is most suitable for routine use.

First, we present an overview of all the optimisations run.

\begin{table}
    \centering
    \begin{tabular}{cccccccc}
        \hline
         & \multicolumn{3}{c}{Aggregated results from all runs} & \multicolumn{4}{c}{Optimised parameters from best run} \\
        Description & $f_\mathrm{best}$ & $n_\mathrm{fev}$ & Time (s) & $\beta$ ($^\circ$) & $g$ (\%) & $\Delta F$ (kHz) & $\tau_\mathrm{p}$ (ms) \\
        \hline
        Initial point & 0.353        & --     & --        & (25.0) & (2.00) & (10.0) & (30.0) \\
        1 parameter   & 0.340--0.343 & 5--10  & 84--168   & 17.5   & (2.00) & (10.0) & (30.0) \\
        2 parameters  & 0.325--0.338 & 12--33 & 205--565  & 16.5   & 1.00   & (10.0) & (30.0) \\
        3 parameters  & 0.320--0.328 & 18--77 & 315--1344 & 17.4   & 1.73   & 15.9   & (30.0) \\
        4 parameters  & 0.316--0.333 & 39--85 & 705--1504 & 13.9   & 1.45   & 15.9   & 36.0   \\
        \hline
    \end{tabular}
    \caption{
        Summary of optimisations on PSYCHE refocusing element.
        Details of the routines used are described in \cref{tbl:psyche1p,tbl:psyche2p,tbl:psyche3p,tbl:psyche4p}.
        Parameters in parentheses were not optimised (they are inherited from the initial point).
        \nfev{}
        \fiveruns{}
        The optimised parameters in each row are taken from the best run, as judged by the lowest value of the cost function found.
    }
    \label{tbl:psyche}
\end{table}

We follow this with details of the individual classes of optimisation.
For the 1-parameter optimisation, the results are summarised in a similar format to previous tables.
However, for the remainder of the optimisations, we present only summaries of $n_\text{fev}$ and the time taken.
For details of the optima found, we point the interested reader in the direction of the \texttt{poise.log} files, as well as the \texttt{parse\_log()} function to help with parsing these (please refer to the documentation for more details).

\begin{table}
    \hbadness=10000
    \centering
    \begin{tabular}{ccccc}
        \toprule
        Entry & Algorithm    & Optimum found (\si{\degree}) & $n_\text{fev}$ & Time taken (\si{\s}) \\
        \midrule
        1     & Nelder--Mead & 15.0--20.0                   & 8--10          & 135--168             \\
        2     & MDS          & 17.5--20.0                   & 8              & 133--136             \\
        3     & BOBYQA       & 18.4--19.9                   & 5              & 84--85               \\
        \bottomrule
    \end{tabular}
    \caption{
        PSYCHE 1-parameter (flip angle) optimisations.
        The \Poise{} routine used was: \mintinline[breaklines]{json}{{"name": "psyche1", "pars": ["cnst20"], "lb": [10.0], "ub": [35.0], "init": [25.0], "tol": [2.0], "cf": "specdiff", "au": "poise_1d"}}.
        Key acquisition parameters: \texttt{NS=2}, \texttt{DS=1}, \texttt{D1=1.5}.
    }
    \label{tbl:psyche1p}
\end{table}

\begin{table}
    \hbadness=10000
    \centering
    \begin{tabular}{ccccc}
        \toprule
        Entry & Algorithm    & $n_\text{fev}$ & Time taken (\si{\s}) \\
        \midrule
        1     & Nelder--Mead & 18--24         & 307--410             \\
        2     & MDS          & 25--33         & 424--565             \\
        3     & BOBYQA       & 12--16         & 205--271             \\
        \bottomrule
    \end{tabular}
    \caption{
        PSYCHE 2-parameter (flip angle and gradient amplitude) optimisations.
        The \Poise{} routine used was: \mintinline[breaklines]{json}{{"name": "psyche2", "pars": ["cnst20", "gpz10"], "lb": [10.0, 0.2], "ub": [35.0, 5.0], "init": [25.0, 2.0], "tol": [2.0, 0.2], "cf": "specdiff", "au": "poise_1d"}}.
        Key acquisition parameters: \texttt{NS=2}, \texttt{DS=1}, \texttt{D1=1.5}.
    }
    \label{tbl:psyche2p}
\end{table}

\begin{table}
    \hbadness=10000
    \centering
    \begin{tabular}{ccccc}
        \toprule
        Entry & Algorithm    & $n_\text{fev}$ & Time taken (\si{\s}) \\
        \midrule
        1     & Nelder--Mead & 33--43         & 576--770             \\
        2     & MDS          & 46--77         & 804--1344            \\
        3     & BOBYQA       & 18--31         & 315--540             \\
        \bottomrule
    \end{tabular}
    \caption{
        PSYCHE 3-parameter (flip angle, gradient amplitude, and bandwidth) optimisations.
        The \Poise{} routine used was: \mintinline[breaklines]{json}{{"name": "psyche3", "pars": ["cnst20", "gpz10", "cnst21"], "lb": [10.0, 0.2, 1000.0], "ub": [35.0, 5.0, 20000.0], "init": [25.0, 2.0, 10000.0], "tol": [2.0, 0.2, 500.0], "cf": "specdiff", "au": "poise_psyche"}}.
        Key acquisition parameters: \texttt{NS=2}, \texttt{DS=1}, \texttt{D1=1.5}.
    }
    \label{tbl:psyche3p}
\end{table}

\begin{table}
    \hbadness=10000
    \centering
    \begin{tabular}{ccccc}
        \toprule
        Entry & Algorithm    & $n_\text{fev}$ & Time taken (\si{\s}) \\
        \midrule
        1     & Nelder--Mead & 40--47         & 733--845             \\
        2     & MDS          & 57--85         & 1006--1504           \\
        3     & BOBYQA       & 39--62         & 705--1130            \\
        \bottomrule
    \end{tabular}
    \caption{
        PSYCHE 4-parameter (flip angle, gradient amplitude, bandwidth, and pulse duration) optimisations.
        The \Poise{} routine used was: \mintinline[breaklines]{json}{{"name": "psyche4", "pars": ["cnst20", "gpz10", "cnst21", "p40"], "lb": [10.0, 0.2, 1000.0, 5000.0], "ub": [35.0, 5.0, 20000.0, 75000.0], "init": [25.0, 2.0, 10000.0, 30000.0], "tol": [2.0, 0.2, 500.0, 2000.0], "cf": "specdiff", "au": "poise_psyche"}}.
        Key acquisition parameters: \texttt{NS=2}, \texttt{DS=1}, \texttt{D1=1.5}.
    }
    \label{tbl:psyche4p}
\end{table}

The best of each of these optimisations are shown in the next figure.

\begin{figure}
    \centering
    \includegraphics[width=\textwidth]{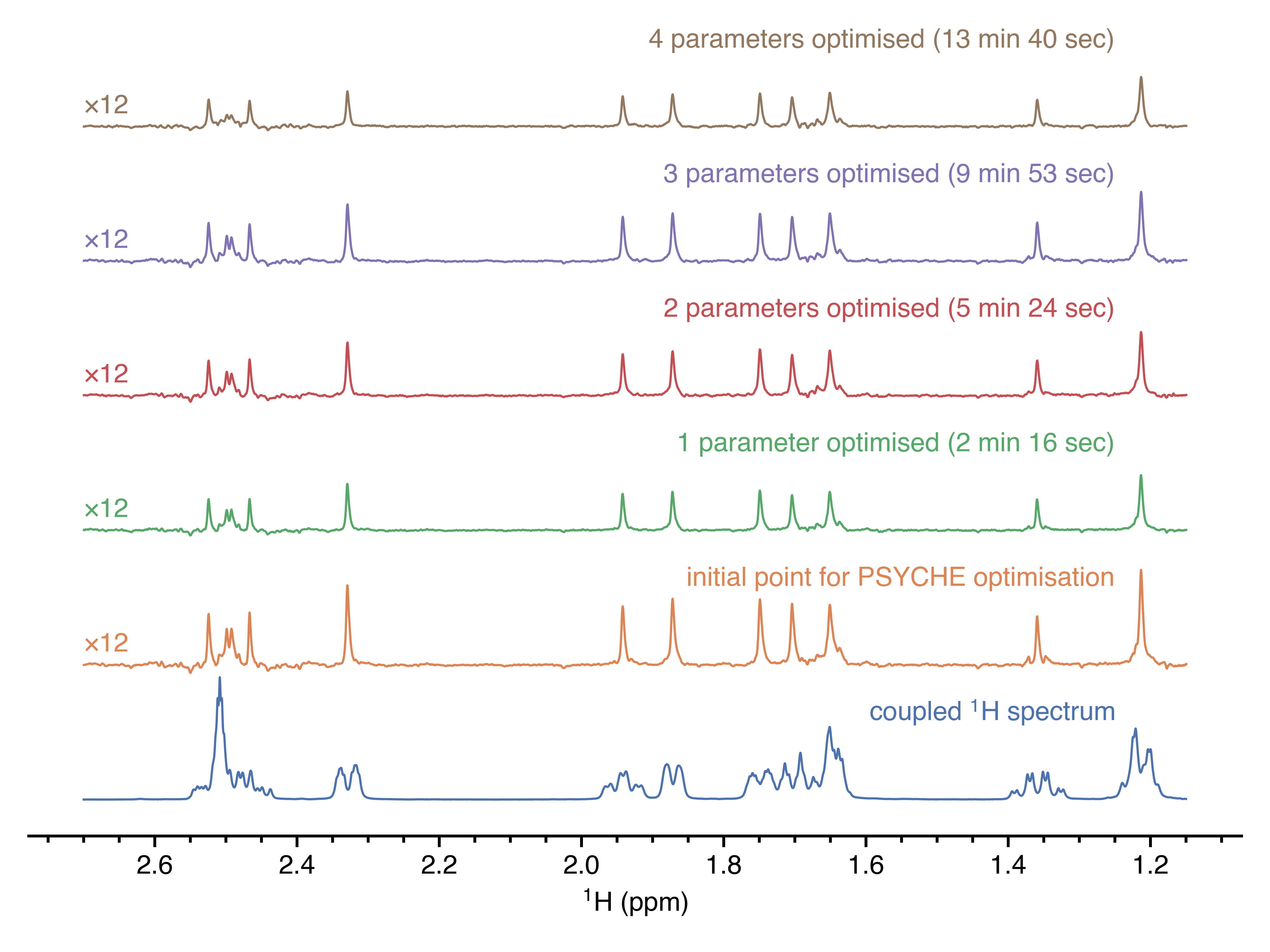}
    \caption{
        Insets of PSYCHE spectra of a sample of \SI{45}{\milli\molar} andrographolide in DMSO-$d_6$, obtained before and after optimisation of up to 4 parameters.
        The original coupled \ce{^1H} spectrum is shown as a reference.
        The time required for each optimisation is indicated on each spectrum.
    }
    \label{fig:psyche_5spec}
\end{figure}

\clearpage

\subsection{AU programme used for optimisations}

The \texttt{poise\_psyche} AU programme was used for all the optimisations above.
It is the same as \texttt{poise\_1d}, except that it calls on a separate Python script to create a new double saltire pulse with the specified parameters prior to acquisition (similar to TopSpin's WaveMaker software):

\begin{minted}{c}
XCMD("sendgui xpy make_double_saltire")
ZG
EFP
APBK
QUIT
\end{minted}

The Python script \texttt{make\_double\_saltire} is described next.

\subsection{Python script for generating double saltire}

Please note that this is a Python script to be run \textit{inside} TopSpin.
It can be downloaded from the link in the docstring.

\begin{minted}{python2}
"""
make_double_saltire.py
----------------------

Download from: https://git.io/JUHOZ

This script reads in a pulse duration (stored in us as P40) and bandwidth
(stored in Hz as CNST21), then creates a double saltire shaped file called
"auto_double_saltire" which can be directly used as the shaped file in
PSYCHE experiments.

This can be included in an AU programme using the line:

    XCMD("sendgui xpy make_double_saltire")

SPDX-License-Identifier: GPL-3.0-or-later
"""

from __future__ import division
from math import sin, cos, pi, sqrt, degrees, atan2


def make_double_saltire(duration, bandwidth):
    """
    Parameters
    ----------
    duration : float
        Time in units of seconds.

    bandwidth : float
        Bandwidth in units of Hz.

    Returns
    -------
    A : list of float
        Amplitudes, scaled between 0 and 100.
    phi : list of float
        Phases, in degrees between -180 and 180.

    These lists can be directly passed to the Bruker SAVE_SHAPE routine, as
    demonstrated in this script. The amplitude of the pulse should be
    automatically calculated in the pulse programme.
    """
    # Our strategy is to generate the first saltire first.

    # Number of points in half of the pulse. We use 1 point per us.
    npoints = int(duration * 1000000 / 2)
    # 20% smoothing.
    s = 0.2
    # This is equivalent to ls = np.linspace(0, 1, npoints), but you can't use
    # numpy in TopSpin...
    ts = [i/(npoints - 1) for i in range(0, npoints)]

    # Construct the amplitude piecewise function for a single chirp.
    A_chirp = [sin((pi * t) / (2 * s)) if t < s
               else 1 if s <= t and t <= 1 - s
               else sin((pi * (1 - t)) / (2 * s))
               for t in ts]
    # These are phases for a single chirp, not a saltire.
    phi_chirp = [pi * (duration/2) * bandwidth * ((t - 0.5) ** 2) for t in ts]
    # Calculate x-coefficients (Cx) for the saltire.
    Cx = [100 * a * cos(phi) for a, phi in zip(A_chirp, phi_chirp)]
    # The Cy's are 100 * a * sin(phi) ..., but that's only for one chirp. For a
    # saltire, the two component chirps have equal and opposite Cy's (because
    # sin(phi) = -sin(-phi)), so the total Cy is simply zero for every point.
    # In principle we could make a list full of zeroes, like Cy = [0 for _ in
    # range(npoints)], but that's a waste of memory and time.
    # Incidentally, the Cx's for both chirps add up since cos(phi) = cos(-phi),
    # but because the amplitudes are scaled to [0, 100], there's no point
    # actually adding them up.

    # Convert back to polar coordinates. Ordinarily A = sqrt(xi * xi + yi *
    # yi), but since all the yi's are zero, we just leave it out of the
    # equation.
    A = [sqrt(xi * xi) for xi in Cx]
    # Likewise here we use atan2(0, xi) instead of atan2(yi, xi).
    phi = [degrees(atan2(0, xi)) for xi in Cx]
    # Finally, we double the lists (because it's a double saltire).
    return A * 2, phi * 2


if __name__ == "__main__":
    # Let the user know what's going on
    SHOW_STATUS("make_double_saltire: generating saltire...")
    # Pulses are by default in us, so must convert first
    duration = float(GETPAR("P 40"))/1000000
    # Bandwidth is specified in Hz.
    bandwidth = float(GETPAR("CNST 21"))
    # Generate the amplitudes and phases.
    A, phi = make_double_saltire(duration, bandwidth)
    # Save it to disk.
    shape_name = "auto_double_saltire"
    SAVE_SHAPE(shape_name, "Excitation", A, phi)
    # Use it as SPNAM40. It's too easy to forget this!
    PUTPAR("SPNAM 40", shape_name)
\end{minted}

\clearpage

\subsection{1D spin echo pulse programme}

\begin{minted}{text}
;psyche_1dopt_jy
;
;download from: https://git.io/JUHOS
;
;90 - 15m delay - double saltire - 15m delay - acquire
;Jonathan Yong - University of Oxford - 26 Aug 2020

;Optimisation parameters:
;cnst20 : flip angle (degrees)
;gpz10  : gradient amplitude (%)
;cnst21 : bandwidth (Hz)
;p40    : saltire + gradient duration (us)

#include <Avance.incl>
#include <Grad.incl>

"d2=15m"
"cnst10=(cnst20/360)*sqrt((2*cnst21)/(p40/2000000))" ; rf amplitude of saltire
"cnst11=1000000/(p1*4)"                              ; rf amplitude hard pulse
"spw40=plw1*(cnst10/cnst11)*(cnst10/cnst11)"         ; power level of saltire

"acqt0=0"
baseopt_echo

1 ze
2 30m
  50u BLKGRAD
  d1 pl1:f1
  50u UNBLKGRAD

  p1 ph1

  d2
  d16
  p16:gp1
  d16

  (center (p40:sp40 ph3):f1 (p40:gp10) )

  d16
  p16:gp1
  d16
  d2
  
  go=2 ph31
  30m mc #0 to 2 F0(zd)

50u BLKGRAD
exit

ph1 = 0
ph3 = 0 1 2 3
ph31= 0 2 0 2


;pl1 : f1 channel - power level for pulse (default)
;p1 : f1 channel -  high power pulse
;p40: shaped pulse in PS element
;p21: gradient during PS element
;d1 : relaxation delay; 1-5 * T1
;d2 : delay on either side of double saltire
;d16: gradient recovery delay
;p16: CTP gradient duration
;ns: 1 * n, total number of scans: NS * TD0

;cnst20: flip angle (degrees)
;cnst21: bandwidth of PSYCHE element (Hz) (ca. 10000)

;gpnam10: RECT.1
;gpz10  : 2%
;gpnam1 : SINE.100
;gpz1   : 50%
\end{minted}

\section{1D NOESY-based solvent suppression}
\label{sec:si_solvsupp}

All spectra in this section were acquired on a \SI{400}{\MHz} Bruker Avance NEO spectrometer, equipped with a broadband Smart probe with a nominal maximum $z$-gradient strength of \SI{50}{\gauss\per\cm}.
A sample of rodent urine in \ce{D2O} was used, kindly provided by Fay Probert and Abi Yates (Department of Chemistry, University of Oxford).

\subsection{Cost function}

The cost function used for all optimisations here was \texttt{zerorealint\_squared}, which attempts to minimise the squared intensity of the real spectrum:

\begin{equation}
    f = \sum_i \mathbf{S}_{\mathrm{real},i}^2
\end{equation}

The aim of this is to strongly penalise large deviations from the baseline, which are more likely to potentially obscure nearby peaks.
This cost function comes pre-installed with \Poise{}.
The region of the spectrum under consideration was restricted to between 4.65 and \SI{4.75}{\ppm} using the \texttt{dpl} function in TopSpin.

\subsection{Transmitter offset reference grid search}

It is not feasible to run a grid search over four parameters at the same time.
As with the PSYCHE section, we therefore only ran one grid search for the first parameter, i.e.\ the transmitter offset \texttt{O1}.

\begin{figure}
    \centering
    \includegraphics[width=0.7\textwidth]{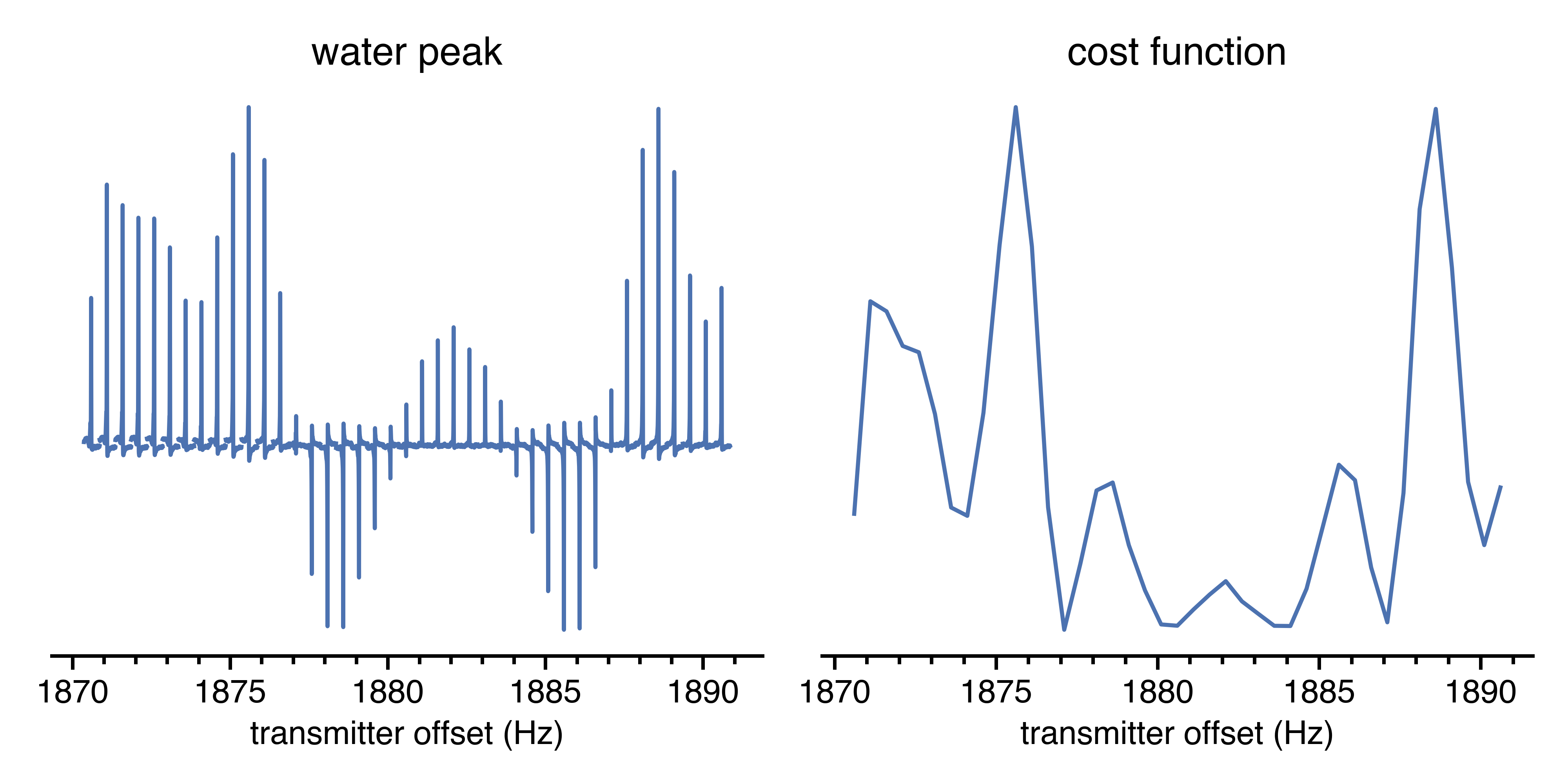}
    \caption{
        Insets of the residual water peak, and the value of the cost function thus calculated, as a function of transmitter offset (\texttt{O1}).
    }
    \label{fig:solvsupp_o1scan}
\end{figure}

It is evident here that there are multiple possible minima which \Poise{} could optimise to.
\Poise{} is not a \textit{global} minimiser; it will only converge to a \textit{local} minimum.
In fact, a similar situation was observed when we ran grid searches over the other three parameters: multiple potential minima were observed, due to an oscillatory dependence on the parameters under optimisation.

\subsection{Optimisations}
\label{sec:si_solvsupp_optim}

It is likely that there are multiple different sets of optimised parameters which give similar minima in the cost function.
We assume that the user does not mind exactly \textit{which} minimum is obtained, as long as there is a sufficient decrease in the cost function.
In other words, we assume that the user is not concerned with exactly which combination of parameters provides the best solvent suppression (as long as they are within reasonable bounds, e.g.\ presaturation power is not too high; these should be set appropriately in the \Poise{} routine).

As with the PSYCHE optimisations (\cref{sec:si_psyche}), we note that optimising more parameters simultaneously takes a longer time and risks being caught in local minima.
However, it is also clear that this has the potential of leading to a greater decrease in the cost function.
In our experience, optimising two parameters (O1 and CNST20) is usually sufficient to give a good decrease in the cost function (i.e.\ good water suppression).
For routine usage this two-parameter optimisation should be considered.

The detailed optimisation results are presented below, in a similar format to the PSYCHE section (\cref{sec:si_psyche_optim}).
On the \SI{400}{\MHz} spectrometer we used, the default \texttt{O1} value of \SI{1880.61}{\Hz} corresponds to exactly \SI{4.70}{\ppm}.

\begin{table}
    \centering
    \begin{tabular}{cccccccc}
        \hline
         & \multicolumn{3}{c}{Aggregated results from all runs} & \multicolumn{4}{c}{Optimised parameters from best run} \\
        Description & $f_\mathrm{best} / 10^{18}$ & $n_\mathrm{fev}$ & Time (s) & \texttt{O1} (\si{\Hz}) & \texttt{CNST20} (\si{\Hz}) & \texttt{D8} (\si{\s}) & \texttt{D1} (\si{\s}) \\
        \hline
        Initial point & 14.7         & --     & --         & (1880.61) & (50.0) & (0.100) & (2.00) \\
        1 parameter   & 1.85--2.49   & 6--12  & 259--520   & 1880.41   & (50.0) & (0.100) & (2.00) \\
        2 parameters  & 1.21--9.68   & 9--21  & 390--911   & 1880.20   & 51.94  & (0.100) & (2.00) \\
        3 parameters  & 1.24--4.20   & 19--26 & 825--1128  & 1879.90   & 47.79  & 0.118   & (2.00) \\
        4 parameters  & 0.165--1.65  & 25--53 & 1143--2314 & 1881.10   & 53.28  & 0.150   & (3.00) \\
        \hline
    \end{tabular}
    \caption{
        Summary of optimisations on 1D NOESY / presaturation pulse sequence for solvent suppression..
        Details of the routines used are described in \cref{tbl:solvsupp1p,tbl:solvsupp2p,tbl:solvsupp3p,tbl:solvsupp4p}.
        Parameters in parentheses were not optimised (they are inherited from the initial point).
        \nfev{}
        \fiveruns{}
        The optimised parameters in each row are taken from the best run, as judged by the lowest value of the cost function found.
    }
    \label{tbl:solvsupp_overview}
\end{table}

\begin{table}
    \hbadness=10000
    \centering
    \begin{tabular}{ccccc}
        \toprule
        Entry & Algorithm    & Optimum found (\si{\Hz}) & $n_\text{fev}$ & Time taken (\si{\s}) \\
        \midrule
        1     & Nelder--Mead & 1880.24--1880.49         & 10--12         & 436--519             \\
        2     & MDS          & 1880.24--1880.36         & 12             & 518--520             \\
        3     & BOBYQA       & 1880.34--1880.47         & 6--7           & 259--303             \\
        \bottomrule
    \end{tabular}
    \caption{
        Solvent suppression 1-parameter (transmitter offset) optimisations.
        The \Poise{} routine used was: \mintinline[breaklines]{json}{{"name": "solvsupp1", "pars": ["o1"], "lb": [1870.61], "ub": [1890.61], "init": [1880.61], "tol": [0.2], "cf": "zerorealint_squared", "au": "poise_1d_noapk"}}.
        Key acquisition parameters: \texttt{NS=2}, \texttt{DS=4}, \texttt{AQ=4.19}.
    }
    \label{tbl:solvsupp1p}
\end{table}

\begin{table}
    \hbadness=10000
    \centering
    \begin{tabular}{ccccc}
        \toprule
        Entry & Algorithm    & $n_\text{fev}$ & Time taken (\si{\s}) \\
        \midrule
        1     & Nelder--Mead & 17--21         & 737--911             \\
        2     & MDS          & 15             & 649--652             \\
        3     & BOBYQA       & 9--14          & 390--608             \\
        \bottomrule
    \end{tabular}
    \caption{
        Solvent suppression 2-parameter (transmitter offset and presaturation power) optimisations.
        The \Poise{} routine used was: \mintinline[breaklines]{json}{{"name": "solvsupp2", "pars": ["o1", "cnst20"], "lb": [1870.61, 10.0], "ub": [1890.61, 55.0], "init": [1880.61, 50.0], "tol": [0.2, 2.5], "cf": "zerorealint_squared", "au": "poise_1d_noapk"}}.
        Key acquisition parameters: \texttt{NS=2}, \texttt{DS=4}, \texttt{AQ=4.19}.
    }
    \label{tbl:solvsupp2p}
\end{table}

\begin{table}
    \hbadness=10000
    \centering
    \begin{tabular}{ccccc}
        \toprule
        Entry & Algorithm    & $n_\text{fev}$ & Time taken (\si{\s}) \\
        \midrule
        1     & Nelder--Mead & 23--26         & 1002--1128           \\
        2     & MDS          & 24--25         & 1034--1080           \\
        3     & BOBYQA       & 19--26         & 825--1133            \\
        \bottomrule
    \end{tabular}
    \caption{
        Solvent suppression 3-parameter (transmitter offset, presaturation power, and mixing time) optimisations.
        The \Poise{} routine used was: \mintinline[breaklines]{json}{{"name": "solvsupp3", "pars": ["o1", "cnst20", "d8"], "lb": [1870.61, 10.0, 0.010], "ub": [1890.61, 55.0, 0.150], "init": [1880.61, 50.0, 0.100], "tol": [0.2, 2.5, 0.010], "cf": "zerorealint_squared", "au": "poise_1d_noapk"}}.
        Key acquisition parameters: \texttt{NS=2}, \texttt{DS=4}, \texttt{AQ=4.19}.
    }
    \label{tbl:solvsupp3p}
\end{table}

\begin{table}
    \hbadness=10000
    \centering
    \begin{tabular}{ccccc}
        \toprule
        Entry & Algorithm    & $n_\text{fev}$ & Time taken (\si{\s}) \\
        \midrule
        1     & Nelder--Mead & 29--40         & 1250--1726           \\
        2     & MDS          & 34--53         & 1487--2314           \\
        3     & BOBYQA       & 25--40         & 1143--1843           \\
        \bottomrule
    \end{tabular}
    \caption{
        Solvent suppression 4-parameter (transmitter offset, presaturation power, mixing time, and presaturation duration) optimisations.
        The \Poise{} routine used was: \mintinline[breaklines]{json}{{"name": "solvsupp4", "pars": ["o1", "cnst20", "d8", "d1"], "lb": [1870.61, 10.0, 0.010, 1.0], "ub": [1890.61, 55.0, 0.150, 3.0], "init": [1880.61, 50.0, 0.100, 2.0], "tol": [0.2, 2.5, 0.010, 0.1], "cf": "zerorealint_squared", "au": "poise_1d_noapk"}}.
        Key acquisition parameters: \texttt{NS=2}, \texttt{DS=4}, \texttt{AQ=4.19}.
    }
    \label{tbl:solvsupp4p}
\end{table}

The best of each of these optimisations are shown in the next figure.

\begin{figure}
    \centering
    \includegraphics[width=\textwidth]{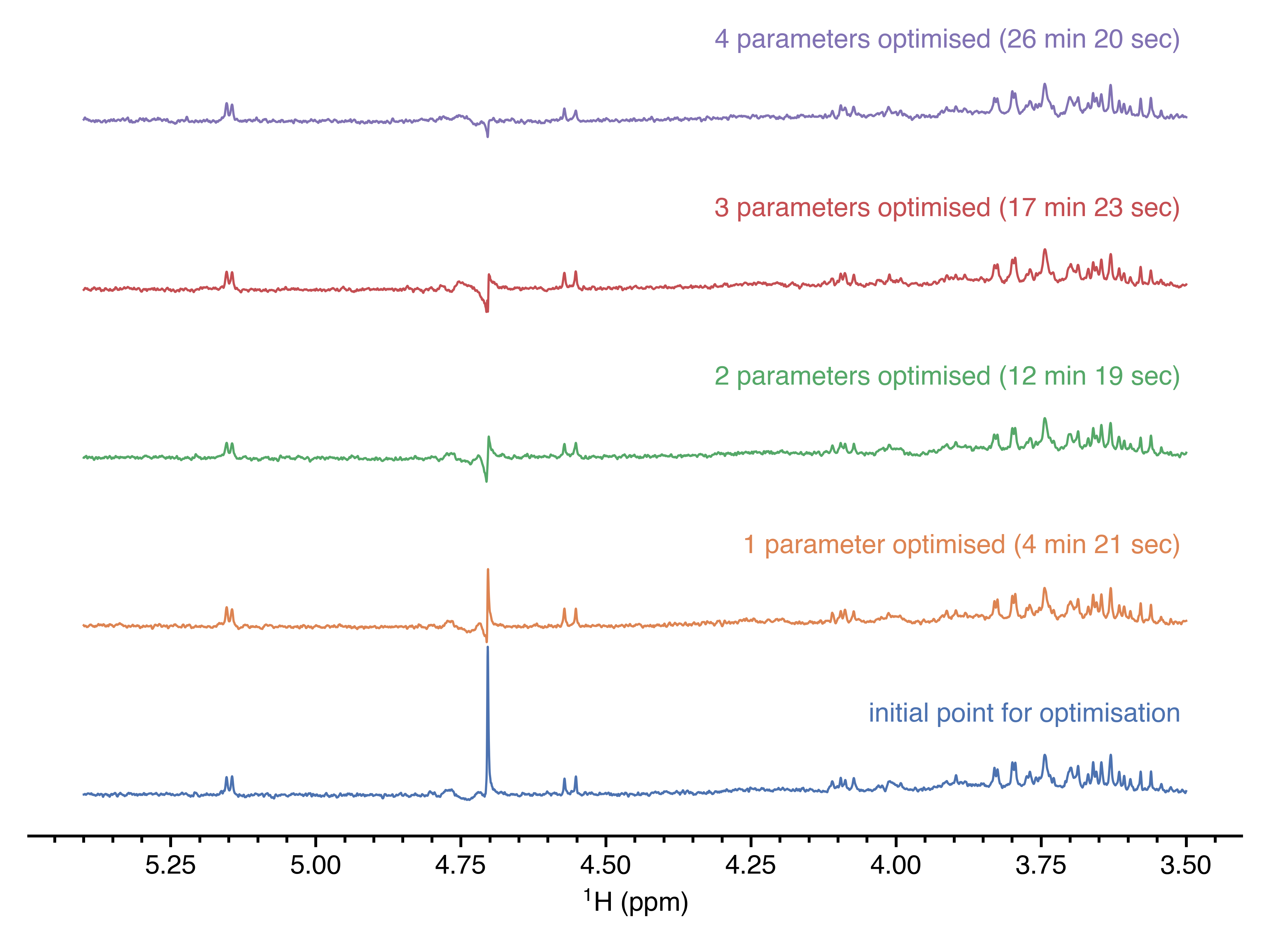}
    \caption{
        Insets of 1D NOESY / presaturation spectra of a sample of rodent urine, obtained before and after optimisation of up to 4 parameters.
        The time required for each optimisation is indicated on each spectrum.
    }
    \label{fig:solvsupp_5spec}
\end{figure}

\clearpage

\subsection{1D NOESY pulse programme}
\label{sec:si_solvsupp_pp}

This pulse programme is almost the same as the Bruker default \texttt{noesygppr1d}, except that the presaturation power is controlled via the \texttt{CNST27} parameter (which represents the RF amplitude in \si{\Hz}).

\begin{minted}{text}
;noesygppr1d_jy
;
;download from: https://git.io/Jnokc
;
;modified from noesygppr1d to use CNST20 as presaturation RF amplitude (in Hz)
;Jonathan Yong - University of Oxford - 20 June 2021
;
;$CLASS=HighRes
;$DIM=1D
;$TYPE=
;$SUBTYPE=
;$COMMENT=
;$RECOMMEND=

#include <Avance.incl>
#include <Grad.incl>

"d12=20u"
"plw9=plw1*(cnst20/(250000/p1))*(cnst20/(250000/p1))"
"acqt0=-p0*2/3.1416"


1 ze
2 30m

#   ifdef FLAG_BLK
  4u BLKGRAD
#   else
  4u
#   endif /*FLAG_BLK*/

  d12 pl9:f1
  d1 cw:f1 ph29
  4u do:f1
  50u UNBLKGRAD
  p16:gp1
  d16 pl1:f1
  p1 ph1
  4u
  p1 ph2
  d12 pl9:f1
  d8 cw:f1
  4u do:f1
  p16:gp2
  d16 pl1:f1

#   ifdef FLAG_BLK
  4u
#   else
  4u BLKGRAD
#   endif /*FLAG_BLK*/

  p0 ph3
  go=2 ph31
  30m mc #0 to 2 F0(zd)

#   ifdef FLAG_BLK
  4u BLKGRAD
#   else
  4u
#   endif /*FLAG_BLK*/

exit


ph1=0 2 
ph2=0 0 0 0 0 0 0 0 2 2 2 2 2 2 2 2
ph3=0 0 2 2 1 1 3 3
ph29=0
ph31=0 2 2 0 1 3 3 1 2 0 0 2 3 1 1 3


;pl1 : f1 channel - power level for pulse (default)
;pl9 : f1 channel - power level for presaturation
;p0 : for any flip angle
;p1 : f1 channel -  90 degree high power pulse
;p16: homospoil/gradient pulse
;d1 : relaxation delay; 1-5 * T1
;d8 : mixing time
;d12: delay for power switching                      [20 usec]
;d16: delay for homospoil/gradient recovery
;ns: 8 * n, total number of scans: NS * TD0
;ds: 4
;cnst20: presaturation power (Hz)


;use gradient ratio:    gp 1 : gp 2
;                         50 :  -10

;for z-only gradients:
;gpz1: 50%
;gpz2: -10%

;use gradient files:
;gpnam1: SMSQ10.100
;gpnam2: SMSQ10.100



                                          ;preprocessor-flags-start
;FLAG_BLK: for BLKGRAD before d1 rather than go
;             option -DFLAG_BLK: (eda: ZGOPTNS)
                                          ;preprocessor-flags-end

\end{minted}

\section{Diffusion NMR}
\label{sec:si_dosy}

All spectra in this section were acquired on a \SI{600}{\MHz} Bruker AVIII spectrometer, equipped with a Prodigy \ce{N2} broadband cryoprobe with a nominal maximum $z$-gradient strength of \SI{66}{\gauss\per\cm}.
The sample used was \SI{45}{\milli\molar} andrographolide in DMSO-$d_6$.

\subsection{Sequential optimisation strategy}

Traditionally, setting up a diffusion experiment is considered to be a tedious process.
At the very least, one must search for a range of gradient amplitudes ($G_\mathrm{min}$ through $G_\mathrm{max}$) which provide an appropriate range of signal attenuation: thus, for example, the increment acquired with $G_\mathrm{max}$ must exhibit a certain level of attenuation with respect to $G_\mathrm{min}$.
If $G_\mathrm{max}$ is too large or too small, then the resulting fitting will be less accurate.
In more complicated cases, the diffusion delay $\Delta$ and/or the gradient pulse duration $\delta$ must also be tweaked.

Here, we use the oneshot DOSY experiment (\textit{Magn.\ Reson.\ Chem.}\ \textbf{2002,} \textit{40} (13), S147--S152, DOI: \href{https://doi.org/10.1002/mrc.1107}{\texttt{10.1002/mrc.1107}}); the process can easily be adapted to all other DOSY variants.
Throughout the optimisation, the script maintains a ``reference'' 1D spectrum acquired with $G_\mathrm{min} = 10\%$.
The entire optimisation can be run using a single wrapper Python script, \texttt{dosy\_opt.py}.

We seek to first optimise $\Delta$ by performing stepwise increases from a small initial value, until a separate spectrum, obtained with $G_\mathrm{max} = 80\%$, displays ``sufficient'' attenuation with respect to the reference spectrum.
Here, we define ``sufficient'' as 75\% attenuation.
To determine this, the script leverages the \texttt{dosy\_aux} routine and cost function.
Since this is not a traditional ``optimisation'' in the sense that the value can only go upwards, we use the wrapper script to check the value of the cost function: thus we set the \texttt{maxfev} option to 1, such that on every function evaluation, control is returned to the wrapper script.
The routine is (note that \texttt{lb}, \texttt{ub}, and \texttt{tol} have no use in this routine and can be set to any valid value):

\begin{minted}{json}
{"name": "dosy_aux", "pars": ["gpz1"], "lb": [79.0], "ub": [81.0], "init": [80.0], "tol": [0.01], "cf": "dosy_aux", "au": "poise_1d"}
\end{minted}

The \texttt{dosy\_aux} cost function is defined by

\begin{equation}
    f_{\text{att}} = \frac{\sum_i \mathbf{S}_i}{\sum_i \mathbf{R}_i} - 0.25
\end{equation}

where $\mathbf{R}$ is the reference spectrum (acquired with a minimum gradient amplitude of 10\%) and $\mathbf{S}$ is the maximally attenuated spectrum (acquired with a gradient amplitude of 80\%).

Once a suitable diffusion delay has been found, we then carry out a simple \Poise{} optimisation to find the actual value of $G_\mathrm{max}$ that provides 75\% attenuation.
This utilises the \texttt{dosy} routine and cost function.
The routine is as follows:

\begin{minted}{json}
{"name": "dosy", "pars": ["gpz1"], "lb": [20.0], "ub": [80.0], "init": [50.0], "tol": [2.0], "cf": "dosy", "au": "poise_1d"}
\end{minted}

and the \texttt{dosy} cost function is simply the absolute value of $f_\mathrm{att}$.

The total time taken for the entire process depends on how many times $\Delta$ must be increased from its initial value (roughly 1 additional minute per increase).
In order to minimise this, it is recommended to start with a sensible value of $\Delta$ based on known precedent.
The second step takes approximately 3 minutes in our hands regardless of optimisation algorithm, for an overall total time of ca.\ 5 minutes (using the parameters \texttt{DS=2}, \texttt{NS=4}).

\begin{figure}
    \centering
    \includegraphics[width=0.7\textwidth]{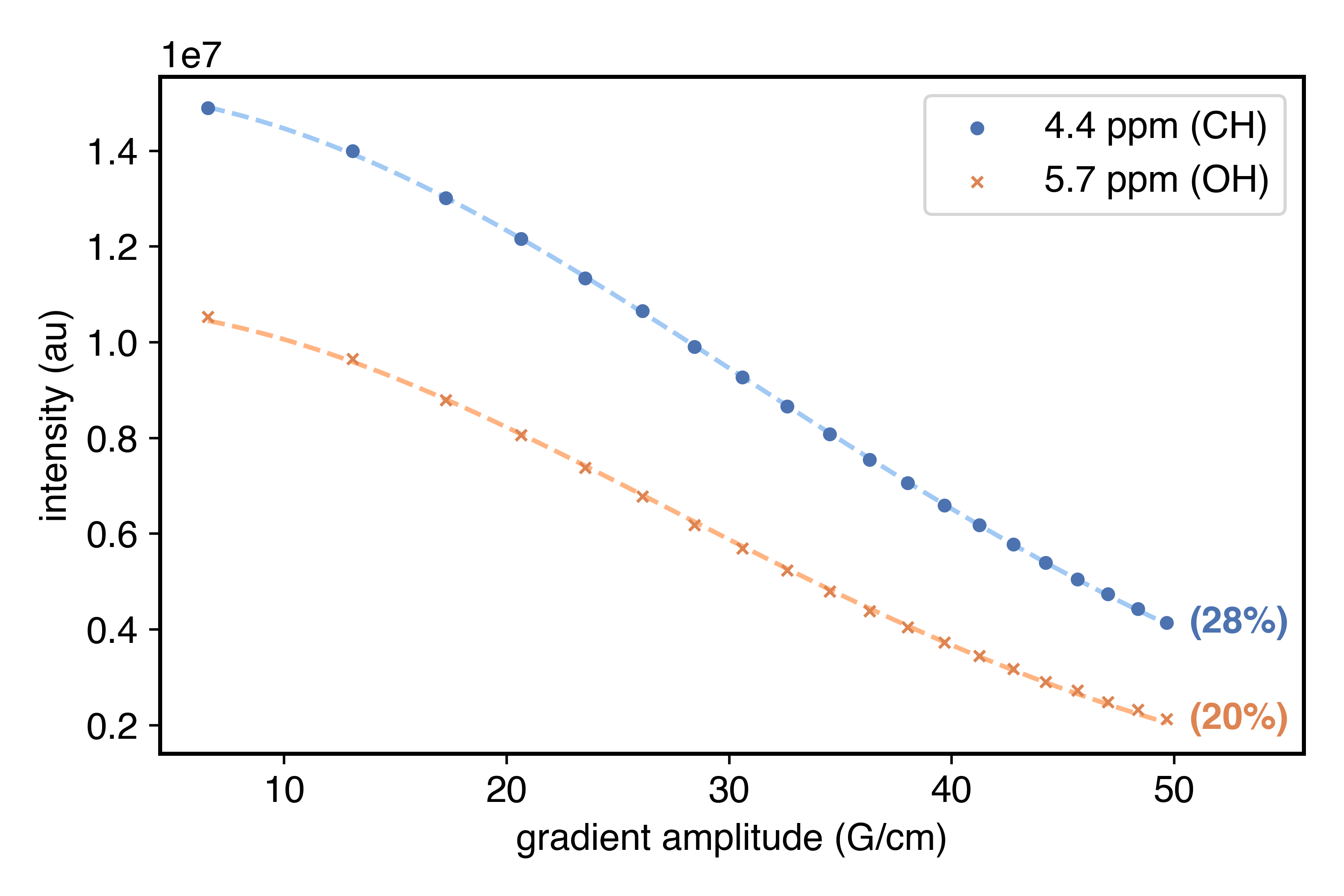}
    \caption{Example diffusion decay profiles obtained using optimised values of $G_\mathrm{max} = 75.6\%$ (equivalent to \SI{49.7}{G\per\cm} on the spectrometer we used) and $\Delta = \SI{100}{\ms}$ in the oneshot DOSY experiment. $G_\mathrm{min}$ is $10\%$, or \SI{6.57}{G\per\cm}. Dashed lines are the Gaussian profiles which have been fitted to the observed intensities. The sample used was \SI{45}{\milli\molar} andrographolide in DMSO-$d_6$.}
    \label{fig:dosy}
\end{figure}

In Figure \ref{fig:dosy}, we plot the typical decay profiles obtained for \ce{CH} and \ce{OH} protons in andrographolide, obtained from a oneshot DOSY experiment using parameters optimised according to the procedure described above.
The two profiles are slightly different because of chemical exchange between the \ce{OH} protons and water present in the sample.
This data is clearly amenable towards further analysis using processing tools such as Bruker's Dynamic Centre software or the DOSY Toolbox.

\subsection{Python script for automation}

The steps to use this script are:

\begin{enumerate}
    \item Make sure that you have enabled the DOSY routines and cost functions. The routines can be found in the \texttt{.../py/user/poise\_backend/example\_routines} directory and should be copied to the \texttt{.../py/user/poise\_backend/routines} directory; the cost functions are in \texttt{.../py/user/poise\_backend/costfunctions.py} and must be uncommented before they can be used.
    \item Download this Python script. There is a GitHub link in the source code below.
    \item Set up a 2D diffusion experiment (most easily done by reading in a parameter set). The initial gradient amplitudes do not matter, but the diffusion delay should be set to something that is not too large (since the script only \textit{increases} the diffusion delay if it is found to be too short).
    \item If the particular DOSY experiment has not been registered in this script yet, then do so by modifying the dictionary near the top of the script. This only needs to be done once.
        (If the DOSY experiment has never been set up before, then you need to find (or create) a 1D version of the diffusion experiment. It will basically be the same as the 2D version, except that the encoding gradient is not incremented.)
    \item Run this Python script (just type \texttt{dosy\_opt} in the TopSpin command line).
\end{enumerate}

\begin{minted}{python}
"""
dosy_opt.py
-----------

Download from: https://git.io/JUHOY

Script that optimises parameters for a diffusion experiment. Currently works
with oneshot DOSY (Pelta et al., Magn. Reson. Chem. 2002, 40 (13), S147-S152,
DOI: 10.1002/mrc.1107), as well as the stegp1s and ledgp2s DOSY variants which
are standard TopSpin pulse programmes.

This is meant to be called from the TopSpin command line on a 2D diffusion
experiment.

Note that this script does not tweak the gradient lengths ("little delta"), so
these should be set to a sensible value (e.g. 1 ms) before running this script.

The algorithm is as follows:

 1. Create 1D versions of the diffusion experiment, one with the minimum
    gradient (10%, the "reference spectrum", expno 99998) and one with the
    maximum gradient (80%, "optimisation spectrum", expno 99999).
 2. Acquires both spectra.
 3. Check whether the opt. spectrum has at least 75% signal attenuation
    relative to the ref. spectrum. If not, increases the diffusion delay
    ("big Delta") and repeats steps 1 and 2.
 4. If the attenuation is sufficient, it then runs a POISE optimisation on the
    opt. spectrum to find the gradient strength at which 75% signal attenuation
    is achieved.
 5. Once this gradient strength has been found, returns to the 2D experiment
    and runs the `dosy` Bruker AU programme using the new parameter.

SPDX-License-Identifier: GPL-3.0-or-later
"""

# Dictionary of DOSY pulse programmes, to be entered in the form
#    "2d_pulprog": ["1d_pulprog", "GRAD_AMPLITUDE_PARAM", "DIFF_DELAY_PARAM"]
# In more detail:
#  - 2d_pulprog is the name of the 2D DOSY pulse sequence.
#  - 1d_pulprog is the name of the corresponding 1D DOSY pulse sequence, which
#     should be just a single increment of the 2D sequence, but with Difframp
#     removed. See the doneshot_nd_jy pulse programmes for an example.
#  - GRAD_AMPLITUDE_PARAM is the parameter name in the 2D DOSY pulse sequence
#     that is multiplied by Difframp. In the 1D DOSY pulse sequence it should
#     not be multiplied by Difframp, so simply corresponds to the gradient
#     strength. Note that you need a space between GPZ and the number.
#  - DIFF_DELAY_PARAM is the parameter name in both 1D and 2D sequences that
#     corresponds to the diffusion delay. Again, a space between D and the
#     number is required.
# Other DOSY variants can be entered here to "register" them and make them
# available for use with this script.
pulprog_dict = {
    # doneshot_nd_jy can be downloaded from https://git.io/JUHOX and
    # https://git.io/JUHOP.
    "doneshot_2d_jy": ["doneshot_1d_jy", "GPZ 1", "D 20"],
    # the following two entries are Bruker standard sequences and already come
    # with TopSpin.
    "stegp1s": ["stegp1s1d", "GPZ 6", "D 20"],
    "ledgp2s": ["ledgp2s1d", "GPZ 6", "D 20"],
}


def main():
    # Look up the 1D pulse programme and associated parameters based on the
    # current 2D pulse programme.
    pp2d = GETPAR("PULPROG")
    if pp2d in pulprog_dict:
        pp1d, gpzparam, dparam = pulprog_dict[pp2d]
    else:
        MSG("dosy_opt: unsupported pulse programme. New pulse programmes can"
            " be added in the dosy_opt.py script.")
        EXIT()

    # Create new routines. We need to do this within this script since the
    # routine depends on the GPZ parameter being optimised; we can't hardcode
    # that in a predefined routine.
    gpzparam_short = gpzparam.replace(" ", "").lower()
    XCMD("poise --create dosy {} 20 80 50 2"
         " dosy poise_1d".format(gpzparam_short))
    XCMD("poise --create dosy_aux {} 79 81 80 0.1"
         " dosy_aux poise_1d".format(gpzparam_short))

    # Get the expno of the 2D dataset, and store the parameter set. This is
    # important because we want the optimisation to inherit most parameters,
    # e.g.  D1, SW, etc. from the parent experiment (which has presumably been
    # set up by the user).
    original_2d_expno = int(CURDATA()[1])
    original_2d_dataset = CURDATA()
    XCMD("wpar dosytemp all")

    # Set up the reference experiment.
    reference_expno = 99998
    reference_dataset = list(original_2d_dataset)
    reference_dataset[1] = str(reference_expno)
    NEWDATASET(reference_dataset, None, "dosytemp")
    RE(reference_dataset)
    PUTPAR("PULPROG", pp1d)
    PUTPAR("PARMODE", "1D")
    PUTPAR("PPARMOD", "1D")
    PUTPAR(gpzparam, "10")
    XCMD("sendgui browse_update_tree")
    # Acquire reference dataset.
    msg_nonmodal("dosy_opt: optimising diffusion delay...")
    XCMD("poise_1d", wait=WAIT_TILL_DONE)

    # Set up optimisation expno.
    optimisation_expno = 99999
    optimisation_dataset = list(original_2d_dataset)   # make a copy
    optimisation_dataset[1] = str(optimisation_expno)
    NEWDATASET(optimisation_dataset, None, "dosytemp")
    RE(optimisation_dataset)
    PUTPAR("PULPROG", pp1d)
    PUTPAR("PARMODE", "1D")
    PUTPAR("PPARMOD", "1D")
    PUTPAR(gpzparam, "80")
    XCMD("sendgui browse_update_tree")

    # Optimise the diffusion delay first.
    while True:
        # Here we use the `--maxfev 1` trick to evaluate a cost function at 80%
        # gradient. The cost function is stored as the `TI` parameter after the
        # experiment has been recorded.
        RE(optimisation_dataset)
        PUTPAR("TI", " ")  # has to be one space, not an empty string (try it)
        XCMD("poise dosy_aux -a nm --maxfev 1 -q", wait=WAIT_TILL_DONE)
        # If TI isn't a valid float, something went wrong with POISE.
        try:
            cf = float(GETPAR("TI"))
        except ValueError:
            EXIT()
        # The dosy_aux cost function is (opt. intensity/ref. intensity) minus
        # 0.25. If the opt. spectrum has been attenuated too little, then this
        # value will be positive.
        if cf > 0:
            # Increase delay, i.e. big Delta, by 100 ms
            current_delay = float(GETPAR(dparam))
            new_delay = current_delay + 0.1
            PUTPAR(dparam, str(new_delay))
            # Set it in the reference spectrum too
            RE(reference_dataset)
            PUTPAR(dparam, str(new_delay))
            # Re-acquire the reference spectrum.
            XCMD("xau poise_1d", wait=WAIT_TILL_DONE)
        # If the cost function is negative, then that means it's sufficiently
        # attenuated, and we can go on to the actual optimisation.
        else:
            break

    # At this point, we know that 80% gradient strength provides at least 75%
    # attenuation. So we can run an optimisation to find the actual value that
    # does provide 75% attenuation (which is somewhere between 10% and 80%).
    RE(optimisation_dataset)
    best_delay = GETPAR(dparam)
    msg_nonmodal("dosy_opt: diffusion delay set to {} seconds.\n"
                 "    now optimising gradient amplitude...".format(best_delay))
    PUTPAR("TI", " ")
    XCMD("poise dosy -a bobyqa -q", wait=WAIT_TILL_DONE)
    # Again, check to make sure that TI is a valid float.
    try:
        float(GETPAR("TI"))
    except ValueError:
        EXIT()
    best_gpz = GETPAR(gpzparam)

    # Jump back to original 2D expt. Set dparam, but not gpzparam as that
    # should always be 100% (the actual gradient amplitude is controlled by
    # Difframp, which will be generated by the `dosy` AU programme).
    RE(original_2d_dataset)
    PUTPAR(dparam, best_delay)
    PUTPAR(gpzparam, "100")
    # Generate Difframp, etc. and then run the AU programme.
    msg_nonmodal("dosy_opt: maximum gradient set to {}%.\n"
                 "    starting 2D experiment with optimised"
                 " parameters...".format(best_gpz))
    XCMD("xau dosy 10 {} 16 l y".format(best_gpz))


def msg_nonmodal(s):
    """
    Show a message box that does not force the user to click OK before the
    programme continues. See also the same function in poise.py (the frontend
    script).
    """
    dialog = mfw.BInfo()
    dialog.setMessage(s)
    dialog.setTitle("dosy_opt")
    dialog.setBlocking(0)  # in ordinary MSG() this is 1
    dialog.show()


if __name__ == "__main__":
    main()
\end{minted}

\subsection{Sequential optimisation details}

The second part of the optimisation ($G_\mathrm{max}$ optimisation) is the only place where using different algorithms can have an effect.
In order to evaluate the effect of using different algorithms, we ran the second optimisation using each of the algorithms, using the diffusion delay $\Delta = \SI{100}{\ms}$ obtained from the first step.
The results are summarised in Table \ref{tbl:dosy}.

\begin{table}
    \centering
    \begin{tabular}{ccccc}
        \toprule
        Entry & Algorithm    & Optimum found (\%) & $n_\text{fev}$ & Time taken (\si{\s}) \\
        \midrule
        1     & Nelder--Mead & 75.00              & 9              & 195--197             \\
        2     & MDS          & 75.00              & 9              & 196--197             \\
        3     & BOBYQA       & 75.00--75.60       & 8--9           & 175--197             \\
        \bottomrule
    \end{tabular}
    \caption{
        Optimisations of maximum gradient amplitude ($G_\mathrm{max}$) for diffusion NMR spectroscopy.
        The \Poise{} routine used here is: \mintinline[breaklines]{json}{{"name": "dosy", "pars": ["gpz1"], "lb": [20.0], "ub": [80.0], "init": [50.0], "tol": [2.0], "cf": "dosy", "au": "poise_1d"}}.
        The pulse programme used was \texttt{doneshot\_1d\_jy} (\cref{sec:si_dosy_pp}).
        Key acquisition parameters: \texttt{NS=4}, \texttt{DS=2}, \texttt{D1=1.5}, \texttt{AQ=1.14}.
        \nfev{}
        \fiveruns{}
    }
    \label{tbl:dosy}
\end{table}

\subsection{Simultaneous optimisation strategy}

We also tried directly performing a simultaneous two-parameter optimisation of both the gradient amplitude and the diffusion delay, instead of the entire procedure above.
The \Poise{} routine and its associated cost function are called \texttt{dosy\_2p}.
The routine is:

\begin{minted}{json}
{"name": "dosy_2p", "pars": ["gpz1", "d20"], "lb": [20.0, 0.05], "ub": [80.0, 0.4], "init": [50.0, 0.05], "tol": [2.0, 0.1], "cf": "dosy_2p", "au": "poise_2pdosy"}
\end{minted}

In order to bias the optimisation towards smaller values of $\Delta$ (which is desirable as it minimises $T_1$ losses), we added a term to the cost function that is proportional to $\Delta$.
This cost function can be found in \texttt{.../py/user/poise\_backend/costfunctions.py} as well, although it is commented out by default.

\begin{equation}
    f = |f_\mathrm{att}| + (\Delta / \mathrm{s})
\end{equation}

Unfortunately, although this optimisation \textit{did} work, it was far slower: for example, using the Nelder--Mead algorithm, this took 36 function evaluations and 1572 seconds (over 26 minutes).
The optimum converged to was $G_\mathrm{max} = 79.3\%$ and $\Delta = \SI{91.6}{\ms}$, which is very similar to that found above.
Part of the reason why this is so slow is that every time the diffusion delay $\Delta$ is changed, the reference spectrum must be reacquired: this means that every function evaluation in this two-parameter optimisation takes twice as long as every function evaluation in the individual optimisation of $G_\mathrm{max}$.

\subsection{Oneshot DOSY pulse programmes}
\label{sec:si_dosy_pp}

\subsubsection{2D Oneshot}

Some modifications from the published version were made so that it could be run on our spectrometers.

\begin{minted}{text}
;doneshot_2d_jy
;
;download from: https://git.io/JUHOX
;
;Jonathan Yong - University of Oxford - 26 Aug 2020
;minor modifications to original oneshot DOSY sequence to make it play better with TopSpin.
;
;modified from:
;Ralph Adams, Juan Aguilar, Robert Evans, Mathias Nilsson and Gareth Morris
;University of Manchester
;Release 1.0c (27Mar2012)
;Source citation:
;M.D. Pelta, G.A. Morris, M.J. Stchedroff, S.J. Hammond, Magn. Reson. Chem. 40 (2002) S147-S152.
;
;Other relevant papers that could be of use include:
;A. Botana, J.A. Aguilar, M. Nilsson, G.A. Morris, J. Magn. Reson. 208 (2011) 270-278.
; 
;2D Doneshot DOSY pulse sequence
;$CLASS=HighRes
;$DIM=2D
;$TYPE=
;$SUBTYPE=
;$COMMENT=

#include <Avance.incl>
#include <Grad.incl>
#include <Delay.incl>

define list<gradient> diff=<Difframp>

;JY - alpha is hardcoded as 0.2
"cnst14=0.2"

"cnst17=(2*p1+d16+p30)*0.000001"; Dtau
;Assuming half-sine gradient pulses [most common on Bruker systems]
"cnst18=0.000001*p30*2*0.000001*p30*2*(d20 - (2*0.000001*p30*(5-3*cnst14*cnst14)/16) - (cnst17*(1-cnst14*cnst14)/2) )" ; Dosytimecubed

"cnst15=1+cnst14"  ; 1 + alpha
"cnst16=1-cnst14"  ; 1 - alpha
"p2=p1*2"
"DELTA1=d20-4.0*p1-4.0*p30-5.0*d16-p19"
"acqt0 = -p1*2/3.1416"

1 ze
2 d1
3 50u UNBLKGRAD
  p19:gp2*-1               ;Spoiler gradient balancing pulse
  d16
  p1 ph1                   ;1st 90
  p30:gp1*diff*cnst16      ;1 - alpha
  d16
  p2 ph2                   ;First 180
  p30:gp1*diff*cnst15*-1   ;1 + alpha
  d16
  p1 ph3                   ;2nd 90
  p30:gp1*diff*cnst14*2    ;Lock refocusing pulse pulse
  d16
  p19:gp2                  ;Spoiler gradient balancing pulse
  d16
  DELTA1
  p30:gp1*diff*cnst14*2    ;Lock refocusing pulse pulse
  d16
  p1 ph4                   ;3rd 90
  p30:gp1*diff*cnst16      ;1 - alpha
  d16
  p2 ph5
  p30:gp1*diff*cnst15*-1   ;1 + alpha
  d16
  4u BLKGRAD
  go=2 ph31
  d1 mc #0 to 2 F1QF(calgrad(diff))
exit

ph1 = 0 2 0 2 1 3 1 3 0 2 0 2 1 3 1 3 0 2 0 2 1 3 1 3 0 2 0 2 1 3 1 3 0 2 0 2 1 3 1 3 0 2 0 2 1 3 1 3 0 2 0 2 1 3 1 3 0 2 0 2 1 3 1 3 0 2 0 2 1 3 1 3 0 2 0 2 1 3 1 3 0 2 0 2 1 3 1 3 0 2 0 2 1 3 1 3 0 2 0 2 1 3 1 3 0 2 0 2 1 3 1 3 0 2 0 2 1 3 1 3 0 2 0 2 1 3 1 3 0 2 0 2 1 3 1 3 0 2 0 2 1 3 1 3 0 2 0 2 1 3 1 3 0 2 0 2 1 3 1 3 0 2 0 2 1 3 1 3 0 2 0 2 1 3 1 3 0 2 0 2 1 3 1 3 0 2 0 2 1 3 1 3 0 2 0 2 1 3 1 3 0 2 0 2 1 3 1 3 0 2 0 2 1 3 1 3 0 2 0 2 1 3 1 3 0 2 0 2 1 3 1 3 0 2 0 2 1 3 1 3 0 2 0 2 1 3 1 3 0 2 0 2 1 3 1 3
ph2 = 0 0 0 0 0 0 0 0 0 0 0 0 0 0 0 0 0 0 0 0 0 0 0 0 0 0 0 0 0 0 0 0 0 0 0 0 0 0 0 0 0 0 0 0 0 0 0 0 0 0 0 0 0 0 0 0 0 0 0 0 0 0 0 0 0 0 0 0 0 0 0 0 0 0 0 0 0 0 0 0 0 0 0 0 0 0 0 0 0 0 0 0 0 0 0 0 0 0 0 0 0 0 0 0 0 0 0 0 0 0 0 0 0 0 0 0 0 0 0 0 0 0 0 0 0 0 0 0 2 2 2 2 2 2 2 2 2 2 2 2 2 2 2 2 2 2 2 2 2 2 2 2 2 2 2 2 2 2 2 2 2 2 2 2 2 2 2 2 2 2 2 2 2 2 2 2 2 2 2 2 2 2 2 2 2 2 2 2 2 2 2 2 2 2 2 2 2 2 2 2 2 2 2 2 2 2 2 2 2 2 2 2 2 2 2 2 2 2 2 2 2 2 2 2 2 2 2 2 2 2 2 2 2 2 2 2 2 2 2 2 2 2 2 2 2 2 2 2 2 2 2 2 2 2 2 2
ph3 = 0 0 0 0 0 0 0 0 0 0 0 0 0 0 0 0 0 0 0 0 0 0 0 0 0 0 0 0 0 0 0 0 2 2 2 2 2 2 2 2 2 2 2 2 2 2 2 2 2 2 2 2 2 2 2 2 2 2 2 2 2 2 2 2 0 0 0 0 0 0 0 0 0 0 0 0 0 0 0 0 0 0 0 0 0 0 0 0 0 0 0 0 0 0 0 0 2 2 2 2 2 2 2 2 2 2 2 2 2 2 2 2 2 2 2 2 2 2 2 2 2 2 2 2 2 2 2 2 0 0 0 0 0 0 0 0 0 0 0 0 0 0 0 0 0 0 0 0 0 0 0 0 0 0 0 0 0 0 0 0 2 2 2 2 2 2 2 2 2 2 2 2 2 2 2 2 2 2 2 2 2 2 2 2 2 2 2 2 2 2 2 2 0 0 0 0 0 0 0 0 0 0 0 0 0 0 0 0 0 0 0 0 0 0 0 0 0 0 0 0 0 0 0 0 2 2 2 2 2 2 2 2 2 2 2 2 2 2 2 2 2 2 2 2 2 2 2 2 2 2 2 2 2 2 2 2
ph4 = 0 0 2 2 0 0 2 2 1 1 3 3 1 1 3 3 2 2 0 0 2 2 0 0 3 3 1 1 3 3 1 1 0 0 2 2 0 0 2 2 1 1 3 3 1 1 3 3 2 2 0 0 2 2 0 0 3 3 1 1 3 3 1 1 0 0 2 2 0 0 2 2 1 1 3 3 1 1 3 3 2 2 0 0 2 2 0 0 3 3 1 1 3 3 1 1 0 0 2 2 0 0 2 2 1 1 3 3 1 1 3 3 2 2 0 0 2 2 0 0 3 3 1 1 3 3 1 1 0 0 2 2 0 0 2 2 1 1 3 3 1 1 3 3 2 2 0 0 2 2 0 0 3 3 1 1 3 3 1 1 0 0 2 2 0 0 2 2 1 1 3 3 1 1 3 3 2 2 0 0 2 2 0 0 3 3 1 1 3 3 1 1 0 0 2 2 0 0 2 2 1 1 3 3 1 1 3 3 2 2 0 0 2 2 0 0 3 3 1 1 3 3 1 1 0 0 2 2 0 0 2 2 1 1 3 3 1 1 3 3 2 2 0 0 2 2 0 0 3 3 1 1 3 3 1 1
ph5 =0 0 0 0 0 0 0 0 0 0 0 0 0 0 0 0 2 2 2 2 2 2 2 2 2 2 2 2 2 2 2 2 0 0 0 0 0 0 0 0 0 0 0 0 0 0 0 0 2 2 2 2 2 2 2 2 2 2 2 2 2 2 2 2 1 1 1 1 1 1 1 1 1 1 1 1 1 1 1 1 3 3 3 3 3 3 3 3 3 3 3 3 3 3 3 3 1 1 1 1 1 1 1 1 1 1 1 1 1 1 1 1 3 3 3 3 3 3 3 3 3 3 3 3 3 3 3 3 0 0 0 0 0 0 0 0 0 0 0 0 0 0 0 0 2 2 2 2 2 2 2 2 2 2 2 2 2 2 2 2 0 0 0 0 0 0 0 0 0 0 0 0 0 0 0 0 2 2 2 2 2 2 2 2 2 2 2 2 2 2 2 2 1 1 1 1 1 1 1 1 1 1 1 1 1 1 1 1 3 3 3 3 3 3 3 3 3 3 3 3 3 3 3 3 1 1 1 1 1 1 1 1 1 1 1 1 1 1 1 1 3 3 3 3 3 3 3 3 3 3 3 3 3 3 3 3
ph31 = 0 2 2 0 1 3 3 1 3 1 1 3 0 2 2 0 2 0 0 2 3 1 1 3 1 3 3 1 2 0 0 2 2 0 0 2 3 1 1 3 1 3 3 1 2 0 0 2 0 2 2 0 1 3 3 1 3 1 1 3 0 2 2 0 2 0 0 2 3 1 1 3 1 3 3 1 2 0 0 2 0 2 2 0 1 3 3 1 3 1 1 3 0 2 2 0 0 2 2 0 1 3 3 1 3 1 1 3 0 2 2 0 2 0 0 2 3 1 1 3 1 3 3 1 2 0 0 2 0 2 2 0 1 3 3 1 3 1 1 3 0 2 2 0 2 0 0 2 3 1 1 3 1 3 3 1 2 0 0 2 2 0 0 2 3 1 1 3 1 3 3 1 2 0 0 2 0 2 2 0 1 3 3 1 3 1 1 3 0 2 2 0 2 0 0 2 3 1 1 3 1 3 3 1 2 0 0 2 0 2 2 0 1 3 3 1 3 1 1 3 0 2 2 0 0 2 2 0 1 3 3 1 3 1 1 3 0 2 2 0 2 0 0 2 3 1 1 3 1 3 3 1 2 0 0 2

;pl1 : f1 channel - power level for pulse (default)
;p1  : f1 channel -  90 degree high power pulse
;p19 : gradient pulse 2 (spoil gradient)
;p30 : gradient pulse (little DELTA*0.5)
;d1  : relaxation delay; 1-5 * T1
;d16 : delay for gradient recovery
;d20 : diffusion time (big DELTA)
;gpnam1: SMSQ10.100
;gpnam2: SMSQ10.100
;gpz1 : diffusion encoding gradient (100%)
;gpz2 : Oneshot CTP gradients

;cnst14 : alpha, typically 0.2
;cnst15 : 1+alpha
;cnst16 : 1-alpha
;cnst17 : Dtau
;cnst18 : dosytimecubed

;NS : 1 * n
;DS : 1 * m
;td1: number of experiments
;FnMODE: QF

;use AU-program dosy to calculate gradient ramp-file Difframp
;        use xf2 and DOSY processing
;        use "setdiffparmUoM" if 'setdiffparm' does not work
;    or  use "setdiffparm STEbp" but this gives slighly distorted diffusion coefficients 
\end{minted}

\subsubsection{1D Oneshot}

\begin{minted}{text}
; doneshot_1d_jy
;
; download from: https://git.io/JUHOP
;
; 1D version of Doneshot
; Jonathan Yong - University of Oxford - 26 Aug 2020

; modified from
; Doneshot
; DOSY pulse sequence
; Ralph Adams, Juan Aguilar, Robert Evans, Mathias Nilsson and Gareth Morris
; University of Manchester
; Release 1.0c (27Mar2012)
; 
; Source citation:
; M.D. Pelta, G.A. Morris, M.J. Stchedroff, S.J. Hammond, Magn. Reson. Chem. 40 (2002) S147-S152.
;
; Other relevant papers that could be of use include:
; A. Botana, J.A. Aguilar, M. Nilsson, G.A. Morris, J. Magn. Reson. 208 (2011) 270-278.
;
;$CLASS=HighRes
;$DIM=1D
;$TYPE=
;$SUBTYPE=
;$COMMENT=

#include <Avance.incl>
#include <Grad.incl>
#include <Delay.incl>

"cnst17=(2*p1+d16+p30)*0.000001"  ; Dtau

;JY - alpha is hardcoded as 0.2
"cnst14=0.2"

;Assuming half-sine gradient pulses [most common on Bruker systems]
"cnst18=0.000001*p30*2*0.000001*p30*2*(d20 - (2*0.000001*p30*(5-3*cnst14*cnst14)/16) - (cnst17*(1-cnst14*cnst14)/2) )" ; Dosytimecubed
"cnst15=1+cnst14"  ; 1 + alpha
"cnst16=1-cnst14"  ; 1 - alpha
"p2=p1*2"
"DELTA1=d20-4.0*p1-4.0*p30-5.0*d16-p19"

;The below line should be commented out for Topspin < 2.0
"acqt0 = -p1*2/3.1416"

1 ze
2 d1
3 50u UNBLKGRAD
  p19:gp2*-1			;Spoiler gradient balancing pulse
  d16
  p1 ph1				;1st 90
  p30:gp1*cnst16 		 	;1 - alpha
  d16
  p2 ph2				;First 180
  p30:gp1*cnst15*-1		;1 + alpha
  d16
  p1 ph3				; 2nd 90
  p30:gp1*cnst14*2		;Lock refocusing pulse pulse
  d16
  p19:gp2				;Spoiler gradient balancing pulse
  d16
  DELTA1
  p30:gp1*cnst14*2 		;Lock refocusing pulse pulse
  d16
  p1 ph4				; 3rd 90
  p30:gp1*cnst16   			; 1 - alpha
  d16
  p2 ph5
  p30:gp1*cnst15*-1			;1 + alpha
  d16
  4u BLKGRAD
  go=2 ph31
  d1 mc #0 to 2 F0(zd)
exit

ph1 = 0 2 0 2 1 3 1 3 0 2 0 2 1 3 1 3 0 2 0 2 1 3 1 3 0 2 0 2 1 3 1 3 0 2 0 2 1 3 1 3 0 2 0 2 1 3 1 3 0 2 0 2 1 3 1 3 0 2 0 2 1 3 1 3 0 2 0 2 1 3 1 3 0 2 0 2 1 3 1 3 0 2 0 2 1 3 1 3 0 2 0 2 1 3 1 3 0 2 0 2 1 3 1 3 0 2 0 2 1 3 1 3 0 2 0 2 1 3 1 3 0 2 0 2 1 3 1 3 0 2 0 2 1 3 1 3 0 2 0 2 1 3 1 3 0 2 0 2 1 3 1 3 0 2 0 2 1 3 1 3 0 2 0 2 1 3 1 3 0 2 0 2 1 3 1 3 0 2 0 2 1 3 1 3 0 2 0 2 1 3 1 3 0 2 0 2 1 3 1 3 0 2 0 2 1 3 1 3 0 2 0 2 1 3 1 3 0 2 0 2 1 3 1 3 0 2 0 2 1 3 1 3 0 2 0 2 1 3 1 3 0 2 0 2 1 3 1 3 0 2 0 2 1 3 1 3
ph2 = 0 0 0 0 0 0 0 0 0 0 0 0 0 0 0 0 0 0 0 0 0 0 0 0 0 0 0 0 0 0 0 0 0 0 0 0 0 0 0 0 0 0 0 0 0 0 0 0 0 0 0 0 0 0 0 0 0 0 0 0 0 0 0 0 0 0 0 0 0 0 0 0 0 0 0 0 0 0 0 0 0 0 0 0 0 0 0 0 0 0 0 0 0 0 0 0 0 0 0 0 0 0 0 0 0 0 0 0 0 0 0 0 0 0 0 0 0 0 0 0 0 0 0 0 0 0 0 0 2 2 2 2 2 2 2 2 2 2 2 2 2 2 2 2 2 2 2 2 2 2 2 2 2 2 2 2 2 2 2 2 2 2 2 2 2 2 2 2 2 2 2 2 2 2 2 2 2 2 2 2 2 2 2 2 2 2 2 2 2 2 2 2 2 2 2 2 2 2 2 2 2 2 2 2 2 2 2 2 2 2 2 2 2 2 2 2 2 2 2 2 2 2 2 2 2 2 2 2 2 2 2 2 2 2 2 2 2 2 2 2 2 2 2 2 2 2 2 2 2 2 2 2 2 2 2 2
ph3 = 0 0 0 0 0 0 0 0 0 0 0 0 0 0 0 0 0 0 0 0 0 0 0 0 0 0 0 0 0 0 0 0 2 2 2 2 2 2 2 2 2 2 2 2 2 2 2 2 2 2 2 2 2 2 2 2 2 2 2 2 2 2 2 2 0 0 0 0 0 0 0 0 0 0 0 0 0 0 0 0 0 0 0 0 0 0 0 0 0 0 0 0 0 0 0 0 2 2 2 2 2 2 2 2 2 2 2 2 2 2 2 2 2 2 2 2 2 2 2 2 2 2 2 2 2 2 2 2 0 0 0 0 0 0 0 0 0 0 0 0 0 0 0 0 0 0 0 0 0 0 0 0 0 0 0 0 0 0 0 0 2 2 2 2 2 2 2 2 2 2 2 2 2 2 2 2 2 2 2 2 2 2 2 2 2 2 2 2 2 2 2 2 0 0 0 0 0 0 0 0 0 0 0 0 0 0 0 0 0 0 0 0 0 0 0 0 0 0 0 0 0 0 0 0 2 2 2 2 2 2 2 2 2 2 2 2 2 2 2 2 2 2 2 2 2 2 2 2 2 2 2 2 2 2 2 2
ph4 = 0 0 2 2 0 0 2 2 1 1 3 3 1 1 3 3 2 2 0 0 2 2 0 0 3 3 1 1 3 3 1 1 0 0 2 2 0 0 2 2 1 1 3 3 1 1 3 3 2 2 0 0 2 2 0 0 3 3 1 1 3 3 1 1 0 0 2 2 0 0 2 2 1 1 3 3 1 1 3 3 2 2 0 0 2 2 0 0 3 3 1 1 3 3 1 1 0 0 2 2 0 0 2 2 1 1 3 3 1 1 3 3 2 2 0 0 2 2 0 0 3 3 1 1 3 3 1 1 0 0 2 2 0 0 2 2 1 1 3 3 1 1 3 3 2 2 0 0 2 2 0 0 3 3 1 1 3 3 1 1 0 0 2 2 0 0 2 2 1 1 3 3 1 1 3 3 2 2 0 0 2 2 0 0 3 3 1 1 3 3 1 1 0 0 2 2 0 0 2 2 1 1 3 3 1 1 3 3 2 2 0 0 2 2 0 0 3 3 1 1 3 3 1 1 0 0 2 2 0 0 2 2 1 1 3 3 1 1 3 3 2 2 0 0 2 2 0 0 3 3 1 1 3 3 1 1
ph5 =0 0 0 0 0 0 0 0 0 0 0 0 0 0 0 0 2 2 2 2 2 2 2 2 2 2 2 2 2 2 2 2 0 0 0 0 0 0 0 0 0 0 0 0 0 0 0 0 2 2 2 2 2 2 2 2 2 2 2 2 2 2 2 2 1 1 1 1 1 1 1 1 1 1 1 1 1 1 1 1 3 3 3 3 3 3 3 3 3 3 3 3 3 3 3 3 1 1 1 1 1 1 1 1 1 1 1 1 1 1 1 1 3 3 3 3 3 3 3 3 3 3 3 3 3 3 3 3 0 0 0 0 0 0 0 0 0 0 0 0 0 0 0 0 2 2 2 2 2 2 2 2 2 2 2 2 2 2 2 2 0 0 0 0 0 0 0 0 0 0 0 0 0 0 0 0 2 2 2 2 2 2 2 2 2 2 2 2 2 2 2 2 1 1 1 1 1 1 1 1 1 1 1 1 1 1 1 1 3 3 3 3 3 3 3 3 3 3 3 3 3 3 3 3 1 1 1 1 1 1 1 1 1 1 1 1 1 1 1 1 3 3 3 3 3 3 3 3 3 3 3 3 3 3 3 3
ph31 = 0 2 2 0 1 3 3 1 3 1 1 3 0 2 2 0 2 0 0 2 3 1 1 3 1 3 3 1 2 0 0 2 2 0 0 2 3 1 1 3 1 3 3 1 2 0 0 2 0 2 2 0 1 3 3 1 3 1 1 3 0 2 2 0 2 0 0 2 3 1 1 3 1 3 3 1 2 0 0 2 0 2 2 0 1 3 3 1 3 1 1 3 0 2 2 0 0 2 2 0 1 3 3 1 3 1 1 3 0 2 2 0 2 0 0 2 3 1 1 3 1 3 3 1 2 0 0 2 0 2 2 0 1 3 3 1 3 1 1 3 0 2 2 0 2 0 0 2 3 1 1 3 1 3 3 1 2 0 0 2 2 0 0 2 3 1 1 3 1 3 3 1 2 0 0 2 0 2 2 0 1 3 3 1 3 1 1 3 0 2 2 0 2 0 0 2 3 1 1 3 1 3 3 1 2 0 0 2 0 2 2 0 1 3 3 1 3 1 1 3 0 2 2 0 0 2 2 0 1 3 3 1 3 1 1 3 0 2 2 0 2 0 0 2 3 1 1 3 1 3 3 1 2 0 0 2

;pl1: f1 channel - power level for pulse (default)
;p1 : f1 channel -  90 degree high power pulse
;p19: gradient pulse 2 (spoil gradient)
;p30: gradient pulse (little DELTA*0.5)
;d1 : relaxation delay; 1-5 * T1
;d16: delay for gradient recovery
;d20: diffusion time (big DELTA)

;cnst14: alpha, typically 0.2
;cnst15: 1+alpha
;cnst16: 1-alpha
;cnst17: Dtau
;cnst18: dosytimecubed

;ns: 1*n
;ds: 1*m
;td1: number of experiments

;gpnam1: SMSQ10.100
;gpnam2: SMSQ10.100
;gpz1: DOSY gradient strength. DO NOT EXCEED 80
;gpz2: Oneshot CTP gradients
\end{minted}

\section{Documentation}

Finally, the current version of the \Poise{} documentation is attached here.
This is meant to act as a user guide, with specific instructions on how to set up and run optimisations.

Please note that the latest version of the documentation can always be found online at

\begin{minted}{text}
https://foroozandehgroup.github.io/nmrpoise
\end{minted}

and a PDF version of this can be downloaded at

\begin{minted}{text}
https://foroozandehgroup.github.io/nmrpoise/poise.pdf
\end{minted}

% Manually insert the final documentation PDF here.
% We could put it in using the pdfpages package, but that causes hyperlinks, etc.
% to be lost.



\section*{Supplementary material}

3. Check if `TI` is equal to that sentinel value. If it is, then quit unceremoniously with an error message of your choice.

Briefly, here is an example of this strategy in action. We show a Python script here, but the AU script is essentially the same, just with different function names.

```python
PUTPAR("TI", "poise")  # here "poise" serves as the sentinel value.
XCMD("poise p1cal -q")

# Here POISE should be done. If it succeeded then TI will no longer be "poise".
if GETPAR("TI") == "poise":
    raise RuntimeError("POISE failed!")

# After this you can continue with whatever you wanted to do.
```

# Chapter 6

# Builtin cost functions

POISE comes with a few, very basic, builtin cost functions. These largely mirror those that are in TopSpin's native `popt` screen.

---

**Note:** Just like in `popt`, it is possible to use the `dpl` command in TopSpin to select a *portion* of the spectrum to be optimised. This stores the left and right region of the currently active view to the parameters `F1P` and `F2P` respectively. This works for all the builtin cost functions except for `noe_1d`.

More generally, any cost function that uses any of the `get1d` or `get2d` functions will respect the bounds placed in `F1P` and `F2P`. See *Custom cost functions* for a more in-depth explanation.

---

## 6.1 minabsint

Seeks to minimise the intensity of the magnitude-mode spectrum. The intensity is measured by integration of the entire spectral region, i.e. summation of every point.

Note that this is different from the `MAGMIN` criterion in `popt`, which (from what the Bruker documentation suggests) seeks to minimise the highest point in the magnitude-mode spectrum.

## 6.2 maxabsint

Seeks to maximise the intensity of the magnitude-mode spectrum.

## 6.3 minrealint

Seeks to minimise the intensity of the real spectrum (this is probably equivalent to `INTMIN` in `popt`).

Note that this does *not* behave in the same way as `minabsint`. Because the real spectrum can have negative peaks, this essentially tries to maximise the intensity of negative peaks.

## 6.4  maxrealint

Seeks to maximise the intensity of the real spectrum (equivalent to `INTMAX`).

## 6.5  zerorealint

Seeks to make the intensity of the real spectrum as close as possible to zero (equivalent to `ZERO`).

## 6.6  noe_1d

Seeks to minimise the intensity of the spectrum, *except* for a region of 50 Hz centred on the parameter `SPOFFS2` (which corresponds to the frequency of the selective pulse).

Since NOE crosspeaks are typically negative (and `apk` typically phases them to be so), this essentially seeks to maximise the intensity of the crosspeaks.

Please see the POISE paper for example usage.

## 6.7  epsi_gradient_drift

Performs EPSI processing on a 1D FID, and seeks to minimise the 'drift' seen in the echo positions. This is only valid when the echo locations are *supposed* to be constant, e.g. when there is no indirect-dimension evolution period (otherwise, echoes will be observed at indirect-dimension frequencies). This is to be used for optimising the positive/negative gradient balance in EPSI acquisitions.

Please see the POISE paper for example usage.

# Chapter 7

# Builtin AU programmes

POISE is bundled with two basic AU programmes for 1D and 2D spectra respectively. These are mostly self-explanatory, so the text is just given below. Of course, you can write your own AU programmes to be used with POISE routines.

These are the "default" AU programmes that POISE uses (for 1D and 2D spectra respectively) if an AU programme is not specified with the routine.

## 7.1 poise_1d

```
ZG
EFP
APBK
QUIT
```

The APBK command on older versions of TopSpin falls back to APK then ABS, so can be safely used.

## 7.2 poise_2d

```
ZG
XFB
XCMD("apk2d")
ABS2
QUIT
```

# Chapter 8

# Custom cost functions

All user-defined cost functions are stored inside the file:

```
$TS/exp/stan/nmr/py/user/poise_backend/costfunctions_user.py
```

where `$TS` is your TopSpin installation path. In order to modify or add cost functions, you will need to edit this file (with your favourite text editor or IDE).

The corresponding file containing builtin cost functions is `costfunctions.py`. You *can* edit this file directly: if you add a cost function there, it will work. However, there are two risks with this. Firstly, if you ever reinstall POISE, this file will be reset to the default (whereas `costfunctions_user.py` will not). Secondly, any cost functions defined in `costfunctions_user.py` will shadow (i.e. take priority over) the cost functions defined in `costfunctions.py` if they have the same name.

## 8.1  The rules for cost functions

Cost functions are defined as a standard Python 3 function which takes no parameters and returns a float (the value of the cost function).

1. **Do write a useful docstring if possible**: this docstring will be shown to the user when they type `poise -1` into TopSpin (which lists all available cost functions and routines).

2. **The spectrum under optimisation, as well as acquisition parameters, can be accessed via helper functions.** These are described more fully below.

3. **Never print anything inside a cost function directly to stdout**. This will cause the optimisation to stop. If you want to perform debugging, use the `log` function described below.

4. **To terminate the optimisation prematurely and return the best point found so far, raise Cost-FunctionError().** See below for more information.

## 8.2 Accessing spectra and parameters

The most primitive way of accessing "outside" information is through the class `_g`, which is imported from `shared.py` and contains a series of global variables reflecting the current optimisation. For example, `_g.p_spectrum` is the path to the procno folder: you can read and parse the `1r` file inside this to get the real spectrum as a `numpy.ndarray` (for example).

**class** `nmrpoise.poise_backend.shared.`**`_g`**

>   Class to store the "global" variables.

>   **Attributes**

>>   **optimiser**  [str from {'nm', 'mds', 'bobyqa'}] The optimiser being used.

>>   **routine_id**  [str] The name of the routine being used.

>>   **p_spectrum**  [`Path`] The path to the procno folder of the spectrum just acquired. (e.g. `/path/to/data/1/pdata/1`)

>>   **p_optlog**  [`Path`] The path to the currently active `poise.log` file.

>>   **p_errlog**  [`Path`] The path to the currently active `poise_err_backend.log` file.

>>   **maxfev**  [int] The maximum number of function evaluations specified by the user. Can be zero, indicating no limit (beyond the hard limit of 500 times the number of parameters).

>>   **p_poise**  [`Path`] The path to the `$TS/exp/stan/nmr/py/user/poise_backend` folder.

>>   **spec_f1p**  [float or tuple of float] The F1P parameter. For a 1D spectrum this is a float. For a 2D spectrum this is a tuple of floats (`indirect`, `direct`) corresponding to the values of `F1P` in both spectral dimensions.

>>   **spec_f2p**  [float or tuple of float] The F2P parameter.

>>   **xvals**  [list of ndarray] The points sampled during the optimisation, in chronological order. These are ndarrays which contain the values of the parameters being optimised at each spectrum acquisition. The parameters are ordered in the same way as specified in the routine.

>>   **fvals**  [ndarray] The values of the cost functions calculated at each stage of the optimisation.

However, this is quite tedious and error-prone, so there are a number of helper methods which use these primitives. All the existing cost functions (inside `costfunctions.py`) only use these helper methods. All of these methods are stored inside `cfhelpers.py` and are already imported by default.

The ones you are likely to use are the following:

`nmrpoise.poise_backend.cfhelpers.`**`make_p_spec`**(*path=None*, *expno=None*, *procno=None*)

>   Constructs a `Path` object corresponding to the procno folder <path>/<expno>/pdata/<procno>. If parameters are not passed, they are inherited from the currently active spectrum (`_g.p_spectrum`).

>   Thus, for example, make_p_spec(expno=1, procno=1) returns a path to the spectrum with EXPNO 1 and PROCNO 1, but with the same name as the currently active spectrum.

>   **Parameters**

>>   **path**  [str or `Path`, optional] Path to the main folder of the spectrum (one level above the expno folders).

>>   **expno**  [int, optional]

>>   **procno**  [int, optional]

**Returns**

**p_spec** [`Path`] Path pointing to the requested spectrum.

nmrpoise.poise_backend.cfhelpers.**get1d_fid**(*remove_grpdly=True*, *p_spec=None*)
    Returns the FID as a `ndarray`.

**Parameters**

**remove_grpdly** [bool, optional] Whether to remove the group delay (to be precise, it is shifted to the end of the FID). Defaults to True.

**p_spec** [`Path`, optional] Path to the procno folder of interest. (The FID is taken from the expno folder two levels up.) Defaults to the currently active spectrum (i.e. `_g.p_spectrum`).

**Returns**

**ndarray** Complex-valued array containing the FID.

nmrpoise.poise_backend.cfhelpers.**get1d_real**(*bounds=''*, *p_spec=None*)
    Return the real spectrum as a `ndarray`. This function accounts for TopSpin's `NC_PROC` variable, scaling the spectrum intensity accordingly.

Note that this function only works for 1D spectra. It does *not* work for 1D projections of 2D spectra. If you want to work with projections, you can use `get2d_rr` to get the full 2D spectrum, then manipulate it using numpy functions as appropriate. A documented example can be found in the `asaphsqc()` function in `costfunctions.py` (commented out by default).

The *bounds* parameter may be specified in the following formats:

- between 5 and 8 ppm: bounds=`"5..8"` OR bounds=`(5, 8)`
- greater than 9.3 ppm: bounds=`"9.3.."` OR bounds=`(9.3, None)`
- less than -2 ppm: bounds=`"..-2"` OR bounds=`(None, -2)`

**Parameters**

**bounds** [str or tuple, optional] String or tuple describing the region of interest. See above for examples. If no bounds are provided, uses the F1P and F2P processing parameters, which can be specified via `dpl`. If these are not specified, defaults to the whole spectrum.

**p_spec** [`Path`, optional] Path to the procno folder of interest. Defaults to the currently active spectrum (i.e. `_g.p_spectrum`).

**Returns**

**ndarray** Array containing the spectrum or the desired section of it (if bounds were specified).

nmrpoise.poise_backend.cfhelpers.**get1d_imag**(*bounds=''*, *p_spec=None*)
    Same as `get1d_real`, except that it reads the imaginary spectrum.

nmrpoise.poise_backend.cfhelpers.**get2d_rr**(*f1_bounds=''*, *f2_bounds=''*, *p_spec=None*)

> Return the real part of the 2D spectrum (the "RR" quadrant) as a 2D `ndarray`. This function takes into account the `NC_PROC` value in TopSpin's processing parameters.
>
> The *f1_bounds* and *f2_bounds* parameters may be specified in the following formats:
>
> - between 5 and 8 ppm: f1_bounds="5..8" OR f1_bounds=(5, 8)
>
> - greater than 9.3 ppm: f1_bounds="9.3.." OR f1_bounds=(9.3, None)
>
> - less than -2 ppm: f1_bounds="..-2" OR f1_bounds=(None, -2)
>
> > **Parameters**
> >
> > > **f1_bounds** [str or tuple, optional] String or tuple describing the indirect-dimension region of interest. See above for examples. If no bounds are provided, uses the `1 F1P` and `1 F2P` processing parameters, which can be specified via `dpl`. If these are not specified, defaults to the whole spectrum.
> > >
> > > **f2_bounds** [str or tuple, optional] String or tuple describing the direct-dimension region of interest. See above for examples. If no bounds are provided, uses the `2 F1P` and `2 F2P` processing parameters, which can be specified via `dpl`. If these are not specified, defaults to the whole spectrum.
> > >
> > > **p_spec** [`Path`, optional] Path to the procno folder of interest. Defaults to the currently active spectrum (i.e. `_g.p_spectrum`).
> >
> > **Returns**
> >
> > > `ndarray` 2D array containing the spectrum or the desired section of it (if *f1_bounds* or *f2_bounds* were specified).

nmrpoise.poise_backend.cfhelpers.**get2d_ri**(*f1_bounds=''*, *f2_bounds=''*, *p_spec=None*)

> Same as `get2d_rr`, except that it reads the '2ri' file.

nmrpoise.poise_backend.cfhelpers.**get2d_ir**(*f1_bounds=''*, *f2_bounds=''*, *p_spec=None*)

> Same as `get2d_rr`, except that it reads the '2ir' file.

nmrpoise.poise_backend.cfhelpers.**get2d_ii**(*f1_bounds=''*, *f2_bounds=''*, *p_spec=None*)

> Same as `get2d_rr`, except that it reads the '2ii' file.

nmrpoise.poise_backend.cfhelpers.**getpar**(*par*, *p_spec=None*)

> Obtains the value of a numeric (acquisition or processing) parameter. Non-numeric parameters (i.e. strings) are not currently accessible! Works for both 1D and 2D spectra (see return type below), but nothing higher.
>
> > **Parameters**

> **par** [str] Name of the parameter.
>
> **p_spec** [`Path`, optional] Path to the procno folder of interest. Defaults to the currently active spectrum (i.e. `_g.p_spectrum`).
>
> **Returns**
>
> **float or** `ndarray` Value(s) of the requested parameter. None if the given parameter was not found.
>
> For parameters that exist for both dimensions of 2D spectra, getpar() returns an ndarray consisting of (f1_value, f2_value). Otherwise (for 1D spectra, or for 2D parameters which only apply to the direct dimension), getpar() returns a float.

nmrpoise.poise_backend.cfhelpers.**getndim**(*p_spec=None*)
 Obtains the dimensionality of the spectrum, i.e. the status value of PARMODE, plus one (becaus PARMODE is 0 for 1D spectra, etc.)

> **Parameters**
>
> **p_spec** [`Path`, optional] Path to the procno folder of interest. Defaults to the currently active spectrum (i.e. `_g.p_spectrum`).
>
> **Returns**
>
> **int** Dimensionality of the spectrum.

nmrpoise.poise_backend.cfhelpers.**getnfev**()
 Returns the number of NMR spectra evaluated so far. This will be equal to 1 (not 0) the first time the cost function is called, since the cost function is only called after the NMR spectrum has been acquired.

## 8.3 Logging

As noted above, printing anything to `stdout` will cause the optimisation to crash. The reason for this is because `stdout` is reserved for communication between the POISE backend (the Python 3 component) and frontend (which runs in TopSpin). Please use `log()` instead, which will print to the `poise.log` file in the expno folder. It works in exactly the same way as the familiar `print()`, and accepts the same kind of arguments.

nmrpoise.poise_backend.cfhelpers.**log**(*\*args*)
 Prints something to the poise.log file.

 If this is called from inside a cost function, the text is printed *before* the cost function is evaluated, so will appear above the corresponding function evaluation.

> **Parameters**
>
> **args** The arguments that `log()` takes are exactly the same as those of `print()`.
>
> **Returns**
>
> **None**

## 8.4 Premature termination

In order to terminate an optimisation prematurely, you can raise any exception you like, for example with `raise ValueError`. POISE will stop the optimisation, and the error will be displayed in TopSpin. The drawback of this naive approach is that *no information from the incomplete optimisation will be retained*. That means that even if you have found a point that is substantially better than the initial point, it will not be saved.

If you want to terminate *and* return the best point found so far, please raise a `CostFunctionError` instead of any other exception (such as `ValueError`). `CostFunctionError` takes up to 2 arguments: the number of arguments to supply depends on whether you want to include the final point which caused the error to be raised.

If you want to discard the last point, i.e. just stop the optimisation right away, then raise a `CostFunctionError` with just 1 argument. The argument should be the message that you want to display to the user:

```python
def cost_function():
    cost_fn_value = foo()    # whatever calculation you want here
    if some_bad_condition:
        raise CostFunctionError("Some bad condition occurred!")
    return cost_fn_value
```

It is possible, and probably advisable, to use this string to show the user helpful information (further steps to take, or the value of the cost function, for example).

Alternatively, you may want the current point (and the corresponding cost function value) to be saved as part of the optimisation. For example, it may be the case that a certain threshold is "good enough" for the cost function and any value below that is acceptable. In that situation, you would want to raise CostFunctionError once the cost function goes below that threshold, but *also* save that point as the best value. To do so, pass the value of the cost function as the *second* parameter when raising CostFunctionError:

```python
def cost_function():
    cost_fn_value = foo()    # whatever calculation you want here
    if cost_fn_value < threshold:
        raise CostFunctionError("The cost function is below the threshold.",
                                cost_fn_value)
    # Note that we still need the return statement, because it will be used
    # if cost_fn_value is greater than the threshold.
    return cost_fn_value
```

The value passed as the second argument can in general be any number. If you want to make sure that the final point (which raised the CostFunctionError) is *always* chosen to be the optimal point, then you can do this:

```python
raise CostFunctionError("A message here", -np.inf)
```

Because `-np.inf` is smaller than any other possible number, it will always be picked as the "best" point.

## 8.5  Examples

There are a number of cost functions which ship with POISE. These can all be found inside the `costfunctions.py` file referred to above. This file also contains a number of more specialised cost functions, which were used for the examples in the POISE paper. (Instead of opening the file, you can also find the source code on GitHub.)

A number of these are thoroughly commented with detailed explanations; do consider checking these out if you want more guidance on how to write your own cost function.

# Chapter 9

# Developer notes

POISE does not contain all that much code. However, if you are thinking of modifying it, it helps to have an understanding of the distinction between the *frontend* and the *backend*. This graphic (taken from an old presentation) is a slightly more technical description of the flowchart presented in the POISE paper, and shows which part is responsible for which step. It's a good starting point for understanding how POISE works internally.

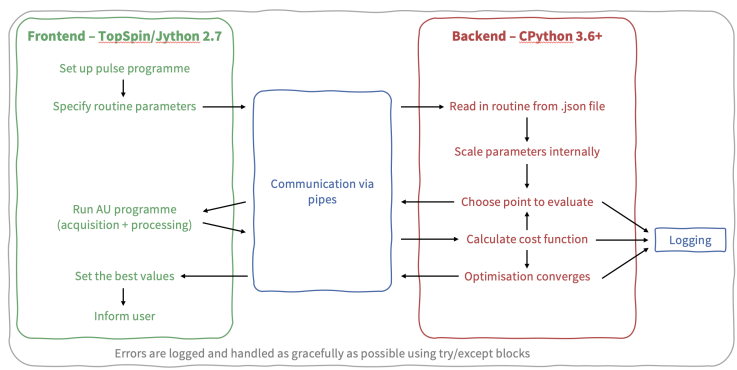

Beyond this, we recommend reading the source code of POISE: start at the frontend script (`poise.py`, which is run from inside TopSpin), then go to the backend script (`poise_backend/backend.py`) at the appropriate time (when the frontend launches it as a subprocess). The source code is quite thoroughly commented.

There are two main things to point out. Probably the most important thing worth mentioning is the location of the relevant files. The backend script is always ran from `$TS/py/user/poise_backend` (putting it here allows the frontend to access it much more easily). The entire `$TS/py/user/poise_backend` folder is treated as if it is a Python package, by virtue of some code near the top of `poise_backend.py`:

```python
if __name__ == "__main__" and __package__ is None:
    __package__ = "poise_backend"
    sys.path.insert(1, str(Path(__file__).parents[1].resolve()))
    __import__(__package__)
```

This allows relative imports of the other files in the same directory, such as `costfunctions.py`, in which the user-defined cost functions reside.

---

**Note:** These backend files *also* reside inside the Python 3 `site-packages` directory, where all packages are installed to. However, these files will *never* be used by POISE. So, there is nothing to be gained by modifying these at all.

---

The only file inside the `site-packages` directory which has any effect is `nmrpoise/__init__.py`, where the `parse_log()` function is defined. This allows you to (for example) run:

```
>>> from nmrpoise import parse_log
```

Python will look inside the `site-packages` directory, *not* `$TS/py/user`, to find this function.

The second thing is that you should never, ever, do anything with the backend's `stdin` and `stdout`, because these are exclusively reserved for communication with the frontend. So you should never print anything from the backend, since the frontend will just interpret that as an error. This applies to all files inside the `$TS/py/user/poise_backend` directory, including cost functions, which is why custom cost functions should always use POISE's `log` function instead of plain old `print`.

**Note:** If you really just want to do some quick-and-dirty debugging, you *can* actually use this behaviour to your advantage. The frontend will echo any "invalid" message it receives from the backend, so if you print some unexpected text from the backend (on purpose), you should see it pop up as a TopSpin message when you run an optimisation. This is slightly less hassle than printing to a file and opening the file.

## 9.1 Testing

Tests are carried out using the excellent pytest and tox tools. To run all tests, simply run:

```
pip install tox
tox
```

from anywhere inside the `nmrpoise` directory. This runs tests on Python 3.6, 3.7, and 3.8. If you only have one of these versions, use:

```
tox -e py38   # or py36 or py37
```

To build the Sphinx documentation, use:

```
tox -e docs
```

The HTML documentation will be built in `docs/dirhtml`, and the PDF documentation in `docs/latex` (this assumes you have a working installation of `pdflatex` on your system).


\end{document}